\ifdefined\XeTeXversion\else\pdfoutput=1\fi
\documentclass{uai2026}
\newif\ifanon
\anonfalse
\usepackage[american]{babel}
\usepackage{natbib}
\usepackage{mathtools}
\usepackage{amssymb,amsthm}
\usepackage{booktabs}
\usepackage{longtable}
\usepackage{float}
\usepackage{placeins}
\usepackage{algorithm}
\usepackage{algpseudocode}
\usepackage{tikz}
\usetikzlibrary{positioning,arrows.meta}
\usepackage{graphicx}
\usepackage{xcolor}
\usepackage{xurl}
\hypersetup{colorlinks=false,pdfborder={0 0 1},linkbordercolor={1 0 0},citebordercolor={0 1 0}}

\newtheorem{theorem}{Theorem}
\newtheorem{lemma}{Lemma}
\newtheorem{proposition}{Proposition}
\newtheorem{corollary}{Corollary}
\theoremstyle{definition}
\newtheorem{definition}{Definition}
\newtheorem{assumption}{Assumption}
\newtheorem{condition}{Condition}

\newtheorem{example}{Example}
\theoremstyle{remark}
\newtheorem{remark}{Remark}
\theoremstyle{plain}

\newcommand{\PP}{\mathbb{P}}
\newcommand{\EE}{\mathbb{E}}
\newcommand{\G}{\mathcal{G}}
\newcommand{\F}{\mathcal{F}}
\newcommand{\Hc}{\mathcal{H}}
\newcommand{\Ic}{\mathcal{I}}
\newcommand{\Jc}{\mathcal{J}}
\newcommand{\Xc}{\mathcal{X}}

\newcommand{\Top}{\mathrm{Top}}
\newcommand{\Unif}{\mathrm{Unif}}
\newcommand{\Exp}{\mathrm{Exp}}

\newcommand{\ind}{\mathbf{1}}

\DeclareMathOperator*{\argmax}{arg\,max}

\newcommand{\EfourteenPtwoMaxEfour}{0.012}

\newcommand{\EfourteenPzeroMax}{0.843}

\newcommand{\EfifteenLowBest}{0.146}

\newcommand{\AudMonoFwerMax}{0.000}
\newcommand{\AudMonoEcdfMax}{1.2\times10^{-4}}

\newcommand{\AudOracleFwerMin}{1.000}
\newcommand{\AudOracleDnMin}{0.050}
\newcommand{\AudOracleDnMax}{0.241}

\newcommand{\AudGthreeFwerMax}{0.000}

\newcommand{\AudNdocs}{9000}
\newcommand{\AudBHundredGthreetwo}{0.188}
\newcommand{\AudBHundredGthreefive}{0.855}

\newcommand{\AudBFourHundredGthreefive}{1.000}
\newcommand{\AudBGoneMax}{0.000}

\newcommand{\TtScanMax}{0.004}

\newcommand{\TtMixFwerMax}{0.002}

\newcommand{\TtOptOneDthousand}{0.586}

\newcommand{\TtOptSevenH}{0.29}

\newcommand{\TtQwenOneDthousand}{0.617}

\newcommand{\TtQwenSevenH}{0.57}

\newcommand{\TaskAuditReportCMax}{0.16\%}
\newcommand{\TaskAuditOriginMax}{no document has}

\newcommand{\RuntimeReferenceN}{1024}
\newcommand{\RuntimePipelineRangeMs}{7.04--36.5}
\newcommand{\RuntimeRepeats}{20}
\newcommand{\RuntimeMinN}{256}
\newcommand{\RuntimeMaxN}{8192}
\newcommand{\RuntimeLambda}{0.5}
\newcommand{\RuntimeTau}{2.5}
\newcommand{\RuntimeAlpha}{0.05}
\newcommand{\RuntimeEta}{1/4}
\newcommand{\RuntimeReportMin}{32}
\newcommand{\RuntimeReportMax}{512}
\newcommand{\RuntimeSignalPi}{0.1}
\newcommand{\RuntimeMixedScanSlope}{2.18}

\newcommand{\RuntimeMixedPipelineSlope}{1.95}
\newcommand{\RuntimeNullScanSlope}{1.97}

\newcommand{\RuntimeNullPipelineSlope}{1.73}

\newcommand{\PEnineReps}{300}
\newcommand{\PEnineAnyFalseA}{0.000}
\newcommand{\PEnineAnyFalseFG}{0.000}
\newcommand{\PEnineMidCovA}{0.996}
\newcommand{\PEnineMidCovFG}{0.794}

\newcommand{\PEoneConfigs}{48}

\newcommand{\FgSTwoWilson}{0.0076}

\newcommand{\TfOptSOne}{0.144}
\newcommand{\TfOptSOneCI}{[0.134, 0.154]}

\newcommand{\TfQwenSOne}{0.218}
\newcommand{\TfQwenSOneCI}{[0.202, 0.235]}

\newcommand{\TfSimKHMax}{0.006}

\newcommand{\TfGapSegMax}{0.004}

\newcommand{\TfOptInsPairs}{154}

\newcommand{\TfOptRetain}{0.405}
\newcommand{\TfRetainRef}{0.410}

\newcommand{\TfQwenInsPairs}{170}

\newcommand{\TfQwenRetain}{0.403}
\newcommand{\TfPairRatioMin}{2.4}
\newcommand{\TfPairRatioMax}{3.8}

\newcommand{\AtCap}{0.622}
\newcommand{\AtOptOneBelow}{0.364}
\newcommand{\AtOptOneOneMinusQ}{0.298}
\newcommand{\AtQwenOneBelow}{0.368}
\newcommand{\AtQwenOneOneMinusQ}{0.309}

\newcommand{\SelDemoSelOpt}{56.4\%}
\newcommand{\SelDemoMixAnyOpt}{40.4\%}
\newcommand{\SelDemoCostOpt}{11.8}
\newcommand{\SelDemoHumanMatchOpt}{11.0}
\newcommand{\SelDemoSelQwen}{70.8\%}
\newcommand{\SelDemoMixAnyQwen}{60.2\%}
\newcommand{\SelDemoCostQwen}{12.0}
\newcommand{\SelDemoHumanMatchQwen}{16.6}
\newcommand{\SelDemoFixedMax}{0.0\%}
\newcommand{\SelDemoCSMax}{0.2\%}
\newcommand{\SelDemoSynNullSel}{96.6\%}
\newcommand{\SelDemoSynCSMax}{0.4\%}

\newcommand{\BOneNumCells}{6}
\newcommand{\BOneNumKeys}{3}

\newcommand{\BOneEquivalenceMargin}{0.02}

\newcommand{\BTwoTemperature}{1.0}
\newcommand{\BTwoQZero}{pass}
\newcommand{\BTwoDropTwo}{0.09}
\newcommand{\BTwoHTwo}{3.03}
\newcommand{\BTwoQStar}{0.610}
\newcommand{\BTwoDOneTwoEight}{0.555}
\newcommand{\BTwoDThousand}{0.542}
\newcommand{\BTwoDThousandCI}{[0.529, 0.555]}

\newcommand{\BTwoDiff}{0.298}
\newcommand{\BTwoDiffCI}{[0.277, 0.319]}

\newcommand{\BTwoCSCover}{0.854}
\newcommand{\BTwoFGCover}{0.556}

\newcommand{\BTwoMixedCOneEvents}{0}
\newcommand{\BTwoMixedDocuments}{500}

\newcommand{\BTwoPurity}{--}

\newcommand{\EtwoOptDThousand}{0.570}

\newcommand{\EtwoOptCoverCS}{0.941}
\newcommand{\EtwoOptCoverFG}{0.702}
\newcommand{\EtwoOptDiff}{0.238}
\newcommand{\EtwoOptDiffCI}{[0.218, 0.260]}
\newcommand{\EtwoOptPurityCS}{0.950}
\newcommand{\EtwoOptPurityFG}{0.995}

\newcommand{\EtwoOptCoverRandCS}{0.010}
\newcommand{\EtwoOptCoverRandFG}{0.003}

\newcommand{\EtwoQwenDThousand}{0.658}

\newcommand{\EtwoQwenCoverCS}{0.956}
\newcommand{\EtwoQwenCoverFG}{0.768}
\newcommand{\EtwoQwenDiff}{0.188}
\newcommand{\EtwoQwenDiffCI}{[0.171, 0.205]}
\newcommand{\EtwoQwenPurityCS}{0.949}
\newcommand{\EtwoQwenPurityFG}{0.991}

\newcommand{\TrfPrimaryAlpha}{0.05}
\newcommand{\TrfPrimaryEta}{0.25}
\newcommand{\TrfOptSOne}{0.0532}
\newcommand{\TrfOptSOneCI}{[0.0509, 0.0553]}

\newcommand{\TrfQwenSOne}{0.0661}
\newcommand{\TrfQwenSOneCI}{[0.0636, 0.0686]}

\newcommand{\TrfOptSTwo}{0.0211}
\newcommand{\TrfOptSTwoCI}{[0.0183, 0.0238]}

\newcommand{\TrfQwenSTwo}{0.0333}
\newcommand{\TrfQwenSTwoCI}{[0.0305, 0.0366]}

\newcommand{\TrfOptMainCapShortASixEfficiency}{0.2109 [0.2076, 0.2142]}

\newcommand{\TrfOptMainCapShortASixCoverage}{0.7902 [0.7790, 0.8011]}

\newcommand{\TrfOptMainCapShortASixPurity}{0.9843 [0.9809, 0.9873]}

\newcommand{\TrfOptMainCapShortAFourEfficiency}{0.5514 [0.5424, 0.5597]}

\newcommand{\TrfOptMainCapShortAFourCoverage}{0.9173 [0.9094, 0.9249]}

\newcommand{\TrfOptMainCapShortAFourPurity}{0.8877 [0.8788, 0.8960]}

\newcommand{\TrfOptMainCapShortAEightEfficiency}{0.5496 [0.5406, 0.5580]}

\newcommand{\TrfOptMainCapShortAEightCoverage}{0.9226 [0.9142, 0.9300]}
\newcommand{\TrfOptMainCapShortAEightPurityValue}{0.8764}

\newcommand{\TrfOptMainCapShortAEightPurity}{0.8764 [0.8675, 0.8856]}

\newcommand{\TrfOptMainCapShortAEightLayerEfficiency}{0.5568 [0.5476, 0.5652]}

\newcommand{\TrfOptMainCapShortAEightLayerCoverage}{0.9202 [0.9119, 0.9279]}

\newcommand{\TrfOptMainCapShortAEightLayerPurity}{0.8632 [0.8531, 0.8733]}

\newcommand{\TrfOptMainCapShortAThreeEfficiency}{0.5247 [0.5165, 0.5323]}

\newcommand{\TrfOptMainCapShortAThreeCoverage}{0.9063 [0.8982, 0.9142]}

\newcommand{\TrfOptMainCapShortAThreePurity}{0.9604 [0.9545, 0.9658]}

\newcommand{\TrfOptMainCapShortATwoEfficiency}{0.5707 [0.5624, 0.5780]}

\newcommand{\TrfOptMainCapShortATwoCoverage}{0.9342 [0.9287, 0.9393]}
\newcommand{\TrfOptMainCapShortATwoPurityValue}{0.9550}

\newcommand{\TrfOptMainCapShortATwoPurity}{0.9550 [0.9489, 0.9604]}

\newcommand{\TrfQwenMainCapShortASixEfficiency}{0.1841 [0.1789, 0.1889]}

\newcommand{\TrfQwenMainCapShortASixCoverage}{0.6919 [0.6729, 0.7096]}

\newcommand{\TrfQwenMainCapShortASixPurity}{0.9867 [0.9838, 0.9896]}

\newcommand{\TrfQwenMainCapShortAFourEfficiency}{0.4836 [0.4737, 0.4932]}

\newcommand{\TrfQwenMainCapShortAFourCoverage}{0.8859 [0.8746, 0.8963]}

\newcommand{\TrfQwenMainCapShortAFourPurity}{0.8779 [0.8696, 0.8867]}

\newcommand{\TrfQwenMainCapShortAEightEfficiency}{0.4737 [0.4638, 0.4836]}

\newcommand{\TrfQwenMainCapShortAEightCoverage}{0.9056 [0.8934, 0.9178]}
\newcommand{\TrfQwenMainCapShortAEightPurityValue}{0.8266}

\newcommand{\TrfQwenMainCapShortAEightPurity}{0.8266 [0.8177, 0.8346]}

\newcommand{\TrfQwenMainCapShortAEightLayerEfficiency}{0.4833 [0.4734, 0.4932]}

\newcommand{\TrfQwenMainCapShortAEightLayerCoverage}{0.9087 [0.8976, 0.9194]}

\newcommand{\TrfQwenMainCapShortAEightLayerPurity}{0.8103 [0.8033, 0.8186]}

\newcommand{\TrfQwenMainCapShortAThreeEfficiency}{0.4646 [0.4564, 0.4728]}

\newcommand{\TrfQwenMainCapShortAThreeCoverage}{0.8758 [0.8661, 0.8838]}

\newcommand{\TrfQwenMainCapShortAThreePurity}{0.9693 [0.9633, 0.9746]}

\newcommand{\TrfQwenMainCapShortATwoEfficiency}{0.5071 [0.4985, 0.5159]}

\newcommand{\TrfQwenMainCapShortATwoCoverage}{0.9097 [0.9012, 0.9175]}
\newcommand{\TrfQwenMainCapShortATwoPurityValue}{0.9575}

\newcommand{\TrfQwenMainCapShortATwoPurity}{0.9575 [0.9512, 0.9635]}

\newcommand{\TrfOptSecondaryCapShortASixEfficiency}{0.0136 [0.0096, 0.0183]}

\newcommand{\TrfOptSecondaryCapShortASixCoverage}{0.0513 [0.0359, 0.0686]}

\newcommand{\TrfOptSecondaryCapShortASixPurity}{0.9890 [0.9745, 0.9993]}

\newcommand{\TrfOptSecondaryCapShortAFourEfficiency}{0.1246 [0.1092, 0.1411]}

\newcommand{\TrfOptSecondaryCapShortAFourCoverage}{0.3809 [0.3354, 0.4286]}

\newcommand{\TrfOptSecondaryCapShortAFourPurity}{0.9469 [0.9331, 0.9593]}

\newcommand{\TrfOptSecondaryCapShortAEightEfficiency}{0.1190 [0.1040, 0.1360]}

\newcommand{\TrfOptSecondaryCapShortAEightCoverage}{0.3695 [0.3230, 0.4196]}

\newcommand{\TrfOptSecondaryCapShortAEightPurity}{0.9317 [0.9164, 0.9466]}

\newcommand{\TrfOptSecondaryCapShortAEightLayerEfficiency}{0.1241 [0.1089, 0.1412]}

\newcommand{\TrfOptSecondaryCapShortAEightLayerCoverage}{0.3767 [0.3303, 0.4259]}

\newcommand{\TrfOptSecondaryCapShortAEightLayerPurity}{0.9271 [0.9107, 0.9423]}

\newcommand{\TrfOptSecondaryCapShortAThreeEfficiency}{0.1172 [0.1009, 0.1336]}

\newcommand{\TrfOptSecondaryCapShortAThreeCoverage}{0.3713 [0.3211, 0.4225]}

\newcommand{\TrfOptSecondaryCapShortAThreePurity}{0.9855 [0.9790, 0.9908]}

\newcommand{\TrfOptSecondaryCapShortATwoEfficiency}{0.1602 [0.1437, 0.1770]}

\newcommand{\TrfOptSecondaryCapShortATwoCoverage}{0.4872 [0.4406, 0.5325]}

\newcommand{\TrfOptSecondaryCapShortATwoPurity}{0.9800 [0.9715, 0.9869]}

\newcommand{\TrfQwenSecondaryCapShortASixEfficiency}{0.0104 [0.0077, 0.0134]}

\newcommand{\TrfQwenSecondaryCapShortASixCoverage}{0.0392 [0.0294, 0.0505]}

\newcommand{\TrfQwenSecondaryCapShortASixPurity}{0.9901 [0.9804, 0.9973]}

\newcommand{\TrfQwenSecondaryCapShortAFourEfficiency}{0.1036 [0.0921, 0.1148]}

\newcommand{\TrfQwenSecondaryCapShortAFourCoverage}{0.3169 [0.2826, 0.3489]}

\newcommand{\TrfQwenSecondaryCapShortAFourPurity}{0.9448 [0.9354, 0.9539]}

\newcommand{\TrfQwenSecondaryCapShortAEightEfficiency}{0.0924 [0.0815, 0.1027]}

\newcommand{\TrfQwenSecondaryCapShortAEightCoverage}{0.2998 [0.2651, 0.3309]}

\newcommand{\TrfQwenSecondaryCapShortAEightPurity}{0.9392 [0.9271, 0.9503]}

\newcommand{\TrfQwenSecondaryCapShortAEightLayerEfficiency}{0.0974 [0.0861, 0.1081]}

\newcommand{\TrfQwenSecondaryCapShortAEightLayerCoverage}{0.3093 [0.2754, 0.3418]}

\newcommand{\TrfQwenSecondaryCapShortAEightLayerPurity}{0.9219 [0.9082, 0.9350]}

\newcommand{\TrfQwenSecondaryCapShortAThreeEfficiency}{0.1068 [0.0947, 0.1188]}

\newcommand{\TrfQwenSecondaryCapShortAThreeCoverage}{0.3438 [0.3045, 0.3805]}

\newcommand{\TrfQwenSecondaryCapShortAThreePurity}{0.9935 [0.9907, 0.9958]}

\newcommand{\TrfQwenSecondaryCapShortATwoEfficiency}{0.1510 [0.1378, 0.1638]}

\newcommand{\TrfQwenSecondaryCapShortATwoCoverage}{0.4696 [0.4321, 0.5065]}

\newcommand{\TrfQwenSecondaryCapShortATwoPurity}{0.9823 [0.9770, 0.9866]}

\newcommand{\TrfOptMainCapShortAFourMinusASixEfficiency}{0.3405 [0.3330, 0.3473]}

\newcommand{\TrfOptMainCapShortAFourMinusASixCoverage}{0.1271 [0.1165, 0.1370]}

\newcommand{\TrfOptMainCapShortAFourMinusASixPurity}{-0.0966 [-0.1058, -0.0875]}

\newcommand{\TrfOptMainCapShortAEightMinusAFourEfficiency}{-0.0017 [-0.0034, -0.0003]}

\newcommand{\TrfOptMainCapShortAEightMinusAFourCoverage}{0.0053 [0.0006, 0.0103]}

\newcommand{\TrfOptMainCapShortAEightMinusAFourPurity}{-0.0113 [-0.0190, -0.0035]}

\newcommand{\TrfOptMainCapShortAThreeMinusAEightEfficiency}{-0.0249 [-0.0279, -0.0219]}

\newcommand{\TrfOptMainCapShortAThreeMinusAEightCoverage}{-0.0163 [-0.0218, -0.0106]}

\newcommand{\TrfOptMainCapShortAThreeMinusAEightPurity}{0.0840 [0.0738, 0.0938]}

\newcommand{\TrfOptMainCapShortATwoMinusAThreeEfficiency}{0.0460 [0.0445, 0.0474]}

\newcommand{\TrfOptMainCapShortATwoMinusAThreeCoverage}{0.0279 [0.0221, 0.0336]}

\newcommand{\TrfOptMainCapShortATwoMinusAThreePurity}{-0.0054 [-0.0084, -0.0023]}

\newcommand{\TrfQwenMainCapShortAFourMinusASixEfficiency}{0.2995 [0.2922, 0.3064]}

\newcommand{\TrfQwenMainCapShortAFourMinusASixCoverage}{0.1940 [0.1804, 0.2093]}

\newcommand{\TrfQwenMainCapShortAFourMinusASixPurity}{-0.1088 [-0.1180, -0.0992]}

\newcommand{\TrfQwenMainCapShortAEightMinusAFourEfficiency}{-0.0098 [-0.0117, -0.0081]}

\newcommand{\TrfQwenMainCapShortAEightMinusAFourCoverage}{0.0197 [0.0134, 0.0256]}

\newcommand{\TrfQwenMainCapShortAEightMinusAFourPurity}{-0.0513 [-0.0602, -0.0430]}

\newcommand{\TrfQwenMainCapShortAThreeMinusAEightEfficiency}{-0.0092 [-0.0124, -0.0055]}

\newcommand{\TrfQwenMainCapShortAThreeMinusAEightCoverage}{-0.0298 [-0.0386, -0.0204]}

\newcommand{\TrfQwenMainCapShortAThreeMinusAEightPurity}{0.1427 [0.1329, 0.1523]}

\newcommand{\TrfQwenMainCapShortATwoMinusAThreeEfficiency}{0.0425 [0.0414, 0.0438]}

\newcommand{\TrfQwenMainCapShortATwoMinusAThreeCoverage}{0.0339 [0.0291, 0.0390]}

\newcommand{\TrfQwenMainCapShortATwoMinusAThreePurity}{-0.0119 [-0.0157, -0.0080]}

\newcommand{\TrfOptSecondaryCapShortAFourMinusASixEfficiency}{0.1109 [0.0960, 0.1263]}

\newcommand{\TrfOptSecondaryCapShortAFourMinusASixCoverage}{0.3296 [0.2873, 0.3726]}

\newcommand{\TrfOptSecondaryCapShortAFourMinusASixPurity}{-0.0421 [-0.0591, -0.0233]}

\newcommand{\TrfOptSecondaryCapShortAEightMinusAFourEfficiency}{-0.0056 [-0.0099, -0.0013]}

\newcommand{\TrfOptSecondaryCapShortAEightMinusAFourCoverage}{-0.0115 [-0.0278, 0.0051]}

\newcommand{\TrfOptSecondaryCapShortAEightMinusAFourPurity}{-0.0152 [-0.0279, -0.0023]}

\newcommand{\TrfOptSecondaryCapShortAThreeMinusAEightEfficiency}{-0.0018 [-0.0090, 0.0051]}

\newcommand{\TrfOptSecondaryCapShortAThreeMinusAEightCoverage}{0.0019 [-0.0247, 0.0268]}

\newcommand{\TrfOptSecondaryCapShortAThreeMinusAEightPurity}{0.0538 [0.0381, 0.0693]}

\newcommand{\TrfOptSecondaryCapShortATwoMinusAThreeEfficiency}{0.0430 [0.0341, 0.0514]}

\newcommand{\TrfOptSecondaryCapShortATwoMinusAThreeCoverage}{0.1159 [0.0811, 0.1514]}

\newcommand{\TrfOptSecondaryCapShortATwoMinusAThreePurity}{-0.0055 [-0.0130, 0.0003]}

\newcommand{\TrfQwenSecondaryCapShortAFourMinusASixEfficiency}{0.0932 [0.0826, 0.1028]}

\newcommand{\TrfQwenSecondaryCapShortAFourMinusASixCoverage}{0.2778 [0.2465, 0.3068]}

\newcommand{\TrfQwenSecondaryCapShortAFourMinusASixPurity}{-0.0453 [-0.0574, -0.0307]}

\newcommand{\TrfQwenSecondaryCapShortAEightMinusAFourEfficiency}{-0.0112 [-0.0151, -0.0079]}

\newcommand{\TrfQwenSecondaryCapShortAEightMinusAFourCoverage}{-0.0172 [-0.0314, -0.0046]}

\newcommand{\TrfQwenSecondaryCapShortAEightMinusAFourPurity}{-0.0056 [-0.0173, 0.0057]}

\newcommand{\TrfQwenSecondaryCapShortAThreeMinusAEightEfficiency}{0.0144 [0.0095, 0.0194]}

\newcommand{\TrfQwenSecondaryCapShortAThreeMinusAEightCoverage}{0.0440 [0.0255, 0.0622]}

\newcommand{\TrfQwenSecondaryCapShortAThreeMinusAEightPurity}{0.0543 [0.0431, 0.0665]}

\newcommand{\TrfQwenSecondaryCapShortATwoMinusAThreeEfficiency}{0.0442 [0.0382, 0.0506]}

\newcommand{\TrfQwenSecondaryCapShortATwoMinusAThreeCoverage}{0.1258 [0.1028, 0.1511]}

\newcommand{\TrfQwenSecondaryCapShortATwoMinusAThreePurity}{-0.0112 [-0.0162, -0.0065]}

\newcommand{\TrfOptMainCapLongAThreeMinusAEightEfficiency}{0.0195 [0.0174, 0.0216]}

\newcommand{\TrfQwenMainCapLongAThreeMinusAEightEfficiency}{0.0328 [0.0306, 0.0351]}

\newcommand{\TrfRfSimMax}{0.004}
\newcommand{\TrfRfSimUpperMax}{0.014}
\newcommand{\TrfRfSimConfigs}{384}
\newcommand{\TrfRfSimDocs}{500}

\newcommand{\EtDocs}{200}
\newcommand{\EtZeroExcess}{0.0026}

\newcommand{\EtZeroCsCover}{0.920}
\newcommand{\EtZeroCsCoverCI}{[0.909, 0.930]}
\newcommand{\EtZeroFgCover}{0.704}
\newcommand{\EtZeroFgCoverCI}{[0.673, 0.732]}

\newcommand{\EtZeroDiff}{0.216}
\newcommand{\EtZeroDiffCI}{[0.188, 0.246]}
\newcommand{\EtZeroBard}{156.1}

\newcommand{\EtZeroKH}{287.9}
\newcommand{\EtTenExcess}{0.0025}
\newcommand{\EtTenNullN}{130{,}736}
\newcommand{\EtTenCsCover}{0.892}
\newcommand{\EtTenCsCoverCI}{[0.878, 0.905]}
\newcommand{\EtTenFgCover}{0.594}
\newcommand{\EtTenFgCoverCI}{[0.556, 0.631]}

\newcommand{\EtTenDiff}{0.298}
\newcommand{\EtTenDiffCI}{[0.265, 0.332]}
\newcommand{\EtTenBard}{123.4}

\newcommand{\EtTenFTwoUnknown}{185}

\newcommand{\EtThirtyExcess}{0.0024}
\newcommand{\EtThirtyNullN}{137{,}137}
\newcommand{\EtThirtyCsCover}{0.844}
\newcommand{\EtThirtyCsCoverCI}{[0.820, 0.867]}
\newcommand{\EtThirtyFgCover}{0.495}
\newcommand{\EtThirtyFgCoverCI}{[0.456, 0.534]}

\newcommand{\EtThirtyDiff}{0.349}
\newcommand{\EtThirtyDiffCI}{[0.316, 0.383]}
\newcommand{\EtThirtyBard}{96.2}

\newcommand{\EtThirtyFTwoUnknown}{192}
\newcommand{\EtThirtyKH}{202.9}
\newcommand{\EtMaxEvents}{0}
\newcommand{\EtWilsonUpperMax}{0.019}

\newcommand{\EtTenCsDrop}{0.028}
\newcommand{\EtTenCsDropCI}{[0.015, 0.041]}
\newcommand{\EtTenFgDrop}{0.110}
\newcommand{\EtTenFgDropCI}{[0.086, 0.136]}
\newcommand{\EtThirtyCsDrop}{0.076}
\newcommand{\EtThirtyCsDropCI}{[0.054, 0.098]}
\newcommand{\EtThirtyFgDrop}{0.209}
\newcommand{\EtThirtyFgDropCI}{[0.177, 0.242]}

\newcommand{\EtThirtyOutputs}{667}
\newcommand{\EtThirtyUnchanged}{28}
\newcommand{\EtThirtyNewline}{193}
\newcommand{\EtThirtyTruncated}{3}

\newcommand{\EtTenTokenFrac}{0.150}

\newcommand{\EtThirtyTokenFrac}{0.301}
\newcommand{\EtThirtyTokenFracDocs}{195}

\newcommand{\BridgeHalfAllQ}{0.247}
\newcommand{\BridgeHalfEightyQ}{0.090}
\newcommand{\BridgeHalfSixtyQ}{0.000}
\newcommand{\BridgeThreeAllQ}{0.585}
\newcommand{\BridgeThreeEightyQ}{0.326}
\newcommand{\BridgeThreeSixtyQ}{0.133}

\newcommand{\CostOPTPred}{14.4}

\newcommand{\CostOPTOneTwoEight}{11.4}
\newcommand{\CostOPTFiveOneTwo}{13.2}

\newcommand{\CostQwenOneTwoEight}{11.3}
\newcommand{\CostQwenFiveOneTwo}{12.8}

\newcommand{\OvUnMatches}{301}

\newcommand{\OvEdMatches}{113}
\newcommand{\OvEdBard}{39}

\newcommand{\Gam}{\operatorname{Gamma}}
\newcommand{\st}{\preceq_{\mathrm{st}}}
\newcommand{\ProtocolN}{300}
\newcommand{\ProtocolH}{180}
\newcommand{\ProtocolBudget}{420}
\newcommand{\ProtocolRetry}{120}
\newcommand{\ProtocolPowerLower}{0.564395}
\newcommand{\ProtocolPowerUpper}{0.145655}

\newcommand{\ProtocolGammaThreshold}{272}
\newcommand{\ProtocolDraftFailureApprox}{0.09565}

\newcommand{\ProtocolMissLower}{0.854}

\newcommand{\SwRho}{0.2}
\newcommand{\SwPriceFrac}{0.4}
\newcommand{\SwPriceCount}{120}
\newcommand{\SwConvM}{84}
\newcommand{\SwConvShortfall}{96}
\newcommand{\SwConvEps}{0.0444}

\newcommand{\SwConvLower}{0.905}

\newcommand{\SwTailCut}{135}
\newcommand{\SwTailFirst}{0.527}
\newcommand{\SwGammaHalfTen}{0.326}
\newcommand{\SwGammaHalfMean}{7.3}
\newcommand{\SwGammaHalfMedian}{3}
\newcommand{\SwGammaThreeMedian}{66}

\newcommand{\SwScanHalfFirst}{0.0030}

\newcommand{\SwScanEligHalf}{1.8e-05}
\newcommand{\SwScanEligThree}{0.0030}
\newcommand{\SwScanHalfFixedFirst}{0.225}

\newcommand{\HXSoneWilsonMax}{2.78\%}
\newcommand{\HXSoneCells}{60}
\newcommand{\HXDocs}{200}
\newcommand{\HXReps}{5000}
\newcommand{\HXStwoPrimeThr}{9.62\%}
\newcommand{\HXStwoPass}{6}
\newcommand{\HXStwoTotal}{8}

\newcommand{\HXOptNzeroPiMatched}{0.22}

\newcommand{\HXOptNtwoDet}{6.5\%}

\newcommand{\HXOptNtwoPiMatched}{0.82}
\newcommand{\HXOptNtwoCapShare}{87\%}

\newcommand{\HXOptEmptyMask}{41}
\newcommand{\HXOptNzeroDefined}{159}
\newcommand{\HXOptSthreeDiff}{$-$1.84}
\newcommand{\HXOptSthreeCI}{[$-$2.67, $-$1.04]}
\newcommand{\HXOptNthreeMultiWinner}{6820}

\newcommand{\HXQwenNzeroPiMatched}{0.28}

\newcommand{\HXQwenNoneCa}{55.7\%}
\newcommand{\HXQwenNoneCaCI}{[53.9, 57.4]}

\newcommand{\HXQwenNtwoDet}{7.0\%}

\newcommand{\HXQwenNtwoPiMatched}{0.77}
\newcommand{\HXQwenNtwoCapShare}{83\%}

\newcommand{\HXQwenNthreeCa}{55.1\%}
\newcommand{\HXQwenNthreeCaCI}{[53.3, 56.9]}

\newcommand{\HXQwenEmptyMask}{4}
\newcommand{\HXQwenNzeroDefined}{196}
\newcommand{\HXQwenSthreeDiff}{$-$0.54}
\newcommand{\HXQwenSthreeCI}{[$-$1.18, +0.12]}
\newcommand{\HXQwenNthreeMultiWinner}{4629}

\newcommand{\HXNtwoCountMax}{0.9\%}
\newcommand{\HXNtwoKappaRange}{39.8--40.2\%}
\newcommand{\HXSepDetMin}{80.5\%}
\newcommand{\HXSepCaMin}{10.2\%}
\newcommand{\HXSepCaMax}{17.6\%}
\newcommand{\HXSepBestMax}{32.2\%}
\newcommand{\HXSepBestMin}{18.0\%}

\newcommand{\HXHalfGapMax}{10.7}
\newcommand{\HXMultiWinnerNoRetryMax}{0.07\%}
\newcommand{\HXMultiWinnerRetryMin}{8.0\%}
\newcommand{\HXMultiWinnerRetryMax}{14.6\%}
\newcommand{\GLOptGoneGL}{28.6\%}
\newcommand{\GLOptGoneHist}{56.8\%}
\newcommand{\GLQwenGoneGL}{27.6\%}
\newcommand{\GLQwenGoneHist}{53.5\%}
\newcommand{\GLGtwoMax}{3.6\%}
\newcommand{\GLGthreeMax}{0.01\%}
\newcommand{\GLWilsonMax}{3.57\%}
\newcommand{\GLToneCells}{10}
\newcommand{\GLTtwoCells}{12}

\newcommand{\RmRhoOne}{20\%}

\newcommand{\RmQTwo}{2/3}

\newcommand{\RmDetUpperTwo}{0.0566}
\newcommand{\RmQThree}{3/4}

\newcommand{\RmDetUpperThree}{0.0501}

\newcommand{\HBRhoA}{10\%}
\newcommand{\HBRhoR}{13.3\%}
\newcommand{\HBKappa}{0.6}
\newcommand{\HBBudget}{1.1}
\newcommand{\HBLowAThree}{0.00\%}
\newcommand{\HBUpAThree}{39.34\%}
\newcommand{\HBLowAThousand}{0.00\%}
\newcommand{\HBUpAThousand}{25.70\%}
\newcommand{\HBLowAThreeK}{0.00\%}
\newcommand{\HBUpAThreeK}{18.94\%}
\newcommand{\HBLowATenK}{0.39\%}
\newcommand{\HBUpATenK}{15.05\%}
\newcommand{\HBLowAHundredK}{6.78\%}
\newcommand{\HBUpAHundredK}{11.64\%}
\newcommand{\HBLowRThree}{0.00\%}
\newcommand{\HBUpRThree}{41.67\%}
\newcommand{\HBLowRThousand}{0.00\%}
\newcommand{\HBUpRThousand}{28.70\%}
\newcommand{\HBLowRThreeK}{0.00\%}
\newcommand{\HBUpRThreeK}{22.04\%}
\newcommand{\HBLowRTenK}{3.50\%}
\newcommand{\HBUpRTenK}{18.26\%}
\newcommand{\HBLowRHundredK}{10.04\%}
\newcommand{\HBUpRHundredK}{14.93\%}
\newcommand{\HBAlpha}{0.05}
\newcommand{\HBLowProb}{0.90}
\newcommand{\HBCexN}{20}
\newcommand{\HBCexBase}{1}
\newcommand{\HBCexMix}{0}
\newcommand{\HBEoneCells}{28}
\newcommand{\HBEoneEvents}{0}
\newcommand{\HBEoneWilson}{0.19\%}
\newcommand{\HBEtwoCells}{336}
\newcommand{\HBMixBetterCells}{237}
\newcommand{\HBRFixed}{6.29\%}
\newcommand{\HBRUnified}{7.58\%}

\newcommand{\HBRminusA}{2.97}
\newcommand{\HBRminusACI}{[2.80,3.13]}
\newcommand{\HBRminusATheory}{3.33}
\newcommand{\HBAMix}{5.49\%}
\newcommand{\HBAFixed}{3.35\%}
\newcommand{\HBACone}{0.00\%}
\newcommand{\HBABinMix}{0.00\%}
\newcommand{\HBAGain}{2.14}
\newcommand{\HBAGainCI}{[2.11,2.17]}
\newcommand{\HBShortMax}{0.0\%}

\newcommand{\HBAMixCells}{24}
\newcommand{\HBAMixBelowBin}{7}
\newcommand{\HBAMixBelowCone}{2}
\newcommand{\HBAMixBelowFixed}{0}
\newcommand{\HBAEoneCells}{10}
\newcommand{\HBAEoneWilson}{0.85\%}

\newcommand{\EfPosAThreeK}{83\%}

\newcommand{\EfTauATenK}{100\%}

\newcommand{\EfMinLenARetry}{$3{,}000$}
\newcommand{\EfMinLenAInterleave}{$300$}

\newcommand{\EfPosRThreeK}{89\%}

\newcommand{\HistSizeUpper}{0.049430252}
\newcommand{\HistPowerLower}{0.564395052}

\newcommand{\HistPowerDisplay}{0.564395052049920}
\newcommand{\HistPowerUpper}{0.145654737}
\newcommand{\HistFailureDisplay}{0.095654736604316}

\newcommand{\EditedComparisons}{4800}

\newcommand{\UneditedGainRange}{0.040--0.060}
\newcommand{\EThreeDocs}{200}

\newcommand{\CellDocs}{200}

\newcommand{\MatchedCoverRange}{0.00010--0.00015}
\newcommand{\MatchedCountRange}{0.016--0.020}
\newcommand{\MatchedPrimaryUnionChanges}{49}

\newcommand{\MatchedPrimaryDocs}{1000}
\newcommand{\MatchedPrimaryReportChanges}{50}
\newcommand{\MatchedEditedDocs}{4000}
\newcommand{\MatchedEditedReportChanges}{2}

\title{Detecting Is Not Counting: CountScan and the Limits of Certifying Watermark Matches}
\ifanon
\author{Anonymous Author(s)}
\hypersetup{pdfauthor={Anonymous}}
\else
\author[ ]{Guanyu Chen}
\affil[ ]{Beijing Normal University}
\newcommand{\PdfAuthors}{Guanyu Chen}
\makeatletter
\let\authordraft@author\@author
\AtBeginDocument{%
  \let\@author\authordraft@author
  \renewcommand{\notice@text}{\textit{Preprint.}}%
}
\makeatother
\hypersetup{pdfauthor={\PdfAuthors}}
\fi
\makeatletter
\AtBeginDocument{\renewcommand{\notice@text}{\textit{Preprint.}}}
\makeatother
\hypersetup{pdftitle={Detecting Is Not Counting: CountScan and the Limits of Certifying Watermark Matches}}

\begin{document}
\maketitle

\begin{abstract}
Writing assistants can watermark the words they suggest. In an audit, when a school or publisher asks the provider of such an assistant about a submitted report, the provider checks how many positions of a passage match tokens its service returned, without having retained request logs. A watermark test shows that such matches exist, not how many. We introduce CountScan, a scan that turns per-token watermark evidence into lower bounds on the number of matches, valid for every passage at once and after key-blind editing by the writer, at a cost logarithmic in the text length. Since different hidden histories can release the same text with different numbers of matches, no valid count can certify more than they all share. In explicit models of the writing interface, in which the writer may request, retry and discard suggestions, retain their own draft and edit under a request cap, we compute this limit exactly and show that a variant of CountScan attains it. In simulations it stays valid and, under a hard history, certifies at least $2.5\%$ of the positions in \EfTauATenK{} of documents with $10^4$ positions.
\end{abstract}

\section{Introduction}\label{sec:intro}
Writing assistants can watermark the tokens they return so that the holder of a key can later test a text \citep{aaronson2023watermarking,li2025framework}. Commercial detectors of text generated by artificial intelligence (AI) already report the share of a document they flag \citep{copyleaks2023faq}; with a key one can ask what share is certifiable. In an audit of a submitted report by the provider of such an assistant, or by an institution it authorizes, the auditor holds the key and the final document but not the request log, while the writer may have taken suggestions word by word, kept those that agree with an own draft, regenerated others and edited the text. After reading, the auditor selects a passage $S$ and asks for a certificate that at least $k$ of its context--token pairs were returned by the assistant. A pair may also match through copying or coincidence, which a key cannot tell apart from model output, so historical matching, not authorship, is the target a key can verify (Appendix~\ref{app:historyaudit}).

Existing tools answer neighbouring questions. Detection tests whether a text carries a watermark, localization and segmentation say where it occurs, proportion estimators estimate a watermarked share in a mixture model, and FissionGuard controls the false discovery rate (FDR) of reported segments. None of these targets is a certified lower bound on the number of returned pairs in a passage chosen after reading. We seek lower confidence bounds $d(S)$ on the realized number $\delta(S)$ of retained historical context--token matches in every interval $S$ of the text, simultaneously:
\begin{equation}\label{eq:tdg}
\PP\big(\forall S:\ d(S)\le\delta(S)\big)\ge1-\alpha,
\end{equation}
where $\alpha\in(0,1)$ is the error level and the probability is over the key and the history. Two questions follow. The writer edits after seeing generated text and the passage is chosen after reading, so a count must first stay reliable. Coverage alone permits $d\equiv0$, and different request histories can produce the same text and key, so the central question is how much any reliable count can certify. Our answer is that a key keeps counts reliable after editing and selection, while how much it can certify is set by the ambiguity of the unobserved request history. In short, detecting is not counting: under the histories of Section~\ref{sec:history-protocol} a binary certificate can detect matches with high probability while no valid count guarantees more than a fixed fraction of them, and some histories defeat detection as well.

\paragraph{Contributions.} \emph{Reliability for every key-blind history} (Section~\ref{sec:method}): we derive, rather than assume, that nonmatch pivots stay conditionally super-uniform after adaptive generation and text-dependent, key-blind editing (Proposition~\ref{prop:K}); CountScan, which runs a start-wise $e$-process and trims its largest terms, gives \eqref{eq:tdg} for all intervals, at a cost of about $\log|T|/\xi^*$ certified matches, where $|T|$ is the number of retained positions and $\xi^*$ a critical evidence value (Section~\ref{sec:price}). \emph{Exact limits in explicit protocols} (Section~\ref{sec:history-protocol}): an indistinguishability lemma explains why hidden histories cap counts; for $r$ fixed menus and for a two-menu protocol with a request budget, two release permissions and free editing, we give matching upper and lower worst-case limits; these optimality statements are specific to the protocols. \emph{A count that attains them}: a single mixture CountScan, valid for all intervals and blind to $\kappa$, $b$ and the log, attains these limits with explicit finite-length bands. \emph{Evidence} (Section~\ref{sec:experiments}): we check validity after editing and selection, show that detection and counting separate under simulated writing policies, and identify the lengths at which the attaining count certifies a useful share.

\paragraph{Related work.}\label{sec:related}
Pivot-based methods address detection, robustness and online evidence \citep{li2025framework,li2025robust,su2026online}, proportion estimators work in mixture models \citep{li2025proportion,chai2026sample}, and localization methods identify where a watermark occurs \citep{li2024segmenting,bonnerjee2026wiser,blanchet2026spot}. FissionGuard splits evidence between localization and validation at a prespecified proportion and controls the FDR of online reports \citep[\S2.2 and Theorem~3.3]{wei2026fissionguard}; our bounds instead cover all interval counts on one event. Closed testing with $e$-values yields simultaneous discovery bounds \citep{goeman2011multiple,goeman2021only,vovk2023confidence}, start-wise scans arise in change detection \citep{shin2024edetectors,saha2026nonpartitioned}, and SeqE-Guard \citep{fischer2025admissible} turns valid sequential $e$-values into interval counts. SeqE-Guard presupposes such $e$-values, whereas text-dependent editing breaks pivot uniformity and hidden histories cap what any count can certify. We supply the calibration that makes watermark pivots valid $e$-values under key-blind editing, use endpoint trimming because the first-order cost of Section~\ref{sec:price} is derived for it (restarted SeqE-Guard gives weakly larger counts and the same validity; Appendices~\ref{app:backend-review} and~\ref{app:related}), and build menu-aware scores that attain the hidden-history limits.

\section{Setting and the Null Condition}\label{sec:setup}
\paragraph{Gumbel-max watermark.} Let $V$ be the vocabulary, $w_1,\dots,w_n$ the final text and $m$ the context width. The context of position $t>m$ is $\chi_t=w_{t-m:t-1}$, and the context $\chi$ has the pseudorandom vector $\zeta_\chi=(U_{\chi,v})_{v\in V}=\mathrm{Hash}(\chi,\mathrm{Key})$, where $\mathrm{Hash}$ is a keyed hash and $\mathrm{Key}$ the secret key. A watermarked step with next-token probability (NTP) distribution $\pi=(\pi_v)_{v\in V}$ in context $\chi$ outputs $\argmax_{v:\pi_v>0}\log U_{\chi,v}/\pi_v$ \citep{aaronson2023watermarking}. When the row is fresh and independent of the information determining $\pi$, the draw has law $\pi$ \citep{li2025framework}; reused rows can have a different conditional law. The auditor computes the pivot
\begin{equation}\label{eq:pivot}
P_t=1-U_{\chi_t,w_t},
\end{equation}
where $U_{\chi_t,w_t}$ is the coordinate of the released token $w_t$ in the row of its context $\chi_t$.

\paragraph{Masking and the target.} Positions with the same context share a pseudorandom vector. We keep the positions whose context occurs once,
$T=\{t>m:\ \chi_s\ne\chi_t\text{ for all }s>m,\ s\ne t\}$, a function of the text. Let $M\subseteq T$ be the retained \emph{history matches}: positions whose pair $(\chi_t,w_t)$ occurs at some watermarked step of the key-wide generation history. Let $N=T\setminus M$ be the null positions and $\delta(S)=|S\cap M|$.

\subsection{What We Need from Null Positions}\label{subsec:editing}
An editor sees the generated text, which depends on the key, so null pivots need not be uniform.

\begin{example}[Editing makes nulls super-uniform]\label{ex:edit}
Let $V=\{0,1\}$ and let the sole watermarked request in context $\chi$ with $\pi=(1/2,1/2)$ output $1$, so $U_{\chi,0}<U_{\chi,1}$. A key-blind editor changes the $1$ into $0$. The edited position has context $\chi$ and token $0$, is null, and has $P=1-U_{\chi,0}$ with $U_{\chi,0}$ the smaller of two independent uniforms. Hence $\PP(P\le p)=p^2\le p$.
\end{example}
The editor does not know the key, but by reacting to generated text it can select nulls whose pivots depend on it. What must be controlled is the law of a null pivot after such selection, given the information the betting rules may use.
Let $\Ic$ be the $\sigma$-field of information that may be used for betting, given explicitly in Proposition~\ref{prop:K}, and $\F_t=\sigma(\Ic,\,P_s\ind\{s\in N\}:m<s\le t)$.

\begin{condition}[Conditional super-uniformity]\label{cond:S}
(S1) The text and $N$ are $\Ic$-measurable. (S2) For every $t>m$ and every $p\in[0,1]$,
\[
\PP\big(P_t\le p\mid\F_{t-1}\big)\le p\qquad\text{almost surely on }\{t\in N\},
\]
where $\F_{t-1}$ is the information the betting rules may use before position $t$ and $N$ is the set of retained nonmatch positions; by (S1), $\{t\in N\}$ is $\F_{t-1}$-measurable. (S3) Almost surely $P_t\in(0,1)$ for all $t\in T$.
\end{condition}
We establish Condition~\ref{cond:S} for adaptive generation followed by text-dependent, key-blind editing. Let $R$ and $Q$ be the key-free generation and editing randomness.

\begin{assumption}[Key-blind generation and editing]\label{ass:Astar}
(K1) The coordinates $U_{\chi,v}$ over all contexts and tokens are independent and identically distributed (i.i.d.)\ $\Unif(0,1)$. (K2) The history has at most $L$ steps; at each step the status, context and NTP distribution are measurable functions of earlier outputs and $R$, and a watermarked step outputs the Gumbel-max token. Nonwatermarked outputs and stopping decisions depend only on earlier outputs and $R$. (K3) $(R,Q)$ is independent of the pseudorandom vectors, and the final text is a measurable function of the history outputs, $R$ and $Q$.
\end{assumption}
(K2) allows adaptive generation but excludes other uses of the key, for example regenerating until a detection score is high. (K3) lets editors see everything generated but not the key. (K1) and (K3) adapt the working hypotheses of \citet[Working Hypotheses 2.1--2.2, p.~328]{li2025framework} to a key-wide history with edits. In the latent-mixture model of \citet[Assumption 4.1(b), p.~10]{blanchet2026spot} an edited token is conditionally independent of the pseudorandom variable; our editors may depend on the generated text.

\begin{proposition}[Key-blind nulls are super-uniform]\label{prop:K}\label{lem:superunif}
Under Assumption~\ref{ass:Astar}, Condition~\ref{cond:S} holds with $\Ic=\Hc$, the $\sigma$-field generated by the history outputs, $R$, $Q$ and the pseudorandom vectors of all contexts other than the null contexts $\{\chi_t:t\in N\}$.
\end{proposition}
\begin{proof}[Proof idea]
Fix the output record and the key-free randomness. The record constrains each context row only through its winner events, so the posterior rows are independent. In a nonmatch row the released token never won, and every winner constraint bounds its coordinate from above. Given the other coordinates of the row, the coordinate is therefore uniform on $(0,B]$ for some $B\le1$, which gives $\PP(P_t\le p\mid\cdot)\le p$. The mask places earlier null pivots in other rows. Appendix~\ref{app:propK} gives the factorization and the conditioning.
\end{proof}

Under $\Ic=\Hc$, the text and all earlier pivots at positions $m<s<t$ are $\F_{t-1}$-measurable (Remark~\ref{rem:auditorinfo}). Betting rules may use this information and nothing else. If the pseudorandom vectors of null positions are unused, as in \citet[Assumption 3.1, p.~6]{wei2026fissionguard}, Condition~\ref{cond:S} holds with equality (Proposition~\ref{prop:U}), and the monotonicity requirement below can be dropped.\label{rem:uniform}

The conditional law also gives, through the randomized distributional transform \citep{brockwell2007universal}, a coupling that extends beyond CountScan.
\begin{corollary}[Coupling with uniform nulls]\label{cor:coupling}
Under Condition~\ref{cond:S}, on a space enlarged by independent uniforms there are $V_t$, $t>m$, i.i.d.\ $\Unif(0,1)$ and independent of $\Ic$, with $V_t\le P_t$ for $t\in N$. Hence a procedure keeps level $\alpha$ if its error indicator depends only on the null pivots, $\Ic$-measurable inputs and independent randomization, is nonincreasing in each null pivot, and has level $\alpha$ when the null pivots are i.i.d.\ uniform and independent of $\Ic$.
\end{corollary}
With $\Ic=\Hc$, tests calibrated under fresh uniform pivots, such as sums of a nonincreasing function of the retained pivots, therefore stay valid after key-blind editing (Appendix~\ref{app:coupling}).

\section{Reliable Counts after Editing and Selection}\label{sec:method}\label{sec:theorem}
\begin{figure}[t]
\centering
\includegraphics[width=\columnwidth]{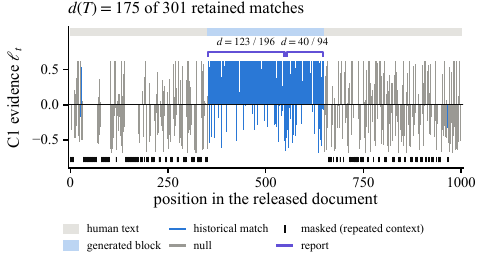}
\caption{CountScan on one prespecified mixed document of OPT-1.3B, an Open Pre-trained Transformer (OPT; $\tau=1.0$): $350$ human tokens, a $300$-token generated block and $350$ human tokens. Bars are C1 evidence at retained positions, blue for retained historical matches and grey for nulls; ticks mark masked positions; brackets are the reports ($\eta=0.25$, $32$--$256$ retained tokens), labelled with $d(S)$ and their retained matches; $\bar d(T)$ sums $d$ over disjoint intervals (Corollary~\ref{cor:select}(c)); edited version in Appendix~\ref{app:real}.}\label{fig:overview}
\end{figure}

\subsection{CountScan}\label{subsec:countscan}

\paragraph{Evidence.}
For each start $a$ and position $t$ take a hedge $\lambda^{(a)}_t\in[0,1]$ and a calibrator $g^{(a)}_t\ge0$ on $(0,1)$ with $\int_0^1g^{(a)}_t\le1$, and set
\[
e^{(a)}_t=1-\lambda^{(a)}_t+\lambda^{(a)}_tg^{(a)}_t(P_t),\qquad \ell^{(a)}_t=\log e^{(a)}_t .
\]
We require: (I1) $\lambda^{(a)}_t$ is $\F_{t-1}$-measurable; (I2) $g^{(a)}_t$ is jointly $\F_{t-1}\otimes\mathcal B$-measurable, where $\mathcal B$ is the Borel $\sigma$-field of $(0,1)$; (I4) $p\mapsto g^{(a)}_t(p)$ is nonincreasing. The default is $g(p)=-\log p$; the prespecified C1 is a truncated version (\S\ref{subsec:real}).

\begin{lemma}[Stochastic dominance]\label{lem:evalue}
Under Condition~\ref{cond:S}, (I1), (I2) and (I4), for every $a$ and $t$, on $\{t\in N\}$,
$\EE[e^{(a)}_t\mid\F_{t-1}]\le1-\lambda^{(a)}_t+\lambda^{(a)}_t\int_0^1g^{(a)}_t\le1$.
\end{lemma}
Indeed, by (S2) the conditional law of $P_t$ given $\F_{t-1}$ puts mass at most $p$ on $[0,p]$, and $g^{(a)}_t$ is nonincreasing, so $\EE[g^{(a)}_t(P_t)\mid\F_{t-1}]\le\int_0^1g^{(a)}_t(p)\,dp\le1$ on $\{t\in N\}$ \citep[Proposition~2.1]{vovk2021evalues}; see Appendix~\ref{app:steps}.

\paragraph{Start-wise scan.}
For each start $a$ and end $b\ge a$, the null-only product $\exp Z_a(b)$ with $Z_a(b)=\sum_{t\in[a,b]\cap N}\ell^{(a)}_t$ is a nonnegative supermartingale starting at one. Allocate $\Ic$-measurable weights $\pi_a\ge0$ with $\sum_a\pi_a\le1$ (I3) and set $c_a=\log(1/(\alpha\pi_a))$. Ville's inequality \citep[eq.~(8)]{ramdas2023safe} and a union over starts give
\[
\Omega_\alpha=\{Z_a(b)<c_a\ \forall a\le b\},\qquad
\PP(\Omega_\alpha)\ge1-\alpha .
\]
Equal weights on nonempty $T$ give $c_a=\log(|T|/\alpha)$ at retained starts.

\paragraph{From evidence to counts.}
$\Omega_\alpha$ concerns the unobservable null evidence. To obtain a conservative count without knowing which positions are historical matches, we remove the largest positive contributions until the remaining evidence falls below the threshold, as in worst-case partial-conjunction testing \citep[eq.~(1)]{benjamini2008screening}. If the evidence still reaches the threshold after the $j$ largest positive terms are removed, then on $\Omega_\alpha$ no $j$ matches can account for it, so at least $j+1$ positions are matches.

\begin{definition}[Trimmed bound]\label{def:trim}
For $S=[a,b]\cap T\ne\varnothing$ with first element $a_0$, let $\Top_j(S)$ be the sum of the $j$ largest positive $\ell^{(a_0)}_t$ in $S$, $R_S(j)=\sum_{t\in S}\ell^{(a_0)}_t-\Top_j(S)$, and
$d(S)=\min\{j\ge0:\ R_S(j)<c_{a_0}\}$.
\end{definition}

\begin{lemma}[Trimming inequality]\label{lem:trim}
For any finite $S$, reals $\ell_t\in[-\infty,\infty)$ and any split $S=S_M\cup S_N$ with $|S_M|=\delta$, $R_S(\delta)\le\sum_{t\in S_N}\ell_t$.
\end{lemma}

\begin{theorem}[Simultaneous validity]\label{thm:main}
Under Condition~\ref{cond:S} and (I1)--(I4), with $S_{ab}=[a,b]\cap T$,
\[
\PP\big(\forall\,a\le b:\ d(S_{ab})\le\delta(S_{ab})\big)\ge1-\alpha .
\]
\end{theorem}
\noindent On $\Omega_\alpha$, Lemma~\ref{lem:trim} gives $R_S(\delta(S))\le Z_{a_0}(b)<c_{a_0}$ for every interval, so $d(S)\le\delta(S)$; Appendix~\ref{app:steps} gives every step.

\begin{corollary}[What the event buys]\label{cor:select}
Under the conditions of Theorem~\ref{thm:main}: (a) for any interval $\hat S$ chosen after seeing all pivots and the scan, $\PP(d(\hat S)>\delta(\hat S))\le\alpha$; (b) the rule ``reject $\delta(S)\le k$ if $d(S)\ge k+1$'' controls the family-wise error rate (FWER) over all $(S,k)$ at level $\alpha$; (c) with probability at least $1-\alpha$, $\bar d(A)\le\delta(A)$ for all $A\subseteq T$, where $\bar d(A)$ is the largest sum of $d$ over disjoint intervals inside $A$.
\end{corollary}

\paragraph{Reports.} Algorithm~\ref{alg:countscan} (Appendix~\ref{app:algo}) reports disjoint intervals of length between $L_{\min}$ and $L_{\max}$ that maximize $\sum_id(S_i)$ subject to $d(S)\ge\lfloor\eta|S|\rfloor+1$, so each report also rejects the partial-conjunction null at proportion $\eta$. For fixed or text-measurable $\lambda$, all $d([a,b])$ are computed in $O(n^2\log n)$ time with a Fenwick tree over the positive terms; median scan-plus-report time is \RuntimePipelineRangeMs{}\,ms at \RuntimeReferenceN{} retained positions (Appendix~\ref{app:algo}).

\subsection{The Price of Covering All Intervals}\label{sec:price}\label{sec:power}\label{subsec:cost}
Simultaneity is needed in practice: in a post hoc check on human-source documents, reporting the highest-count interval made a single-interval bound overclaim on \SelDemoSelOpt{} (OPT) and \SelDemoSelQwen{} (Qwen) of documents, against at most \SelDemoCSMax{} for CountScan (Appendix~\ref{app:seldemo}). Covering all intervals raises the threshold from $\log(1/\alpha)$ to $\log(|T|/\alpha)$, by $\Delta=\log|T|$. Write $d_c(S)=\min\{j:R_S(j)<c\}$. On every realization, raising the threshold by $\Delta$ moves the stop by $\Delta$ over the evidence of the last trimmed tokens (Appendix~\ref{app:price}, Figure~\ref{fig:costgeom}). For positive evidence values $x_1\ge\dots\ge x_r>0$ let $b=d_c(S)$ and $a=d_{c+\Delta}(S)$.

\begin{proposition}[Deterministic sandwich]\label{prop:pricesandwich}
(i) If $1\le a<b$, then $\Delta/x_a-1<b-a<\Delta/x_{b-1}+1$. (ii) If $a=b\ge1$, then $\Delta<x_a$. (iii) If $a=0<b$ and $b\ge2$, then $b<\Delta/x_{b-1}+1$.
\end{proposition}
Under i.i.d.\ evidence $x_a$ and $x_{b-1}$ both converge to the slope $\xi^*$ at the zero crossing of the trimmed evidence curve (Appendix~\ref{app:price}).

\begin{corollary}[First-order cost]\label{cor:price}
Let the evidence in an interval $S_L$ of length $L$ be i.i.d.\ with distribution function $F$ and generic value $\ell$, $\EE|\ell|<\infty$, $0<q^*<\PP(\ell>0)$, $\xi^*=F^{-1}(1-q^*)>0$, where $q^*$ is the limit in Proposition~\ref{prop:qstar} (Appendix~\ref{app:proofs}), and let $F$ have a continuous positive density near $\xi^*$. If $c_L/L\to0$ and $\Delta_L/L\to0$, then
\[
d_{c_L}(S_L)-d_{c_L+\Delta_L}(S_L)=\Delta_L/\xi^*+o_p(\Delta_L)+O_p(1),
\]
where $o_p$ and $O_p$ denote terms that vanish and stay bounded in probability.
\end{corollary}
The level of the count follows from known results: $d/L$ converges to the root $q^*$ of the trimmed integral $\int_0^{1-q}F^{-1}(u)\,du$, a power condition of \citet[App.~B.2, eq.~(30)]{wei2026fissionguard}, with fluctuations from buffered-probability estimation \citep[Theorem~2.1]{mafusalov2018estimation}; Corollary~\ref{cor:price} isolates the shift $\Delta/\xi^*$ that simultaneity adds. A post hoc reanalysis of the cached holdout pivots checks it (Appendix~\ref{app:price}, Figure~\ref{fig:cost}). The loss rises from \CostOPTOneTwoEight{} matches at $L=128$ to \CostOPTFiveOneTwo{} at $L=512$ for OPT (Qwen: \CostQwenOneTwoEight{} and \CostQwenFiveOneTwo{}), approaching $\Delta/\xi^*=\CostOPTPred$ from below. The shortfall is predicted at finite length: with $q_u$ the population count proportion at threshold $u$ and $\xi_u=F^{-1}(1-q_u)$, $\Delta/\xi_{c+\Delta}\le L(q_c-q_{c+\Delta})\le\Delta/\xi_c\le\Delta/\xi^*$ (Appendix~\ref{app:price}).

\section{What a Hidden History Allows}\label{sec:history-protocol}
Section~\ref{sec:method} guarantees that counts do not overstate, but a reliable count may still report zero. We now ask how much any valid count can certify when the auditor knows the key but not which requests occurred. Section~\ref{subsec:fundamental} shows why the answer is limited, Sections~\ref{subsec:single-winner}--\ref{subsec:budget} compute the limit in two protocols, Section~\ref{subsec:unified} attains it, and Section~\ref{subsec:limits-end} marks where nothing can be certified.

\subsection{The fundamental problem}\label{subsec:fundamental}
Fix a region $S$ of $n_S$ retained positions, set by the released text and public randomness, and write $H=\delta(S)$. The auditor sees the text, the key, the public model and the draft, but not the requests, their labels or the policy. A count $D$ is valid for a class $\mathcal G$ of writing policies if $\PP_G(D>H)\le\alpha$ for every $G\in\mathcal G$. Power is evaluated on policies with $H\ge h$: a low count is unsurprising when matches are few, and only a failure to certify many true matches shows that the hidden history itself limits the audit. The auditor is not told $h$.
\begin{lemma}[Indistinguishable histories]\label{lem:indist}
Let $D$ be valid for $\mathcal G$, and let $G_0,G_1\in\mathcal G$ run on the same key, key-free randomness and count randomization, with match counts $H_0$ and $H_1$ and counts $D_0$ and $D_1$. If $H_0\le j$ and $A$ is an event on which the two policies release the same text, then $\PP(D_1>j)\le\alpha+\PP(A^c)$.
\end{lemma}
\begin{proof}
On $A$ the count receives the same input under both policies, so $D_1=D_0$, and $\{D_0>j\}\subseteq\{D_0>H_0\}$; hence $\{D_1>j\}\subseteq A^c\cup\{D_0>H_0\}$, whose probability is at most $\PP(A^c)+\alpha$.
\end{proof}
Every upper limit below applies the lemma to a low policy, with few matches, and a high policy with $H_1\ge h$ that releases the same text on a likely event; every lower limit is a count that is valid for the whole class. Table~\ref{tab:layers} summarizes the results.

\begin{table}[t]
\centering
\caption{Settings and what they allow. In every row the auditor sees the final text, the full key and the public model or menus, but not the requests, their number or the matches. Proportion: the largest proportion of the $n_S$ positions that valid counts can guarantee asymptotically, uniformly over histories with match fraction $\kappa$; Detect: whether some valid binary certificate has power bounded away from $\alpha$.}\label{tab:layers}
\footnotesize\setlength{\tabcolsep}{1.5pt}
\begin{tabular}{@{}p{2.55cm}p{1.3cm}ccc@{}}
\toprule
Menus; writer may & Release & Detect & Proportion & Result\\
\midrule
any; key-blind editing & any & -- & valid & Thm.~\ref{thm:main}\\
$r$ fixed, draft prob.\ $\le s$; each menu once & draft or returned & if $\kappa>q_r$ & $\rho_r$ & Thm.~\ref{thm:sw-sharp}\\
two symmetric; $\lfloor bn\rfloor$ requests, interleave, discard, select & any (A); draft or returned (R) & if $\rho>0$ & $\rho_A$, $\rho_R$ & Thm.~\ref{thm:budget}\\
one per position; keep the draft & draft & no & $0$ & Prop.~\ref{prop:draft-only}\\
two, retries; \ProtocolBudget{} requests & draft & $\le\ProtocolPowerUpper$ & $0$ & Thm.~\ref{thm:full-key}\\
\bottomrule
\end{tabular}
\end{table}

\subsection{Single winners and fixed menus}\label{subsec:single-winner}
Let $X=-\sum_{t\in S\cap T}\log U_{\chi_t,w_t}$ and $C_x=\ind\{X\le x\}$ for a threshold $x$, and write $F_Y$ for the cumulative distribution function (CDF) of a variable $Y$ and $\Gam(k,1)$ for a unit-rate Gamma variable with shape $k$.

\begin{figure*}[t]
\centering
\includegraphics[width=\textwidth]{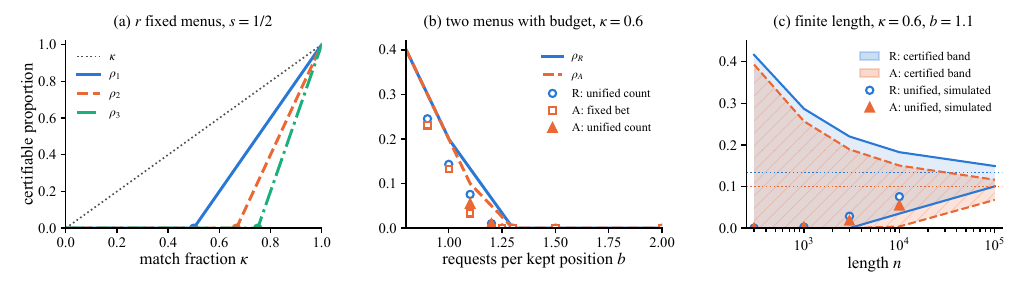}
\caption{Certifiable proportions under hidden histories. (a) $\rho_r$ of Theorem~\ref{thm:sw-sharp} at $s=1/2$ for $r=1,2,3$ fixed menus; dots mark $\kappa=q_r$, below which no valid count detects. (b) $\rho_A$ and $\rho_R$ of Theorem~\ref{thm:budget} at $\kappa=\HBKappa$ against the request budget $b$; markers: mean certified share $D_U([n])/n$ of the count of Theorem~\ref{thm:unified} at $n=10^4$ under the registered hard history (a) of Appendix~\ref{app:budget-sim}, and under A also CountScan with the fixed bet. (c) Finite length at $\kappa=\HBKappa$, $b=\HBBudget$: each band runs from the share that $D_U$ guarantees with probability at least $\HBLowProb$ under every high-match history to an upper share at or below which, for every valid count, some high-match history keeps it with probability at least $0.85$ (Appendix~\ref{app:budget-finite}); dotted lines: $\rho_R$, $\rho_A$; markers: simulated means. Markers are simulation means at one hard history; finite-length bounds are certified at the five displayed lengths, the connecting lines and shading are guides to the eye, and the curves in (a) and (b) are theorems.}\label{fig:hhbudget}
\end{figure*}

\begin{proposition}[Validity and power]\label{prop:gamma-power}
Under Assumption~\ref{ass:Astar}, $\PP(C_x>H)\le F_{\Gam(n_S,1)}(x)$ for any fixed $x$. Fix $1\le h\le n_S$. If $H\ge h$ almost surely, each retained context has at most one distinct historical winner, and every historical request probability in those rows is at most $1/2$, then
\[
\PP(C_x=1)\ge F_{\Gam(n_S-h,1)+\frac12\Gam(n_S,1)}(x),
\]
with independent unit-rate Gamma variables and the shape-zero law a point mass at zero (Appendix~\ref{hp:section}).
\end{proposition}
For $n_S=\ProtocolN$ and $h=\ProtocolH$, the threshold $x=\ProtocolGammaThreshold$ gives size at most $\HistSizeUpper$ and power at least \ProtocolPowerLower. Detection, however, does not say how many matches can be certified. To answer this, let the token probabilities be at most $s$, a further condition that Gumbel-max sampling does not imply, and use for each $n_S$ the public model of Appendix~\ref{app:swprice} with $r$ menus per context, each giving the draft token probability $s$ and the rest to competitors that differ across menus; a policy requests each menu at most once per context and releases the draft token or a returned one ($r=1$: one winner per context). Let
\[
q_r=1-\int_0^\infty e^{-x}\big(1-e^{-x(1-s)/s}\big)^r\,dx ,
\]
the probability that the draft token wins at least one of the $r$ menus, so $q_1=s$ and $q_r=r/(r+1)$ at $s=1/2$.
\begin{theorem}[Counts with $r$ menus]\label{thm:sw-sharp}
Let $n_S\to\infty$ with $h/n_S\to\kappa\in(0,1)$, and put $\rho_r=(\kappa-q_r)_+/(1-q_r)$, where $(y)_+=\max\{y,0\}$. (i) For every $q<\rho_r$ there is a count, valid for every history under Assumption~\ref{ass:Astar}, that certifies at least $qn_S$ matches with probability tending to one, uniformly over the policies with $H\ge h$. (ii) For every $0\le j\le h$ there is a policy with $H\ge h$ on every realization, not depending on the count, under which every count valid over the policies has $\PP(D>j)\le\alpha+\PP\{\mathrm{Bin}(n_S-j,q_r)<h-j\}$, where $\mathrm{Bin}(n,p)$ is a binomial variable. Hence no proportion above $\rho_r$ can be guaranteed, and if $\kappa<q_r$, every valid count has $\PP(D\ge1)\le\alpha+o(1)$ under that policy.
\end{theorem}
\begin{proof}[Proof idea]
(i) inverts $\#\{t:U_{\chi_t,w_t}<u\}$ against $\mathrm{Bin}(n_S-k+1,u)$: a kept draft token that lost its menus has this tail at most $u/(1-q_r)$, and a match has tail $o(u)$. (ii) couples a policy requesting only the last $j$ positions with one that tries the menus until the draft token wins and keeps losing draft tokens (Appendix~\ref{app:rmenu}).
\end{proof}
With $\kappa=0.6$, one menu allows $\rho_1=\RmRhoOne$, but $q_2=\RmQTwo>\kappa$, so with two menus every valid count detects a match with probability at most \RmDetUpperTwo{} under the policy of (ii) (Figure~\ref{fig:hhbudget}a).

\subsection{Two menus, a request budget and editing}\label{subsec:budget}
Real writers do more than Theorem~\ref{thm:sw-sharp} allows: they interleave requests, reread, discard drafts and choose which positions to keep, all under a request cap the auditor does not observe. We capture this in a two-menu protocol, a suggestion and one alternative per context with free editing (Appendix~\ref{app:budget}). Each address, a context, has an independent key row and public labels $h$ (draft), $a_1$, $a_2$ (menu alternatives) and $o$ (outside edit); menu $j\in\{1,2\}$ is the equal-weight Gumbel-max race between $h$ and $a_j$, sharing the coordinate of $h$, so the counterfactual winner pair $\omega=(W_1,W_2)$ takes the values $(h,h),(h,a_2),(a_1,h),(a_1,a_2)$ with probabilities $1/3,1/6,1/6,1/3$. The writer may request any menu at any address, in any order and adaptively, sees only the returned labels, may discard what it requested, and finally keeps $n$ distinct addresses in any order; every request counts against a budget of $\lfloor bn\rfloor$ requests, discarded ones included, where $b$ is the number of requests per kept position. Under permission A the writer may release any label, under R only $h$ or a label returned at that address. The match $H_i$ of kept position $i$ indicates that its released token was returned there. The auditor sees the final text, the full key and the menus, not the requests, their number or $H$.
\begin{theorem}[Budget and editing]\label{thm:budget}
Let $n\to\infty$, $\kappa\in(0,1)$, $b>\kappa$, and call a policy high if $H([n])\ge\lceil\kappa n\rceil$ on every realization. Under permission $P\in\{A,R\}$, the largest proportion that counts valid over all intervals and all policies of the protocol can guarantee, with probability tending to one uniformly over the high policies, is
\begin{align*}
\rho_A&=\max\{0,\ 2\kappa-b,\ \tfrac32\kappa-\tfrac12b-\tfrac14,\ 3\kappa-2\},\\
\rho_R&=\max\{0,\ 2\kappa-b,\ 2\kappa-\tfrac23b-\tfrac13,\ 3\kappa-2\}.
\end{align*}
It is attained by the count $D_U$ of Section~\ref{subsec:unified}, which uses neither $\kappa$, $b$ nor the request log.
\end{theorem}
\begin{proof}[Proof idea]
Lower limit: a score of $(\omega,w)$ with nonpositive conditional mean at nonmatches, and a potential $V\le0$ on local request states with $V(\emptyset)=0$, make $\sum_xV(s_x)$ plus a cost per request a submartingale along all requests, discarding included. Upper limit: explicit pairs of single-row strategies with equal law of $(\omega,w)$ at the vertices of the pieces, a hard-budget fallback and Lemma~\ref{lem:indist} (Appendix~\ref{app:budget}).
\end{proof}
At $b=\kappa$, along feasible integer lengths, every request must yield a match and $\rho_P=\kappa$; once $b\ge2$ both limits equal $(3\kappa-2)_+=\rho_2$ of Theorem~\ref{thm:sw-sharp} at $s=1/2$, so the budget interpolates between a log that is almost revealed and the single-use menu limit. At $\kappa=\HBKappa$ and $b=\HBBudget$, $\rho_A=\HBRhoA$ and $\rho_R=\HBRhoR$ (Figure~\ref{fig:hhbudget}b).

\subsection{A count that attains the limits}\label{subsec:unified}
Given the key, the auditor computes $Z_i=(\omega_i,w_i)$ at each kept position $i$, and $Y_i=\ind\{w_i\in\{W_{i,1},W_{i,2}\}\}$ indicates that the released token wins some menu. Under R an unreturned release is $h$, which wins some menu with conditional probability at most $2/3$ given the hidden record, so $f_R(Z_i)=3Y_i-2\in\{-2,1\}$ has nonpositive conditional mean at nonmatches. Under A, where the writer may also release an unrequested $a_j$ or $o$, two scores $f_{A1},f_{A2}\in[-2,1]$ do the same (Appendix~\ref{app:budget-scores}); every score has $f_j\le1$. With fixed weights $w_j$ ($w_R=1$, $w_{A1}=w_{A2}=1/2$), let
\begin{align*}
\mathcal E(S,k)&=\sum_j w_j\,2\int_0^{1/2}\frac{\prod_{i\in S}\{1+\lambda f_j(Z_i)\}}{(1+\lambda)^{k}}\,d\lambda,\\
D_U(S)&=\max\{k\in\{1,\dots,|S|\}:\ \mathcal E(S,k-1)\ge n/\alpha\},
\end{align*}
where $\mathcal E(S,k)$ is a mixture $e$-value against $H(S)\le k$, $\lambda$ is the bet, and $D_U(S)=0$ if no $k\ge1$ qualifies. $D_U$ is the CountScan of Section~\ref{sec:method} with two changes: its evidence $1+\lambda f_j(Z_i)$ uses the menus, not only the pivot, and its hedge is mixed over $\lambda$; the start-wise argument is unchanged.
\begin{theorem}[Unified count]\label{thm:unified}
Under either permission, (i) $\PP\{\exists S:D_U(S)>H(S)\}\le\alpha$ for every policy of the protocol, editing and selection included; (ii) $D_U([n])/n\ge\rho_P-\varepsilon$ with probability tending to one uniformly over the high policies, for every $\varepsilon>0$.
\end{theorem}
\begin{proof}[Proof idea]
Given the complete record, $\prod_{i=a}^{b}\{1+\lambda f_j(Z_i)\}/(1+\lambda)^{H_i}$ is a nonnegative supermartingale from each start; Ville's inequality over the $n$ starts gives (i), and the prior mass near $\lambda=(\log n/n)^{1/2}$ gives (ii) (Appendix~\ref{app:budget-lower}).
\end{proof}
Under R, $D_U$ is a beta-binomial mixture count of the indicators $Y_i$ \citep{howard2021confidence}, so this simple count already attains the limit; under A, mixing $f_{A1}$ and $f_{A2}$ costs only $\log2$ in evidence (Appendix~\ref{app:budget-lower}). In multiplicative form the same potential gives explicit finite-length guarantees, which close in on the limits as $n$ grows: at $n=10^5$ the bands are \HBLowAHundredK--\HBUpAHundredK{} (A) and \HBLowRHundredK--\HBUpRHundredK{} (R) (Figure~\ref{fig:hhbudget}c). For one menu, pivot-based CountScan with a two-level factor already attains $\rho_1$ (Appendix~\ref{sw:envelope-sec}).

\subsection{Histories that hide every match}\label{subsec:limits-end}\label{subsec:draft-only}\label{subsec:retries}
The limits above measure how much of the history a release reveals, and two histories mark the end of that range. A writer who requests everywhere but releases the own draft (Proposition~\ref{prop:draft-only}), or who retries under a budget at which the limits of Theorem~\ref{thm:budget} vanish (Theorem~\ref{thm:full-key}), releases a text with exactly (draft retention) or nearly (retries, with the explicit failure probability of Theorem~\ref{thm:full-key}) the law of a writer without matches, so every valid count is zero with probability at least \ProtocolMissLower{} (Appendix~\ref{hn:statements}). Table~\ref{tab:layers} thus runs from fully visible to invisible matches.

\section{Experiments}\label{sec:experiments}

The experiments ask three questions in the order of the paper: whether counts stay valid after editing and what simultaneity costs (Section~\ref{sec:method}); whether detection and counting separate under natural writing policies (Section~\ref{sec:history-protocol}); and how close the attaining count comes to the limit, and at what length its certificate becomes useful (Section~\ref{subsec:unified}). Real-data experiments use OPT-1.3B \citep{zhang2022opt} and Qwen2.5-1.5B \citep{qwen2025technical} at temperature $\tau=1$ on books from the PG-19 corpus \citep{rae2019compressive}, with a new key per cell and the calibrator C1 fixed before generation, $g_{\mathrm{C1}}(p)=\min\{-\log p,2.5\}/(1-e^{-2.5})$ with evidence $\ell_t=\log(1+\tfrac12[g_{\mathrm{C1}}(P_t)-1])$, where $2.5$ is a truncation point. Reported intervals resample documents at a fixed key and history.

\paragraph{Are the counts reliable, and what does simultaneity cost?}\label{subsec:real}\label{subsec:synthetic}
No setting overcounted beyond the nominal level (Table~\ref{tab:checks}): synthetic documents with Zipf next-token distributions (Appendix~\ref{app:synthetic}), real mixed documents with random, rare-token, deletion and insertion edits, the simulated writing policies and paraphrasing by a key-blind instruction model (Appendix~\ref{app:ethree}). Paraphrasing $30\%$ of the sentences removes about a third of the retained matches, and the certified count falls in step; among the remaining matches CountScan's coverage falls slowly, while FissionGuard's falls faster but keeps higher purity, because it reports only intervals it validates on held-out evidence. Random replacement of $20\%$ of the tokens lowers the count further by breaking pairs at and after edited positions (Figure~\ref{fig:edits-main}). On documents of $1000$ new tokens generated once at temperature one, the mean $d/1000$ is \TtOptOneDthousand{} for OPT and \TtQwenOneDthousand{} for Qwen, and the loss to simultaneity approaches $\Delta/\xi^*$ from below as $L$ grows (Figure~\ref{fig:cost}); lower temperatures repeat more contexts and certify less (Table~\ref{tab:real}). A new key with Qwen2.5-7B and news-like text from the Colossal Clean Crawled Corpus (C4) give similar results (Appendices~\ref{app:newkeylarger} and~\ref{app:etwo}). At a common $256$-token cap, CountScan's cover$^\ast$, the share of retained matches inside its reports, exceeds FissionGuard's by \TfOptSOne{} \TfOptSOneCI{} (OPT) and \TfQwenSOne{} \TfQwenSOneCI{} (Qwen) on PG-19 and by \EtwoOptDiff{} and \EtwoQwenDiff{} on C4, with lower purity (Appendix~\ref{app:fgcomparison}); it also certifies a larger share than a dyadic reference family \citep{durand2018posthocspatial} and an all-interval union bound (Appendix~\ref{app:reffamily}).

\begin{table}[t]
\centering
\caption{Reliability, editing and the price of simultaneity. Validity: largest frequency of documents with any overcount. Editing: paraphrase of $0\%$ and $30\%$ of the sentences by a key-blind instruction model (\EtDocs{} Qwen2.5-1.5B documents); $H$: retained historical matches per document; $\bar d(T)$: certified count over disjoint intervals; cover$^\ast$: share of retained matches inside reports of CountScan (CS) and FissionGuard (FG), capped at $256$ tokens. Simultaneity: matches lost per block (Figure~\ref{fig:cost}).}\label{tab:checks}
\footnotesize\setlength{\tabcolsep}{3pt}
\resizebox{\columnwidth}{!}{%
\begin{tabular}{@{}llr@{}}
\toprule
Check & Setting & Result\\
\midrule
Validity & synthetic, \PEoneConfigs{} configurations & 0.70
\unskip\%\\
 & real mixed documents & \TtMixFwerMax\\
 & simulated writing policies (Wilson bound) & $\le$\HXSoneWilsonMax\\
 & paraphrased documents & $\EtMaxEvents/\EtDocs$\\
\midrule
Editing & $H$, $0\%\to30\%$ & \EtZeroKH{}$\to$\EtThirtyKH\\
 & $\bar d(T)$, $0\%\to30\%$ & \EtZeroBard{}$\to$\EtThirtyBard\\
 & cover$^\ast$ CS, $0\%\to30\%$ & \EtZeroCsCover{}$\to$\EtThirtyCsCover\\
 & cover$^\ast$ FG, $0\%\to30\%$ & \EtZeroFgCover{}$\to$\EtThirtyFgCover\\
\midrule
Simultaneity & OPT, $L=128$ and $512$ & \CostOPTOneTwoEight, \CostOPTFiveOneTwo\\
 & first-order cost $\Delta/\xi^*$ & \CostOPTPred\\
\bottomrule
\end{tabular}}
\end{table}

\begin{figure}[t]
\centering
\includegraphics[width=\columnwidth]{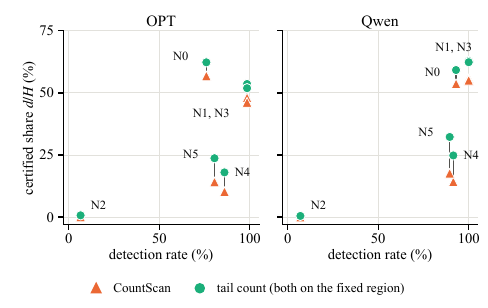}
\caption{Detection versus counting in simulated writing policies (\HXDocs{} documents per model and policy; Appendix~\ref{app:hhexp}). Horizontal: share of documents in which the Gamma test of Proposition~\ref{prop:gamma-power} detects at $\alpha=0.05$. Vertical: mean per-document share $d/H$ of the historical matches certified by CountScan on the fixed region and by the tail count of Proposition~\ref{sw:tail}. Policies as in Section~\ref{sec:experiments}. The gap is empirical for these methods; the general limits are those of Section~\ref{sec:history-protocol}.}\label{fig:hhexp}
\end{figure}

\paragraph{How large is the gap between detection and counting?}
We simulate writers like those of Section~\ref{sec:history-protocol} on PG-19 passages, using the next $300$ book tokens as the draft and a log hidden from the audit to record $H$ (Figure~\ref{fig:hhexp}; Appendix~\ref{app:hhexp}). Six policies vary adoption and retries: N0 adopts every suggestion, N1 half, N3 half after up to three retries, N4 a quarter, N5 half at half of the positions, and N2 none. Policies, thresholds and four tested statements were registered before generation. All counts stayed valid: the largest Wilson upper endpoint of an overcount frequency over \HXSoneCells{} cells is \HXSoneWilsonMax. Three regimes appear. Releasing only the draft (N2) leaves \HXNtwoKappaRange{} of positions as matches, yet the Gamma test rejects in \HXOptNtwoDet{} and \HXQwenNtwoDet{} of documents, inside the registered band \HXStwoPrimeThr{}, and none of the five evaluated counts certifies on average more than \HXNtwoCountMax{} of $H$, as Proposition~\ref{prop:draft-only} predicts. Keeping most of the draft (N4, N5) separates detection from counting: the test detects at least \HXSepDetMin{} of documents, while fixed-region CountScan certifies on average \HXSepCaMin--\HXSepCaMax{} of the matches and the tail count \HXSepBestMin--\HXSepBestMax. Adopting half of the suggestions (N1, N3) costs little: CountScan stays within \HXHalfGapMax{} points of N0, and for Qwen it certifies \HXQwenNoneCa{} and \HXQwenNthreeCa{}, above the registered bound of $50\%$, so the registered separation held in \HXStwoPass{} of \HXStwoTotal{} cells. In Figure~\ref{fig:hhexp} each vertical segment joins CountScan and the tail count of Theorem~\ref{thm:sw-sharp}(i), both on the same fixed region; the gap is evidence that the pivot sum leaves unused and that a count built on the menu structure, as in Section~\ref{subsec:unified}, can take.

\paragraph{How close does the unified count come to the limit?}
In the protocol of Theorem~\ref{thm:budget} we simulated three registered high histories (vertex strategies, retries, and interleaving with discarding and selection) at lengths $300$ to $10^4$ (Appendix~\ref{app:budget-sim}). Validity held: \HBEoneEvents{} overcounts in \HBEoneCells{} registered cells, and an overcount envelope below \HBAEoneWilson{} in a later follow-up. At $\kappa=\HBKappa$, $b=\HBBudget$, $n=10^4$ and the vertex history, $D_U$ certifies \HBRUnified{} of the positions under R and, in the follow-up, \HBAMix{} under A, against \HBAFixed{} for the fixed bet (paired difference \HBAGain{} points, \HBAGainCI) and \HBACone{} for C1. This is about half of $\rho_R=\HBRhoR$ and $\rho_A=\HBRhoA$, inside the bands of Figure~\ref{fig:hhbudget}c. \emph{When is the certificate useful?} Under the vertex history $D_U$ certifies at least one match in \EfPosAThreeK{} (A) and \EfPosRThreeK{} (R) of documents of $3000$ positions and at least $2.5\%$ of the positions in \EfTauATenK{} at $10^4$; under the retry and interleaving histories this share is reached with probability at least $0.9$ from \EfMinLenARetry{} and \EfMinLenAInterleave{} positions (thresholds set after the runs; Appendix~\ref{app:budget-usable}). Short fixed passages inside a long document remain out of reach at this budget; all cells and registered comparisons are in Appendix~\ref{app:budget-sim}.

\section{Discussion and Scope}\label{sec:conclusion}
Detecting is not counting: a key certifies counts that stay valid after editing, but a hidden history limits how much they can certify, by an exact amount that one count attains. A certified count concerns historical context--token matches, which copying and coincidence can also create, not a share of AI-written text.

\paragraph{Scope.} The protocols abstract key permissions of the audit; the exact formulas depend on the menu structure. The auditor holds the key and the model, so it can recompute the menus the assistant would have offered. A cap on requests per document can be enforced without a log. R models accept-or-reject interfaces and A free editing; the A count is valid under both. The pivot count of Section~\ref{sec:method} needs no menus; all guarantees assume Gumbel-max sampling and key-blind editors; score-based editing, green-list watermarking \citep{kirchenbauer2023watermark} and positional keys need their own calibrations; adaptive menus remain open.

\FloatBarrier

\newpage
\onecolumn
\title{Detecting Is Not Counting: CountScan and the Limits of Certifying Watermark Matches\\(Supplementary Material)}
\maketitle
\appendix
\noindent\textbf{Guide to the appendix.} Each section proves or supports one part of the main text. Appendices~\ref{app:validity}--\ref{app:proofs} prove the results of Sections~\ref{sec:setup} and~\ref{sec:method}; Appendices~\ref{hp:section}--\ref{hn:section} prove those of Section~\ref{sec:history-protocol}, in the order of its subsections; Appendices~\ref{app:algosec} and~\ref{app:related} give the algorithm and the comparisons with existing methods; and the remaining appendices give the designs and full results of the experiments in Section~\ref{sec:experiments}.
\section{Proofs for Sections~\ref{sec:setup} and~\ref{sec:method}: Validity}\label{app:validity}

This appendix supports Sections~\ref{sec:setup}--\ref{sec:theorem}. Section~\ref{app:steps} proves Theorem~\ref{thm:main} step by step and then Corollary~\ref{cor:select}, Section~\ref{app:propK} proves Proposition~\ref{prop:K}, Section~\ref{app:uniform} treats the uniform case, and Section~\ref{app:counter} collects the counterexamples. Standard results are cited from the author's version of \citet{durrett2019probability} (Version 5, January 11, 2019; page numbers refer to that version).

\paragraph{Conventions.} We write $\log0=-\infty$ and $e^{-\infty}=0$, and a finite sum containing $-\infty$ equals $-\infty$. A start with $\pi_a=0$ has $c_a=+\infty$ and is disabled; the default weights are $\pi_a=\ind\{a\in T\}/|T|$ if $T\ne\varnothing$ and $\pi_a\equiv0$ otherwise. An interval $S=[a,b]\cap T$ uses $\ell^{(\min S)}$ and $c_{\min S}$, and $d(\varnothing)=0$. By (S3) we work on the event that $P_t\in(0,1)$ for every $t\in T$, so that each $e^{(a)}_t$ lies in $[0,\infty)$. For a nonnegative random variable $X$ that is not known in advance to be integrable, $\EE[X\mid\mathcal D]$ denotes the monotone limit of $\EE[X\wedge k\mid\mathcal D]$; the properties of conditional expectation used below \citep[Thms.~4.1.9, 4.1.13 and 4.1.14, pp.~210--212]{durrett2019probability} extend to this case by monotone convergence.

\subsection{Proof of Theorem~\ref{thm:main}}\label{app:steps}\label{app:info}
The proof has two parts. Part~I shows $\PP(\Omega_\alpha)\ge1-\alpha$; Part~II shows that on $\Omega_\alpha$ every interval count is at most its truth. Theorem and page numbers of \citet{durrett2019probability} refer to the version stated above.

\paragraph{Setting.} Let $w_{1:n}$ be the final text, $m$ the context width and $\chi_t=w_{t-m:t-1}$ the context of position $t>m$. The retained positions are $T=\{t>m:\ \chi_s\ne\chi_t\text{ for all }s>m,\ s\ne t\}$, the history matches are $M=\{t\in T:\ (\chi_t,w_t)\text{ was returned by some watermarked request}\}$, the null positions are $N=T\setminus M$, and $\delta(S)=|S\cap M|$. The pivot is $P_t=1-U_{\chi_t,w_t}$, where $U_{\chi,v}$ is the coordinate of token $v$ in the pseudorandom vector of context $\chi$. The information that the betting rules may use is the $\sigma$-field
\[
\Ic=\Hc=\sigma\Big(A_{1:L},\ R,\ Q,\ \big(\zeta_\chi\ind\{\chi\notin\Xc_N\}\big)_{\chi\in V^m}\Big),
\]
where $A_{1:L}$ are the outputs of all $L$ history steps, $R$ and $Q$ are the key-free generation and editing randomness of Assumption~\ref{ass:Astar}, $\zeta_\chi$ is the pseudorandom vector of context $\chi$, and $\Xc_N=\{\chi_t:t\in N\}$ is the set of null contexts. Thus $\Ic$ contains the whole generation and editing record and every pseudorandom vector except those read by null positions; it plays the role of $\mathcal G=\sigma(W_{1:\infty},(\zeta_t)_{t\in M})$ in \citet[Appendix~A.1]{wei2026fissionguard}, enlarged by the generation record and editing randomness because our editors may react to generated text, as in the joint generation--editing filtration of \citet[Assumption~4.1]{blanchet2026spot}. The text is a function of $(A_{1:L},R,Q)$, so $n$, $w_{1:n}$, $T$ and $N$ are $\Ic$-measurable, which is (S1); in the uniform case of Section~\ref{app:uniform}, $\Ic=\G$ instead. The filtration is
\[
\F_t=\sigma\big(\Ic,\ P_s\ind\{s\in N\}:\ m<s\le t\big)\quad(t>m),\qquad \F_t=\Ic\quad(t\le m),
\]
so $\F_t$ adds to $\Ic$ the pivots of the null positions up to $t$; it increases in $t$, and $\{t\in N\}\in\Ic\subseteq\F_{t-1}$. Condition (S2) states that, for every $t>m$ and $p\in[0,1]$,
\begin{equation}\label{eq:S2cond}
\PP\big(P_t\le p\mid\F_{t-1}\big)\le p\qquad\text{almost surely on }\{t\in N\}.
\end{equation}

\paragraph{Part I, step 1: evidence and the event $\Omega_\alpha$.} For a start $a$ and a position $t$,
\[
e^{(a)}_t=1-\lambda^{(a)}_t+\lambda^{(a)}_tg^{(a)}_t(P_t),\qquad \ell^{(a)}_t=\log e^{(a)}_t,
\]
where (I1) $\lambda^{(a)}_t\in[0,1]$ is $\F_{t-1}$-measurable; (I2) $g^{(a)}_t\ge0$ on $(0,1)$ is $\F_{t-1}\otimes\mathcal B$-measurable as a function of $(\omega,p)$ and $\int_0^1g^{(a)}_t(p)\,dp\le1$; and (I4) $p\mapsto g^{(a)}_t(p)$ is nonincreasing. By (I3), the start weights $\pi_a\ge0$ are $\Ic$-measurable with $\sum_a\pi_a\le1$, and $c_a=\log\{1/(\alpha\pi_a)\}$. Put
\[
Z_a(b)=\sum_{t\in[a,b]\cap N}\ell^{(a)}_t,\qquad Z_a(a-1)=0,\qquad B_a=\Big\{\max_{a\le b\le n}Z_a(b)\ge c_a\Big\},\qquad \Omega_\alpha=\bigcap_aB_a^c,
\]
so that $\Omega_\alpha=\{Z_a(b)<c_a\ \text{for all}\ a\le b\}$ is the event of Section~\ref{sec:method}.

\paragraph{Part I, step 2: conditional mean of the null evidence (Lemma~\ref{lem:evalue}).} Fix $a$ and $t>m$, and write
\[
f(\omega,p)=1-\lambda^{(a)}_t(\omega)+\lambda^{(a)}_t(\omega)g^{(a)}_t(\omega,p),\qquad e^{(a)}_t=f(\cdot,P_t).
\]
By (I1), (I2) and (I4), $f\ge0$, $f$ is $\F_{t-1}\otimes\mathcal B$-measurable, and $p\mapsto f(\omega,p)$ is nonincreasing for every $\omega$.

(2a) \emph{Conditional law.} Since $P_t$ is real-valued, it has a regular conditional distribution $\mu(\omega,\cdot)$ given $\F_{t-1}$ \citep[Thm.~4.1.17, p.~215]{durrett2019probability}: for each Borel set $A$, $\mu(\cdot,A)$ is a version of $\PP(P_t\in A\mid\F_{t-1})$, and $\mu(\omega,\cdot)$ is a probability measure. For each rational $p\in[0,1]$, \eqref{eq:S2cond} gives $\mu(\omega,[0,p])\le p$ for almost every $\omega\in\{t\in N\}$. Removing the countably many exceptional null sets leaves a null set $\Lambda$ outside which the bound holds for all rational $p$; for real $p$ take rationals $p_k\downarrow p$, so that $\mu(\omega,[0,p])=\lim_k\mu(\omega,[0,p_k])\le\lim_kp_k=p$. Hence
\begin{equation}\label{eq:rcdbound}
\mu(\omega,[0,p])\le p\qquad\text{for all }p\in[0,1]\text{ and all }\omega\in\{t\in N\}\setminus\Lambda .
\end{equation}

(2b) \emph{Integration against the conditional law.} For every nonnegative $\F_{t-1}\otimes\mathcal B$-measurable $h$,
\begin{equation}\label{eq:disint}
\EE\big[h(\cdot,P_t)\mid\F_{t-1}\big](\omega)=\int h(\omega,p)\,\mu(\omega,dp)\qquad\text{almost surely}.
\end{equation}
For $h(\omega,p)=\ind_B(\omega)\ind_C(p)$ with $B\in\F_{t-1}$ and Borel $C$, the right side is $\ind_B\mu(\cdot,C)$, a version of $\ind_B\PP(P_t\in C\mid\F_{t-1})=\EE[\ind_B\ind_C(P_t)\mid\F_{t-1}]$ \citep[Thm.~4.1.14, p.~212]{durrett2019probability}. The sets $E\in\F_{t-1}\otimes\mathcal B$ for which \eqref{eq:disint} holds with $h=\ind_E$ form a $\lambda$-system, by linearity and monotone convergence of conditional expectations \citep[Thm.~4.1.9, p.~210]{durrett2019probability}; it contains the $\pi$-system of product sets, so it is all of $\F_{t-1}\otimes\mathcal B$ by the $\pi$-$\lambda$ theorem \citep[Thm.~A.1.4, p.~456]{durrett2019probability}. Linearity extends \eqref{eq:disint} to nonnegative simple $h$, and monotone convergence to nonnegative $h$.

(2c) \emph{Level sets.} For $y>0$ let $\rho_y(\omega)=\int_0^1\ind\{f(\omega,p)>y\}\,dp\in[0,1]$. If $f(\omega,p_0)>y$, then $f(\omega,u)\ge f(\omega,p_0)>y$ for every $u\in(0,p_0]$, because $f(\omega,\cdot)$ is nonincreasing; hence $\rho_y(\omega)\ge\int_0^{p_0}1\,du=p_0$. Therefore
\begin{equation}\label{eq:levelincl}
\{p:\ f(\omega,p)>y\}\subseteq[0,\rho_y(\omega)] .
\end{equation}

(2d) \emph{Integration over levels.} Every $x\ge0$ satisfies $x=\int_0^\infty\ind\{x>y\}\,dy$. Fix $\omega\in\{t\in N\}\setminus\Lambda$. By Tonelli's theorem, then \eqref{eq:levelincl}, then \eqref{eq:rcdbound} with $p=\rho_y(\omega)$, and Tonelli's theorem again,
\[
\int f(\omega,p)\,\mu(\omega,dp)=\int_0^\infty\mu\big(\omega,\{p:f(\omega,p)>y\}\big)\,dy
\le\int_0^\infty\mu\big(\omega,[0,\rho_y(\omega)]\big)\,dy
\le\int_0^\infty\rho_y(\omega)\,dy
=\int_0^1f(\omega,p)\,dp .
\]
By (I1) and (I2), $\int_0^1f(\omega,p)\,dp=1-\lambda^{(a)}_t+\lambda^{(a)}_t\int_0^1g^{(a)}_t(p)\,dp\le1$. Combining with \eqref{eq:disint} for $h=f$,
\begin{equation}\label{eq:nullmean}
\EE\big[e^{(a)}_t\mid\F_{t-1}\big]\le1-\lambda^{(a)}_t+\lambda^{(a)}_t\int_0^1g^{(a)}_t(p)\,dp\le1\qquad\text{almost surely on }\{t\in N\},
\end{equation}
which is Lemma~\ref{lem:evalue}; with $\lambda^{(a)}_t=1$ it reads $\EE[g^{(a)}_t(P_t)\mid\F_{t-1}]\le\int_0^1g^{(a)}_t(p)\,dp\le1$. This is the stochastic-order step of \citet[Proposition~2.1]{vovk2021evalues}, carried out under the conditional law.

\paragraph{Part I, step 3: a supermartingale for each start.} Fix $a$. For $t\ge a$ let
\[
\Gamma^{(a)}_t=\begin{cases}e^{(a)}_t,&t\in N,\\ 1,&t\notin N,\end{cases}
\qquad\text{that is,}\qquad \Gamma^{(a)}_t=1+\ind\{t\in N\}\big(e^{(a)}_t-1\big).
\]
$\Gamma^{(a)}_t$ is $\F_t$-measurable: $\ind\{t\in N\}$ is $\Ic$-measurable, $\lambda^{(a)}_t$ and $g^{(a)}_t$ are $\F_{t-1}$-measurable, and on $\{t\in N\}$ the pivot $P_t=P_t\ind\{t\in N\}$ is $\F_t$-measurable. Since $\{t\notin N\}\in\F_{t-1}$, \eqref{eq:nullmean} gives $\EE[\Gamma^{(a)}_t\mid\F_{t-1}]\le1$ almost surely. Put $M^{(a)}_{a-1}=1$ and, for $b\ge a$,
\[
M^{(a)}_b=\prod_{t=a}^b\Gamma^{(a)}_t=\prod_{t\in[a,b]\cap N}e^{(a)}_t=\exp Z_a(b),\qquad M^{(a)}_b=M^{(a)}_{b-1}\Gamma^{(a)}_b .
\]
For $t>n$ we have $t\notin N$, so $\Gamma^{(a)}_t=1$ and $M^{(a)}_b=M^{(a)}_n$ for $b\ge n$. Since $M^{(a)}_{b-1}\ge0$ is $\F_{b-1}$-measurable, \citet[Thm.~4.1.14, p.~212]{durrett2019probability} gives
\[
\EE\big[M^{(a)}_b\mid\F_{b-1}\big]=M^{(a)}_{b-1}\,\EE\big[\Gamma^{(a)}_b\mid\F_{b-1}\big]\le M^{(a)}_{b-1}.
\]
Thus $(M^{(a)}_b)_{b\ge a-1}$ is a nonnegative supermartingale with respect to $(\F_b)_{b\ge a-1}$, starting at one. Its values are finite, because $P_t\in(0,1)$ by (S3) and $g^{(a)}_t$ is finite on $(0,1)$.

\paragraph{Part I, step 4: Ville's inequality and the start budget.}
\begin{lemma}[Finite-horizon Ville inequality]\label{lem:ville}
Let $(Y_k)_{k=0}^K$ be nonnegative, finite and adapted to a filtration $(\F'_k)$, with $Y_0=1$ and $\EE[Y_k\mid\F'_{k-1}]\le Y_{k-1}$ for $k=1,\dots,K$. If $C$ is $\F'_0$-measurable with values in $[1,\infty]$, then
\[
\PP\Big(\max_{0\le k\le K}Y_k\ge C\ \Big|\ \F'_0\Big)\le\frac1C .
\]
\end{lemma}
\begin{proof}
Let $\tau=\min\{k:Y_k\ge C\}$, with $\tau=K$ if no such $k$ exists. Since $C$ is $\F'_0$-measurable, $\{\tau\le k-1\}=\bigcup_{j\le k-1}\{Y_j\ge C\}\in\F'_{k-1}$, and so $\{\tau\ge k\}\in\F'_{k-1}$. The stopped process satisfies
\[
Y_{\tau\wedge k}=\ind\{\tau\le k-1\}\,Y_{\tau\wedge(k-1)}+\ind\{\tau\ge k\}\,Y_k .
\]
Taking conditional expectations given $\F'_{k-1}$ and pulling out the $\F'_{k-1}$-measurable factors, we obtain
\[
\EE[Y_{\tau\wedge k}\mid\F'_{k-1}]\le\ind\{\tau\le k-1\}\,Y_{\tau\wedge(k-1)}+\ind\{\tau\ge k\}\,Y_{k-1}=Y_{\tau\wedge(k-1)},
\]
because $\tau\wedge(k-1)=k-1$ on $\{\tau\ge k\}$. Iterating this inequality with the tower property \citep[Thm.~4.1.13, p.~212]{durrett2019probability} gives $\EE[Y_\tau\mid\F'_0]\le Y_0=1$. On the event $\{\max_kY_k\ge C\}\cap\{C<\infty\}$ we have $Y_\tau\ge C$, while the event $\{\max_kY_k\ge C\}$ is empty on $\{C=\infty\}$ because each $Y_k$ is finite. Hence $C\,\PP(\max_kY_k\ge C\mid\F'_0)\le\EE[Y_\tau\mid\F'_0]\le1$, which is the claim.
\end{proof}
Fix $a$ and apply Lemma~\ref{lem:ville} with $Y_k=M^{(a)}_{a-1+k}$, $\F'_k=\F_{a-1+k}$ and $C=1/(\alpha\pi_a)=e^{c_a}$, which is $\Ic$-measurable, hence $\F'_0$-measurable, with values in $[1,\infty]$ because $\alpha<1$ and $\pi_a\le1$. For every $K$,
\[
\PP\Big(\max_{0\le k\le K}M^{(a)}_{a-1+k}\ge\frac1{\alpha\pi_a}\ \Big|\ \F_{a-1}\Big)\le\alpha\pi_a .
\]
Since $M^{(a)}_b$ is constant for $b\ge n$, the events on the left increase in $K$ to $\{\max_{b\ge a-1}M^{(a)}_b\ge1/(\alpha\pi_a)\}$, which contains $B_a$ because $M^{(a)}_b=\exp Z_a(b)$. Monotone convergence \citep[Thm.~4.1.9, p.~210]{durrett2019probability} therefore gives
\[
\PP(B_a\mid\F_{a-1})\le\alpha\pi_a\qquad\text{almost surely}.
\]
By the tower property \citep[Thm.~4.1.13, p.~212]{durrett2019probability} and because $\pi_a$ is $\Ic$-measurable, $\PP(B_a\mid\Ic)=\EE[\PP(B_a\mid\F_{a-1})\mid\Ic]\le\alpha\pi_a$. A union over starts and (I3) give
\[
\PP(\Omega_\alpha^c\mid\Ic)=\PP\Big(\bigcup_aB_a\ \Big|\ \Ic\Big)\le\sum_a\PP(B_a\mid\Ic)\le\alpha\sum_a\pi_a\le\alpha ,
\]
and taking expectations, $\PP(\Omega_\alpha)\ge1-\alpha$. Starts $a>n$ have empty $B_a$, so the union is over $a\le n$.

\paragraph{Part II: from $\Omega_\alpha$ to all intervals.} Fix $\omega\in\Omega_\alpha$, so that $Z_a(b)<c_a$ for all $1\le a\le b\le n$. Take an interval $[a,b]$ and $S=[a,b]\cap T$. If $S=\varnothing$, then $d(S)=0=\delta(S)$. Otherwise let $a_0=\min S$. Since $N\subseteq T$, $[a,b]\cap N=[a_0,b]\cap N=S\cap N$, and therefore
\[
\sum_{t\in S\cap N}\ell^{(a_0)}_t=Z_{a_0}(b).
\]
Write $S_M=S\cap M$ and $S_N=S\cap N$, so $S=S_M\cup S_N$ and $|S_M|=\delta(S)$. Recall from Definition~\ref{def:trim} that $\Top_j(S)$ is the sum of the $j$ largest positive values $\ell^{(a_0)}_t$, $t\in S$ (all of them if fewer than $j$ are positive), that $R_S(j)=\sum_{t\in S}\ell^{(a_0)}_t-\Top_j(S)$, and that
\[
d(S)=\min\{j\ge0:\ R_S(j)<c_{a_0}\},
\]
the smallest number of largest evidence values whose removal brings the remaining evidence below the threshold. Each $\ell^{(a_0)}_t\le(\ell^{(a_0)}_t)^+$, and the $\delta(S)$ values $(\ell^{(a_0)}_t)^+$, $t\in S_M$, sum to at most the $\delta(S)$ largest positive values in $S$; hence $\sum_{t\in S_M}\ell^{(a_0)}_t\le\Top_{\delta(S)}(S)$ (Lemma~\ref{lem:trim}). Consequently
\[
R_S(\delta(S))=\sum_{t\in S_M}\ell^{(a_0)}_t+\sum_{t\in S_N}\ell^{(a_0)}_t-\Top_{\delta(S)}(S)\le\sum_{t\in S_N}\ell^{(a_0)}_t=Z_{a_0}(b)<c_{a_0}.
\]
So removing as many values as there are true matches already meets the stopping rule, and the minimum satisfies $d(S)\le\delta(S)$. As the interval was arbitrary,
\[
\Omega_\alpha\subseteq\big\{\forall\,a\le b:\ d(S_{ab})\le\delta(S_{ab})\big\},\qquad
\PP\big(\forall\,a\le b:\ d(S_{ab})\le\delta(S_{ab})\big)\ge\PP(\Omega_\alpha)\ge1-\alpha .
\]
This proves Theorem~\ref{thm:main}.

\begin{proof}[Proof of Corollary~\ref{cor:select}]
All three statements hold on $\Omega_\alpha$. For (a), $\hat S(\omega)$ is an interval for every $\omega$, so $\{d(\hat S)>\delta(\hat S)\}\subseteq\Omega_\alpha^c$. For (b), rejecting a true hypothesis $\delta(S)\le k$ requires $d(S)\ge k+1>\delta(S)$, which happens only on $\Omega_\alpha^c$. For (c), for disjoint intervals $S_i\subseteq A$ we have $\sum_id(S_i)\le\sum_i\delta(S_i)=|(\bigcup_iS_i)\cap M|\le\delta(A)$ on $\Omega_\alpha$, and taking the maximum over such collections gives $\bar d(A)\le\delta(A)$.
\end{proof}

\subsection{Proof of Proposition~\ref{prop:K}}\label{app:propK}

We first make Assumption~\ref{ass:Astar} precise. Let $\Xc=V^m$. (K1) All coordinates $U_{\chi,v}$, $\chi\in\Xc$, $v\in V$, are i.i.d.\ $\Unif(0,1)$. (K2) There are a random element $R$ independent of $\zeta=(\zeta_\chi)_\chi$ and a deterministic $L$ such that, for $s=1,\dots,L$, measurable functions of $(A_{1:s-1},R)$ give a status $W_s\in\{0,1,2\}$, a context $\kappa_s$ and an NTP distribution $\pi_s$, and $A_s=\argmax_{v\in\mathrm{supp}\,\pi_s}\log U_{\kappa_s,v}/\pi_{s,v}$ if $W_s=1$, $A_s=u_s(A_{1:s-1},R)$ if $W_s=0$, and $A_s=\bot$ if $W_s=2$. (K3) There is $Q$ with $(R,Q)$ independent of $\zeta$, and the final text is a measurable function of $(A_{1:L},R,Q)$. History matches are $\theta_t=\ind\{\exists s:W_s=1,\kappa_s=\chi_t,A_s=w_t\}$. Write $\Xc_N=\{\chi_t:t\in N\}$ and $\Hc=\sigma(A_{1:L},R,Q,(\zeta_\chi\ind\{\chi\notin\Xc_N\})_{\chi\in\Xc})$.

We factor the history likelihood by context and use monotonicity in a coordinate that never won. The following elementary inequality supplies the tail bound.

\begin{lemma}\label{lem:avg}
If $\psi:[0,1]\to[0,\infty)$ is bounded, measurable and nonincreasing, then $\int_u^1\psi\le(1-u)\int_0^1\psi$ for every $u\in[0,1]$.
\end{lemma}
\begin{proof}
Since $\psi$ is nonincreasing, $\int_0^u\psi\ge u\psi(u)$ and $\int_u^1\psi\le(1-u)\psi(u)$. Hence $(1-u)\int_0^1\psi-\int_u^1\psi=(1-u)\int_0^u\psi-u\int_u^1\psi\ge0$.
\end{proof}

\begin{proof}[Proof of Proposition~\ref{prop:K}]
We verify (S1), (S3) and then (S2).

The status, the contexts and the outputs of the history are functions of $(A_{1:L},R)$ by (K2), and the text is a function of $(A_{1:L},R,Q)$ by (K3). Hence $\chi_t$, $\theta$, $T$, $M$, $N$ and $\Xc_N$ are $\sigma(A_{1:L},R,Q)$-measurable, which proves (S1). Each $U_{\chi,v}$ is uniform by (K1) and there are finitely many of them, which proves (S3).

To prove (S2), we first describe the history as a function of $\zeta$. Iterating (K2) gives a measurable map $\Phi$ with $A_{1:L}=\Phi(\zeta,R)$, valued in the finite set $(V\cup\{\bot\})^L$. Fix a value $a$ and $r$, and let $W_s(a,r)$, $\kappa_s(a,r)$ and $\pi_s(a,r)$ denote the status, context and distribution of step $s$ computed from $(a_{1:s-1},r)$. Let $\mathcal R(a)$ be the set of $r$ for which, at every step, $W_s(a,r)=0$ implies $a_s=u_s(a_{1:s-1},r)$, $W_s(a,r)=2$ implies $a_s=\bot$, and $W_s(a,r)=1$ implies $a_s\in\mathrm{supp}\,\pi_s(a,r)$. Once the first $s-1$ outputs equal $a_{1:s-1}$, step $s$ uses $W_s(a,r)$, $\kappa_s(a,r)$ and $\pi_s(a,r)$; induction on $s$ therefore gives
\begin{equation}\label{eq:factor}
\ind\{\Phi(z,r)=a\}=\ind\{r\in\mathcal R(a)\}\prod_{\chi\in\Xc}\varphi^{a,r}_\chi(z_\chi),
\end{equation}
where $\varphi^{a,r}_\chi(z_\chi)=\prod_{s:\,W_s(a,r)=1,\,\kappa_s(a,r)=\chi}\ind\{a_s=\argmax_{v\in\mathrm{supp}\,\pi_s(a,r)}\log z_{\chi,v}/\pi_{s,v}(a,r)\}$ depends only on $z_\chi$, and $(z,r)\mapsto\varphi^{a,r}_\chi(z_\chi)$ is jointly measurable.

Next, we show that each factor is monotone in the coordinate of a token that never won. Suppose that $a_s\ne h$ for every $s$ with $W_s(a,r)=1$ and $\kappa_s(a,r)=\chi$. If the other coordinates of $z_\chi$ are held fixed, then $x=z_{\chi,h}\mapsto\varphi^{a,r}_\chi$ is nonincreasing. Indeed, if $h\notin\mathrm{supp}\,\pi_s$, the corresponding factor does not depend on $x$; otherwise, lowering $x$ lowers only the score of $h\ne a_s$ and leaves all other scores unchanged, so a winner $a_s$ remains the winner. With $\mu_1=\Unif([0,1]^V)$, define
\[
\mathrm{num}_u=\int\varphi^{a,r}_\chi\,\ind\{z_{\chi,h}\ge u\}\,d\mu_1,\qquad \mathrm{den}=\int\varphi^{a,r}_\chi\,d\mu_1 .
\]
Integrating first in $z_{\chi,h}$, Lemma~\ref{lem:avg} applies for every value of the other coordinates, and integrating these out gives
\begin{equation}\label{eq:numden}
\mathrm{num}_u\le(1-u)\,\mathrm{den},\qquad u\in[0,1].
\end{equation}

We now pass to the filtration. Fix $t>m$ and let $\Jc_t=\sigma(A_{1:L},R,Q,(\zeta_\chi\ind\{\chi\ne\chi_t\})_{\chi\in\Xc})$, the information other than the row of the context $\chi_t$. We claim that $\ind\{t\in N\}Z$ is $\Jc_t$-measurable for every $\F_{t-1}$-measurable $Z$. Since $\{t\in N\}$ is $\Jc_t$-measurable, it suffices to check the generators of $\F_{t-1}$. On $\{t\in N\}$ we have $\chi_t\in\Xc_N$, hence $\zeta_\chi\ind\{\chi\notin\Xc_N\}=(\zeta_\chi\ind\{\chi\ne\chi_t\})\ind\{\chi\notin\Xc_N\}$. Moreover, for $m<s<t$ we have $\chi_s\ne\chi_t$ on $\{s\in N\}\cap\{t\in N\}$ by the definition of $T$, hence $P_s\ind\{s\in N\}=\ind\{s\in N\}\big(1-\sum_\chi\ind\{\chi_s=\chi\}(\zeta_\chi\ind\{\chi\ne\chi_t\})_{w_s}\big)$. Both are $\Jc_t$-measurable on $\{t\in N\}$, which proves the claim.

On $\{t\in N\}$, evaluate the quantities of \eqref{eq:numden} at $a=A_{1:L}$, $r=R$, $\chi=\chi_t$ and $h=w_t$, and let $\Gamma_u=\mathrm{num}_u/\mathrm{den}$ if $\mathrm{den}>0$ and $\Gamma_u=1-u$ otherwise; let $\Gamma_u=0$ off $\{t\in N\}$. Then $\Gamma_u$ is $\sigma(A_{1:L},R,Q)$-measurable. By the definition of $\theta$, no watermarked step in context $\chi_t$ produced $w_t$, so the monotonicity above applies and \eqref{eq:numden} gives $\Gamma_u\le1-u$. We show that
\begin{equation}\label{eq:Gamma}
\PP\big(U_{\chi_t,w_t}\ge u\mid\Jc_t\big)=\Gamma_u\qquad\text{almost surely on }\{t\in N\},
\end{equation}
that is, by the defining property of conditional expectation, $\PP(E\cap\{U_{\chi_t,w_t}\ge u\})=\EE[\ind_E\Gamma_u]$ for every $E\in\Jc_t$ contained in $\{t\in N\}$.
Consider first $E=\{A_{1:L}=a\}\cap\{(R,Q)\in B\}\cap\{t\in N\}\cap\bigcap_\chi\{\zeta_\chi\ind\{\chi\ne\chi_t\}\in C_\chi\}$. By (K2) and (K3), the joint law of $(\zeta,(R,Q))$ is $\mu_1^{\otimes\Xc}\otimes\PP_{(R,Q)}$, and $\{A_{1:L}=a\}=\{\Phi(\zeta,R)=a\}$. For fixed $(r,q)$, the context $\chi^\ast=\chi_t$, the token $h=w_t$ and the event $\{t\in N\}$ are fixed. Fubini's theorem and \eqref{eq:factor} then express the left side of \eqref{eq:Gamma} as
\[
\int\PP_{(R,Q)}(dr,dq)\,\ind_B\,\ind\{t\in N\}\,\ind\{r\in\mathcal R(a)\}\,\ind\{0\in C_{\chi^\ast}\}\prod_{\chi\ne\chi^\ast}\Big[\int\varphi^{a,r}_\chi\ind_{C_\chi}\,d\mu_1\Big]\,\mathrm{num}_u ,
\]
and the right side as the same integral with $\mathrm{num}_u$ replaced by $\mathrm{den}\cdot\Gamma_u$. The two coincide, since $\mathrm{den}\cdot\Gamma_u=\mathrm{num}_u$ when $\mathrm{den}>0$ and both vanish when $\mathrm{den}=0$. Sets of this form are closed under finite intersections and generate the trace of $\Jc_t$ on $\{t\in N\}$, because $A_{1:L}$ takes finitely many values; both sides of \eqref{eq:Gamma} are finite measures in $E$. The $\pi$-$\lambda$ theorem \citep[Thm.~A.1.4, p.~456]{durrett2019probability} therefore extends the identity to all $E\in\Jc_t$ contained in $\{t\in N\}$, which is \eqref{eq:Gamma}.

Finally, fix $p\in[0,1]$. Since $P_t\le p$ if and only if $U_{\chi_t,w_t}\ge1-p$, for every bounded $\F_{t-1}$-measurable $Z$ the claim above and \eqref{eq:Gamma} give $\EE[\ind\{t\in N\}Z\ind\{P_t\le p\}]=\EE[\ind\{t\in N\}Z\,\Gamma_{1-p}]$. Hence, by the tower property and $\Gamma_{1-p}\le p$,
\[
\PP\big(P_t\le p\mid\F_{t-1}\big)=\EE\big[\Gamma_{1-p}\mid\F_{t-1}\big]\le p\qquad\text{almost surely on }\{t\in N\},
\]
which is (S2). This completes the proof of Proposition~\ref{prop:K}.
\end{proof}

\begin{remark}[Admissible betting information]\label{rem:auditorinfo}
Under $\Ic=\Hc$, the text and the pivots at earlier positions $m<s<t$ are $\F_{t-1}$-measurable: for $s\in N$ by definition, and for $s\in M$ or masked $s$ because then $\chi_s\notin\Xc_N$. Betting rules may use only this information. A key holder could compute other quantities, such as pivots of prompt-padded positions $s\le m$, hash coordinates of tokens that do not occur, or current and future pivots; these are not admissible merely because they are computable. Choosing which interval to report after seeing all scores is protected by Corollary~\ref{cor:select}(a); choosing calibrators or weights after seeing the scores is not.
\end{remark}

\ifanon\else
\begin{remark}[Relation to the earlier draft]
An earlier version stated the editing result with the information $\G=\sigma(\theta,w_{1:n},(\zeta_s)_{s\notin N})$, which includes prompt-padded positions $s\le m$. If such a position shares its context with a null position, $\G$ reveals that null pivot, and a $\G$-measurable nonincreasing calibrator can then exceed the level: with $m=2$, prompt $(A,B)$ and edited text $(C,D,A,B,E,B,C,F)$, a calibrator concentrating on the revealed $P_5,P_8$ gives error $\alpha(1-\log\alpha)\approx0.20$ at $\alpha=0.05$. The statement above, with $\Ic=\Hc$ and Remark~\ref{rem:auditorinfo}, corrects this. The fixed C1 calibrator with uniform weights, used in all experiments, is admissible under both statements.
\end{remark}
\fi

\subsection{Proof of Corollary~\ref{cor:coupling}}\label{app:coupling}
Let $(W_t)_{t>m}$ be i.i.d.\ $\Unif(0,1)$, independent of everything else, and $\mathcal G_t=\sigma(\Ic,\,P_s\ind\{s\in N\},\,W_s:m<s\le t)$.

First, the added randomization does not change the conditional law of $P_t$: since $W$ is independent of all original variables, $\PP(P_t\le p\mid\mathcal G_{t-1})=\PP(P_t\le p\mid\F_{t-1})$ almost surely, and by Condition~\ref{cond:S}(S2) the right side is at most $p$ on $\{t\in N\}$.

Second, on $\{t\in N\}$ let $F_t$ be a regular conditional distribution function of $P_t$ given $\mathcal G_{t-1}$. For each rational $p$, the first step gives $F_t(p)\le p$ almost surely; taking the countable intersection over rational $p$ and using right continuity yields $F_t(p)\le p$ for all $p$ simultaneously, almost surely.

Third, set $V_t=F_t(P_t-)+W_t\{F_t(P_t)-F_t(P_t-)\}$ for $t\in N$ and $V_t=W_t$ otherwise, the randomized distributional transform with conditional distribution functions \citep[eq.~(1) and Theorem~1]{brockwell2007universal}; we give the one-step argument for a general filtration. On $N$, $V_t\le F_t(P_t)\le P_t$, and every $V_t$ is $\mathcal G_t$-measurable. Given $\mathcal G_{t-1}$, $P_t$ has distribution function $F_t$ and $W_t$ is uniform and independent of $P_t$. For $u\in(0,1)$ let $x=\inf\{y:F_t(y)\ge u\}$, so $F_t(x-)\le u\le F_t(x)$. Up to a null set, $\{V_t\le u\}$ consists of $\{P_t<x\}$ and, if $F_t$ jumps at $x$, of $\{P_t=x,\ W_t\le (u-F_t(x-))/(F_t(x)-F_t(x-))\}$; its conditional probability is $u$. Hence $V_t$ is uniform given $\mathcal G_{t-1}$ on $\{t\in N\}$; off $N$ this holds because $W_t$ is independent of $\mathcal G_{t-1}$, and $\{t\in N\}\in\mathcal G_{t-1}$.

Fourth, for $m<t_1<\dots<t_k$ the variables $V_{t_1},\dots,V_{t_{j-1}}$ are $\mathcal G_{t_j-1}$-measurable, and the tower property gives $\PP(V_{t_1}\le u_1,\dots,V_{t_k}\le u_k\mid\Ic)=\prod_ju_j$. The whole sequence $(V_t)_{t>m}$ is therefore i.i.d.\ uniform and independent of $\Ic$; since $N$ is $\Ic$-measurable, the conditional law of $(V_t)_{t\in N}$ given $\Ic$ is the product of uniforms.

Finally, let the error indicator be $\Psi(P_N,Y,\xi)$ with $Y$ $\Ic$-measurable, $\xi$ independent of the data and of $W$, and $\Psi$ nonincreasing in each null pivot for fixed $(Y,\xi)$. The level-$\alpha$ hypothesis refers to the same procedure with the same $Y$ and $\xi$. Since $V_t\le P_t$,
\[
\PP(\text{error})=\EE\,\Psi(P_N,Y,\xi)\le\EE\,\Psi(V,Y,\xi)=\EE\big[\EE\{\Psi(V,Y,\xi)\mid\Ic\}\big]\le\alpha .
\]
Under $\Ic=\Hc$, the pivots of matched and masked positions are $\Ic$-measurable (Remark~\ref{rem:auditorinfo}). With calibrators, hedges and start weights fixed given $\Ic$, CountScan's all-interval error event $\{\exists S:d(S)>\delta(S)\}$ is of this form, because its evidence is nonincreasing in each pivot and the trimmed count is nondecreasing in each evidence value; reports selected afterwards are covered because their errors lie in this event. With hedges that depend on earlier null pivots, monotonicity need not hold, and validity follows instead from Theorem~\ref{thm:main}. This completes the proof of Corollary~\ref{cor:coupling}.

\subsection{The Uniform Case}\label{app:uniform}

\begin{assumption}[Unused pseudorandomness]\label{ass:A}
Let $\G=\sigma(\theta,w_{1:n},(\zeta_{\chi_s})_{s>m,\,s\notin N})$. (i) For distinct $t_1,\dots,t_k$ and Borel $C$, $\PP\big((\zeta_{\chi_{t_i}})_{i\le k}\in C\mid\G\big)=\mu_1^{\otimes k}(C)$ almost surely on $\{t_1,\dots,t_k\in N\}$, where $\mu_1=\Unif([0,1]^V)$. (ii) Each $U_{\chi_t,v}$, $t>m$, is marginally uniform.
\end{assumption}
This is the conditional form of \citet[Assumption 3.1, p.~6]{wei2026fissionguard}, restricted to positions $s>m$. Related conditions are \citet[Working Hypotheses 2.1--2.2, p.~328]{li2025framework}, \citet[Assumption 1(i), p.~9]{su2026online} and \citet[Assumption 4.1(a)(b), p.~10]{blanchet2026spot}. The following standard lemma allows us to integrate out a vector that is independent of the conditioning information.

\begin{lemma}[Freezing]\label{lem:freeze}
Let $\mathcal D$ be a sub-$\sigma$-field, $D_0\in\mathcal D$, $X$ a $[0,1]^k$-valued random vector and $\mu$ a probability measure such that $\PP(X\in C\mid\mathcal D)=\mu(C)$ almost surely on $D_0$ for all Borel $C$. If $f\ge0$ is $\mathcal D\otimes\mathcal B$-measurable, then $\EE[f(\cdot,X)\mid\mathcal D]=\int f(\cdot,x)\,\mu(dx)$ almost surely on $D_0$.
\end{lemma}
\begin{proof}
For $f=\ind_B\ind_C$ with $B\in\mathcal D$ the claim is the hypothesis multiplied by $\ind_B$. The sets $E\in\mathcal D\otimes\mathcal B$ for which the identity holds with $f=\ind_E$ form a $\lambda$-system containing the $\pi$-system of product sets, hence all of $\mathcal D\otimes\mathcal B$ by the $\pi$-$\lambda$ theorem. Linearity and monotone convergence extend the identity to nonnegative $f$.
\end{proof}

\begin{proposition}[Unused randomness gives equality]\label{prop:U}
Under Assumption~\ref{ass:A}, Condition~\ref{cond:S} holds with $\Ic=\G$ and with equality in (S2).
\end{proposition}
\begin{proof}
Conditions (S1) and (S3) are immediate from the definition of $\G$ and from (ii). For distinct null positions $t_1,\dots,t_k$, Lemma~\ref{lem:freeze} with $f(\omega,x)=\prod_i\ind\{1-x_{i,w_{t_i}(\omega)}\le p_i\}$ gives $\PP(P_{t_i}\le p_i,\,i\le k\mid\G)=\prod_ip_i$ on $\{t_1,\dots,t_k\in N\}$, because the selected coordinates are distinct. Fix $t$ and a set $J\subseteq\{m+1,\dots,t-1\}$. On the $\G$-measurable event $\{t\in N,\ N\cap\{m+1,\dots,t-1\}=J\}$, the pivots $(P_s)_{s\in J}$ and $P_t$ are therefore i.i.d.\ uniform given $\G$, and $\F_{t-1}$ is generated there by $\G$ and $(P_s)_{s\in J}$. Hence $\PP(P_t\le p\mid\F_{t-1})=p$ on this event, and since there are finitely many $J$, almost surely on $\{t\in N\}$.
\end{proof}

\begin{corollary}[The two cases]\label{cor:mono}
(a) Under Assumption~\ref{ass:Astar} and (I1)--(I4) with $\Ic=\Hc$, Theorem~\ref{thm:main} holds. (b) Under Assumption~\ref{ass:A} and (I1)--(I3) with $\Ic=\G$, Theorem~\ref{thm:main} holds without (I4).
\end{corollary}
\begin{proof}
Part (a) follows from Proposition~\ref{prop:K} and Theorem~\ref{thm:main}. For (b), Proposition~\ref{prop:U} gives equality in (S2), so the conditional law of $P_t$ given $\F_{t-1}$ is uniform on $\{t\in N\}$. Lemma~\ref{lem:freeze} then gives $\EE[e^{(a)}_t\mid\F_{t-1}]=1-\lambda^{(a)}_t+\lambda^{(a)}_t\int_0^1g^{(a)}_t\le1$ without monotonicity, which replaces Lemma~\ref{lem:evalue}; the rest of the proof of Theorem~\ref{thm:main} is unchanged.
\end{proof}

\subsection{Counterexamples}\label{app:counter}
\emph{Predictability (I1).} With $g=-\log p$ and $\lambda_t=\ind\{P_t\le1/2\}$ under a uniform null, $\EE e_t=\tfrac12+\int_0^{1/2}(-\log p)\,dp=1+\tfrac12\log2>1$.
\emph{Weights (I3).} With all positions null, $g=\alpha^{-1}\ind\{p\le\alpha\}$, $\lambda=1$, and all weight on the first $a^*$ with $P_{a^*}\le\alpha$, the singleton has evidence $\log(1/\alpha)=c_{a^*}$, so $d(\{a^*\})=1>0$; the error probability is $1-(1-\alpha)^n$.
\emph{Monotonicity (I4).} The calibrator $g=2\cdot\ind\{p>1/2\}$ has $\int g=1$, but under the super-uniform density $2p$ of Example~\ref{ex:edit}, $\EE g=2(1-\tfrac14)=\tfrac32$. The example shows that super-uniformity alone gives no uniform guarantee over calibrators lacking (I4). In a block-edit simulation the interval family-wise error of this calibrator reaches \AudBHundredGthreefive{} (Appendix~\ref{app:audit}).
\emph{Key access (K3).} An editor who keeps, among $k\ge2$ candidates with independent uniform pivots, the one with the smallest pivot produces a null with $\PP(P\le p)=1-(1-p)^k>p$. In simulation the family-wise error on pure human text is \AudOracleFwerMin{} (Appendix~\ref{app:audit}).

\section{Proofs for Section~\ref{sec:price}: The Cost of Simultaneity}\label{app:price}
This appendix proves Proposition~\ref{prop:pricesandwich} and Corollary~\ref{cor:price}.
\begin{figure}[h]
\centering
\includegraphics[width=\textwidth]{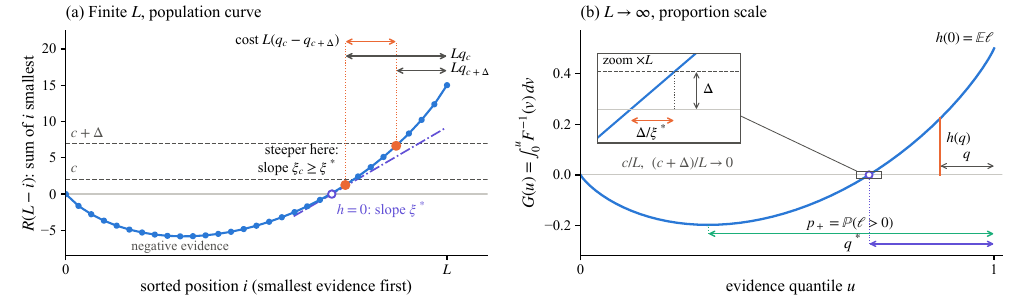}
\caption{Geometry of the price of simultaneity, for an illustrative evidence law (not data). The $L$ evidence values are sorted and summed from the smallest, so trimming $j$ tokens reads the curve $j$ places from the right end, and $h(q)=\int_0^{1-q}F^{-1}(u)\,du$. (a) The population curve at finite $L$: the thresholds $c$ and $c+\Delta$ cut it where its slope exceeds $\xi^*$, so the population cost $L(q_c-q_{c+\Delta})$ is below $\Delta/\xi^*$; sample counts fluctuate around it and are integers. (b) Divided by $L$, both thresholds tend to $0$ and $d/L\to q^*$; magnified by $L$ near the crossing, the cost is $\Delta/\xi^*$.}\label{fig:costgeom}
\end{figure}
This appendix supports Section~\ref{subsec:cost}: it proves the deterministic sandwich and the first-order cost of simultaneity, and the detection length needed for a nonzero count.

Throughout, for an interval $S$ of $L=|S|\ge1$ retained positions and $c>0$, let $A_S(j)$ be the sum of the $L-j$ smallest evidence values in $S$ and
\[
d_c(S)=\min\{j\in\{0,\dots,L\}:A_S(j)<c\}.
\]
Since $A_S(L)=0<c$, $d_c(S)$ exists. Theorem~\ref{thm:main} reports $d(S)=d_{c_a}(S)$, where $a$ is the start of $S$; each result below holds with $c$ equal to the threshold actually used for $S$. Removing positive values lowers $A_S$, and once they are all removed $A_S\le0$, so $d_c(S)\le j$ if and only if $A_S(j)<c$. For a distribution $F$ let $p_+(F)=\PP_F(\ell>0)$, $h_F(q)=\int_0^{1-q}F^{-1}(u)\,du$ and
\[
q_c(F,L)=\sup\big(\{q\in[0,p_+(F)]:h_F(q)>c/L\}\cup\{0\}\big).
\]

Fix an interval $S$ with finite evidence, positive values $x_1\ge\dots\ge x_r>0$ and $R(j)=\sum_{t\in S}\ell_t-\sum_{i\le j}x_i$. For $c,\Delta>0$ let $b=d_c(S)$ and $a=d_{c+\Delta}(S)$ be the trimmed bounds at the two thresholds, so $0\le a\le b\le r$.

\begin{proof}[Proof of Proposition~\ref{prop:pricesandwich}]
The proof uses only the definition of the count. Since $R(j)$ is nonincreasing in $j$ and $R(r)\le0$, the count $d_c(S)$ is the first index at which $R$ falls below $c$; hence $0\le a\le b\le r$, and
\begin{equation}\label{eq:pr-crossing}
R(a-1)\ge c+\Delta\ \ (a\ge1),\qquad R(a)<c+\Delta,\qquad R(b-1)\ge c\ \ (b\ge1),\qquad R(b)<c .
\end{equation}
For (i), the first and last inequalities in \eqref{eq:pr-crossing} give $\sum_{j=a}^bx_j=R(a-1)-R(b)>\Delta$, and since each $x_j\le x_a$ we obtain $(b-a+1)x_a>\Delta$. Similarly, the second and third inequalities give $\sum_{j=a+1}^{b-1}x_j=R(a)-R(b-1)<\Delta$, and since each $x_j\ge x_{b-1}$ we obtain $(b-a-1)x_{b-1}<\Delta$. For (ii), $x_a=R(a-1)-R(a)>(c+\Delta)-c$. For (iii), $\sum_{i<b}x_i=R(0)-R(b-1)<\Delta$ and the left side is at least $(b-1)x_{b-1}$.
\end{proof}
Parts (i) and (ii) together give $\Delta/x_a-1<b-a<\Delta/x_{b-1}+1$ whenever $a\ge1$ and $b\ge2$.

\begin{proof}[Proof of Corollary~\ref{cor:price}]
Write $b_L=d_{c_L}(S_L)$ and $a_L=d_{c_L+\Delta_L}(S_L)$. First, both thresholds are $o(L)$, so Proposition~\ref{prop:qstar} gives $a_L/L\to q^*$ and $b_L/L\to q^*$ almost surely; since $q^*>0$, the events $\{a_L\ge1\}$ and $\{b_L\ge2\}$ have probability tending to one. Second, with $u^*=1-q^*$ and $F_L$ the empirical distribution function of the evidence in $S_L$, $x_j=\ell_{(L-j+1)}=F_L^{-1}((L-j+1)/L)$. Since $F$ is strictly increasing near $\xi^*$, for every $\varepsilon>0$ we have $F(\xi^*-\varepsilon)<u^*<F(\xi^*+\varepsilon)$, and the Glivenko--Cantelli theorem \citep[Thm.~2.4.9, p.~79]{durrett2019probability} yields $\delta>0$ such that eventually $F_L^{-1}(u)\in(\xi^*-\varepsilon,\xi^*+\varepsilon]$ for $|u-u^*|<\delta$. As $(L-a_L+1)/L$ and $(L-b_L+2)/L$ converge to $u^*$ almost surely, $x_{a_L}$ and $x_{b_L-1}$ converge to $\xi^*>0$ almost surely.

On $\{a_L<b_L\}$, Proposition~\ref{prop:pricesandwich}(i) yields $|b_L-a_L-\Delta_L/\xi^*|\le\Delta_L\max\{|1/x_{a_L}-1/\xi^*|,|1/x_{b_L-1}-1/\xi^*|\}+1=o_p(\Delta_L)+1$. On $\{a_L=b_L\}$, Proposition~\ref{prop:pricesandwich}(ii) gives $\Delta_L<x_{a_L}$. If $\Delta_L\to\infty$, this event has probability tending to zero; if $\Delta_L$ is bounded, the difference $b_L-a_L=0$ equals $\Delta_L/\xi^*+O(1)$. In general, we apply these two cases along subsequences. This completes the proof of Corollary~\ref{cor:price}.
\end{proof}
For a prespecified growing interval, the corollary compares two thresholds on the same evidence for the same trimming rule. Equal start weights give $\Delta=\log|T|$. The disjoint-interval aggregate in Proposition~\ref{prop:subset} is a separate count.

At finite length the population target gives a local version. Write $q_s=q_s(F,L)$ and $\xi_s=F^{-1}(1-q_s)$. If $q_s$ stays interior and $F$ has a positive density near $\xi_s$ for all $s\in[c,c+\Delta]$, then $h_F(q_s)=s/L$ gives $-d(Lq_s)/ds=1/\xi_s$. Since $\xi_s$ is nondecreasing in $s$ and $\xi_c\ge\xi^*$,
\[
\frac{\Delta}{\xi_{c+\Delta}}\le L(q_c-q_{c+\Delta})\le\frac{\Delta}{\xi_c}\le\frac{\Delta}{\xi^*}.
\]
The inequalities bound the change in the population target $Lq_c$. It is consistent with the smaller slope of the synthetic cost curve in Appendix~\ref{app:seldemo}, which is descriptive.

\begin{proposition}[Necessary detection length]\label{prop:detlength}
Let $1\le L\le n$, $m=\lfloor n/L\rfloor$ and $B_1,\dots,B_m$ disjoint blocks of length $L$. Under $P_0$ the pivots are i.i.d.\ uniform. Under $P_k$ all pivots are independent, those in $B_k$ have density $f$ with $0<\chi^2=\chi^2(f\|\Unif)<\infty$ and the others are uniform; equivalently $dP_k/dP_0=\prod_{t\in B_k}f(P_t)$. Only the pivots and independent external randomization are observed. If a count procedure $\tilde d$ is simultaneously valid at level $\alpha$ and $\bar q=m^{-1}\sum_kP_k\{\tilde d(B_k)\ge1\}>\alpha$, then
\[
L\ \ge\ \frac{\log\{1+4m(\bar q-\alpha)^2\}}{\log(1+\chi^2)} .
\]
\end{proposition}
\begin{proof}
Let $\phi=\ind\{\exists S:\tilde d(S)\ge1\}$, so $P_0\phi\le\alpha$ and $Q\phi\ge\bar q$ for $Q=m^{-1}\sum_kP_k$. The likelihood ratio $\Lambda=m^{-1}\sum_k\prod_{t\in B_k}f(P_t)$ has $\EE_0\Lambda=1$ and, by independence across disjoint blocks, $\mathrm{Var}_0\Lambda=\{(1+\chi^2)^L-1\}/m$. Hence $\bar q-\alpha\le\mathrm{TV}(Q,P_0)\le\frac12(\mathrm{Var}_0\Lambda)^{1/2}$.
\end{proof}
This is the classical mixture second-moment argument for a signal at an unknown location; \citet[\S2]{chan2013scan} discuss the resulting detection boundaries for the Gaussian scan problem. The bound is for the abstract product-pivot model. Its $f$ ranges over densities that need not be realizable by autoregressive generation. Since $m$ depends on $L$, the condition is implicit. When $m\to\infty$ and $\bar q-\alpha$ stays above a positive constant, the numerator is of order $\log(n/L)$. The proposition bounds detection length at an unknown location. Corollary~\ref{cor:price} quantifies count loss within the trimming construction.

\paragraph{Sufficient detection length.}\label{app:detlength}
Under Assumption~\ref{ass:power} let $S=[a,b]\cap T$ with $|S|=L$, $A=\sum_{S\cap M}D_t-|S\cap N|\kappa_0$ and $V=\sum_{S\cap M}v_t+|S\cap N|v_0$, where $D_t,v_t$ are the mean and variance of $\ell_t$, and $\kappa_0,v_0$ are minus the null mean and the null variance. If $A>c_a$, then $\PP(d(S)\ge1)\ge1-V/(V+(A-c_a)^2)$. If $S\subseteq M$, $D_t\ge\underline D>0$ and $v_t\le\bar v$, then $L\ge\big((k_\beta+\sqrt{k_\beta^2+4\underline Dc_a})/(2\underline D)\big)^2=c_a/\underline D+O(\sqrt{c_a})$, with $k_\beta=\sqrt{\bar v(1-\beta)/\beta}$, gives $\PP(d(S)\ge1)\ge1-\beta$.

\emph{Proof.} $d(S)\ge1$ iff $Z=\sum_S\ell_t\ge c_a$. Apply Cantelli. For the second part, $A\ge L\underline D$ and $V\le L\bar v$, so it suffices that $L\underline D-c_a\ge k_\beta\sqrt L$, a quadratic inequality in $\sqrt L$. \qed

\begin{figure}[h]
\centering
\includegraphics[width=0.6\textwidth]{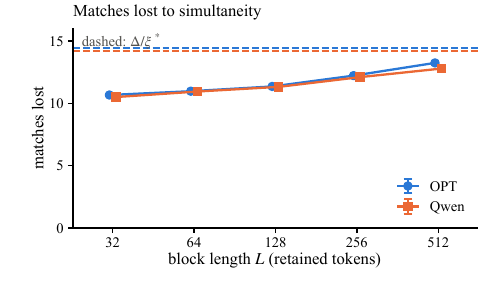}
\caption{Matches lost to simultaneity per block of $L$ retained tokens on the temperature-one holdout (C1, $\alpha=0.05$, $500$ documents per model; blocks fixed before scoring), with $95\%$ document-bootstrap intervals, against the first-order cost $\Delta/\xi^*$ (dashed; $\Delta$ the mean $\log|T|$, $\xi^*$ from the pooled evidence). The geometry behind the sandwich is in Appendix~\ref{app:price}.}\label{fig:cost}
\end{figure}

\section{The Level of the Count}\label{app:proofs}\label{app:powerstatements}
This appendix proves the limit $d/L\to q^*$ used in Section~\ref{sec:price} and gives its finite-length expansion.

Let the $\ell_t$ in a segment of length $L$ be i.i.d.\ with distribution function $F$ and $\EE|\ell|<\infty$, $p_+=\PP(\ell>0)$ and $h(q)=\int_0^{1-q}F^{-1}(u)\,du$, the mean of $\ell$ with the top $q$ fraction removed.
\begin{proposition}[Limit of the certified proportion]\label{prop:qstar}
Let $q^*=\sup\{q\in[0,p_+]:h(q)>0\}$, with $q^*=0$ if $h(0)\le0$. If $c_L>0$ and $c_L/L\to0$, then $d(S_L)/L\to q^*$ almost surely.
\end{proposition}
The same trimmed integral appears in \citet[App.~B.2, eq.~(30)]{wei2026fissionguard} as a sufficient condition for exponential power of a partial-conjunction test at a fixed proportion, with an example showing that power vanishes when it is negative \citep[p.~29]{wei2026fissionguard}. Here it is the almost-sure limit of the certified count proportion. Since $d\le\delta$ on the event of Theorem~\ref{thm:main}, $q^*$ cannot exceed the fraction of retained history matches in the segment when the power model and the validity conditions hold together.

\subsection{Power Model and Finite-Length Expansion}\label{app:rate}
For power analysis we use a model of independent pivots.
\begin{assumption}[Power model]\label{ass:power}
Given the labels and the sequence $(\pi_t)$, the $P_t$ are independent, $\Unif(0,1)$ on $N$, and on $M$ have the Gumbel-max density $f_{\pi_t}(p)=\sum_{v:\pi_{t,v}>0}(1-p)^{1/\pi_{t,v}-1}$.
\end{assumption}
This model describes power only; validity follows from the conditional null law of Section~\ref{sec:setup}. Appendix~\ref{app:condpower} analyses power conditional on the full history, with independent retained rows and context-specific laws.

\begin{figure}[ht]
\centering
\includegraphics[width=.6\linewidth]{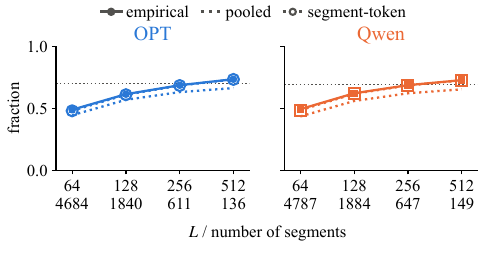}
\caption{Certified proportion on the holdout ($\tau=1.0$, C1, $c=\log(1000/0.05)$). Solid: empirical mean $d/L$ over contiguous segments of kept tokens; dotted: plug-in $q_c(L)$ from the pooled distribution of $\ell$; open markers: $q_c(L)$ from the tokens inside the segments; grey line: pooled $q^*$. Bars are $95\%$ document-bootstrap intervals.}\label{fig:dl}
\end{figure}

With $h$ and $q^*$ as above, let $\xi^*=F^{-1}(1-q^*)$ and $\sigma^{*2}=\mathrm{Var}\min(\ell,\xi^*)$.
\begin{theorem}[Expansion of $d/L$]\label{thm:rate}
If $\EE\ell^2<\infty$, $0<q^*<\PP(\ell>0)$, $F$ has a continuous positive density near $\xi^*$, and $0<c_L=O(\sqrt L)$, then
\[
\frac{d(S_L)}{L}=q^*-\frac{c_L}{\xi^*L}+\frac{\sigma^*}{\xi^*}\frac{Z_L}{\sqrt L}+o_p(L^{-1/2}),
\]
where $Z_L\Rightarrow\mathcal N(0,1)$. The same holds with $q^*-c_L/(\xi^*L)$ replaced by $q_c(L)=\sup\{q:h(q)>c/L\}$.
\end{theorem}
The threshold bias $c/(\xi^*L)$ makes finite-length design depend on both $q^*$ and $\xi^*$; designs that maximize $q^*$ can have $d=0$ at $L=128$. In simulation $q_c(L)$ predicts the empirical $d/L$ to within \EfourteenPtwoMaxEfour{} at all lengths, while $q^*$ alone errs by up to \EfourteenPzeroMax{} (Appendix~\ref{app:synthetic}). On the holdout, long unmasked runs contain higher-entropy tokens than the pool of retained positions; using the evidence distribution within those runs brings $q_c(L)$ within \TfGapSegMax{} of the observed mean at longer lengths (Figure~\ref{fig:dl}).

\subsection{Limit and finite-length expansion}
Let $F_L$ be the empirical distribution of the $L$ evidence values, $\Psi_L(v)=\int_0^vF_L^{-1}(u)\,du$, and $\Psi(v)=\int_0^vF^{-1}(u)\,du$.

\begin{lemma}[Representation of the count]\label{lem:qs-rep}
For $L_+=|\{t:\ell_t>0\}|$, $d(S_L)\le L_+$ and, for $0\le j\le L_+$,
\begin{equation}\label{eq:qs-rep}
d(S_L)\le j\quad\Longleftrightarrow\quad \Psi_L(1-j/L)<c_L/L .
\end{equation}
\end{lemma}
\begin{proof}
For $j\le L_+$, the remainder after trimming is the sum of the $L-j$ smallest terms, which equals $L\Psi_L(1-j/L)$. The remainder is nonincreasing and is nonpositive at $L_+$, whereas $c_L>0$.
\end{proof}

\begin{lemma}[Uniform convergence]\label{lem:qs-unif}
Almost surely, $\sup_{v\in[0,1]}|\Psi_L(v)-\Psi(v)|\to0$ and $L_+/L\to p_+$.
\end{lemma}
\begin{proof}
The supremum is bounded by $W_1(F_L,F)$ by the quantile representation \citep[Theorem~2.10]{bobkov2019empirical}. This distance tends to zero almost surely by the empirical $W_1$ law \citep[Theorem~2.13]{bobkov2019empirical}, since $\EE|\ell|<\infty$. The strong law for $\ind\{\ell_t>0\}$ gives the other limit.
\end{proof}

\begin{lemma}[Shape of $h$]\label{lem:qs-shape}
The function $h$ is continuous on $[0,1]$ and strictly decreasing on $[0,p_+]$, with $h(p_+)\le0$. In particular $h>0$ on $[0,q^*)$ and $h<0$ on $(q^*,p_+]$.
\end{lemma}
\begin{proof}
Integrability of $F^{-1}$ gives continuity. The quantile is positive for $u>1-p_+$ and nonpositive for $u\le1-p_+$.
\end{proof}

\begin{proof}[Proof of Proposition~\ref{prop:qstar}]
On the event of Lemma~\ref{lem:qs-unif}, \eqref{eq:qs-rep} eventually gives
\[
d>\lfloor Lq\rfloor\quad(0\le q<q^*),\qquad
d\le\lceil Lq\rceil\quad(q^*<q<p_+),
\]
since $h(q)$ is respectively positive or negative and $c_L/L\to0$. These indices are eventually at most $L_+$. Let $q\to q^*$; at either boundary use $0\le d\le L_+$. The almost-sure limit implies $\PP(d\ge(q^*-\varepsilon)L)\to1$ for every $\varepsilon>0$, and $\PP(d\ge1)\to1$ when $q^*>0$.
\end{proof}

\begin{remark}
The proof also applies whenever $\Psi_L\to\Psi$ uniformly and $L_+/L\to p_+$ almost surely. For a stationary ergodic integrable sequence these follow from the ergodic theorem for empirical distribution functions, absolute first moments and positive indicators, using \citet[Theorems~2.6 and~2.10]{bobkov2019empirical} for the quantile integral.
\end{remark}

\paragraph{Finite-length expansion (Theorem~\ref{thm:rate}).}
\begin{proof}
By the sign change of $h$ and Lemma~\ref{lem:qs-unif}, with probability tending to one the empirical roots $h_L(\widehat q_0)=0$ and $h_L(\widehat q_c)=c_L/L$, where $h_L(q)=\Psi_L(1-q)$, lie in $(0,L_+/L)$ and converge to $q^*$. The upper-tail mean of $-\ell$ over mass $1-q$ is $-h(q)/(1-q)$ \citep[eq.~(2.2)]{mafusalov2018estimation}; hence $1-q^*$ is its zero-threshold buffered probability of exceedance, and $1-\widehat q_0$ its empirical estimate. The assumptions give $\EE(-\ell)<0<\operatorname{ess\,sup}(-\ell)$ and a unique critical quantile. Apply \citet[Theorem~2.1(iii), eqs.~(2.15)--(2.16)]{mafusalov2018estimation} with coefficient $1/\xi^*$ and loss $(1-\ell/\xi^*)_+=1-\min(\ell,\xi^*)/\xi^*$:
\[
\sqrt L(\widehat q_0-q^*)\Rightarrow
\mathcal N(0,\sigma^{*2}/\xi^{*2}).
\]
Glivenko--Cantelli and local continuity and strict increase of $F$ give $F_L^{-1}=\xi^*+o_p(1)$ uniformly between the two root quantiles, so
\[
\frac{c_L}{L}=
\int_{1-\widehat q_0}^{1-\widehat q_c}F_L^{-1}(u)\,du
=(\widehat q_0-\widehat q_c)(\xi^*+o_p(1)).
\]
Thus $c_L=O(\sqrt L)$ gives $\widehat q_c=\widehat q_0-c_L/(L\xi^*)+o_p(L^{-1/2})$. On the interior-root event, strict crossing in \eqref{eq:qs-rep} gives $0<d/L-\widehat q_c\le1/L$, proving the first expansion. Finally, $h(q_c(L))=c_L/L$ and $h'(q^*)=-\xi^*$ give $q_c(L)=q^*-c_L/(L\xi^*)+o(L^{-1/2})$, proving the second centring.
\end{proof}

A threshold bias of at most $\varepsilon$ requires $L\ge c/(\xi^*\varepsilon)$, and a fluctuation standard deviation of at most $\varepsilon$ requires $L\ge\sigma^{*2}/(\xi^*\varepsilon)^2$.

\paragraph{Checking the density condition for C1.} For C1 the evidence has an atom at its cap $\ell_{\max}=\AtCap$, so the density condition requires $1-q^*<\PP(\ell<\ell_{\max})$. On the real holdout this holds at $\tau=1.0$ ($\AtOptOneOneMinusQ<\AtOptOneBelow$ for OPT, $\AtQwenOneOneMinusQ<\AtQwenOneBelow$ for Qwen) and fails at $\tau=0.7$, where $\xi^*$ lies on the atom; the comparisons of \S\ref{subsec:real} use $q_c(L)$, which does not need it.

\subsection{Conditional Power Given the Full History}\label{app:condpower}
Validity conditions on $\mathcal H$, which fixes the random vectors of non-null contexts; given $\mathcal H$ a signal pivot is a constant. Power analysis instead conditions on $\mathcal W$, the $\sigma$-field generated by the complete generation and editing record $\Gamma$ of Appendix~\ref{hp:conditioning} (the history outputs and the key-free randomness $R,Q$). This $\sigma$-field does not reveal the random vectors, but the historical winner events carry information about them. Independence given $\mathcal W$ does not imply independence given a smaller $\sigma$-field such as the final text alone, and quantities computed given $\mathcal W$ generally require the original prompts, NTP distributions and key-wide history.

Work under Assumption~\ref{ass:Astar} and suppose the evidence is $\ell_t=\varphi_t(P_t)$ for maps $\varphi_t$ that are $\mathcal W$-measurable and do not read other pivots, as for fixed C1. Intervals below consist of retained positions, $S\subseteq T$.

\begin{lemma}[Conditional laws]\label{lem:condlaw}
Given $\mathcal W$, the random vectors of distinct contexts are independent, each distributed as the prior restricted to the intersection of its historical winner events. Hence the retained pivots and evidence values are conditionally independent. Suppose the queried context was used at most once at a watermarked step in the key-wide history; if it was used, let $s$ be that step and $W$ its winner, and let $h$ be the queried token. The bad tail $b_h(r)=\PP(P_h>1-r\mid\mathcal W)$ is
\[
b_h(r)=\begin{cases}
r,&\text{unused context, or }h\ne W,\ \pi_{s,h}=0,\\
r^{1/p},&h=W,\ p=\pi_{s,W}\in(0,1],\\
\dfrac{r-p\,r^{1/p}}{1-p}\ \ge r,&h\ne W,\ p=\pi_{s,h}\in(0,1).
\end{cases}
\]
In the second case $P_h\sim\mathrm{Beta}(1,1/p)$.
\end{lemma}
\begin{proof}
Write $\mathcal A=A_{1:L}$ for the history outputs. As in the factorization \eqref{eq:factor} of Appendix~\ref{app:propK}, given $R$ the event $\{\mathcal A=a\}$ is a product of winner constraints over context rows. On this fibre the contexts and NTP distributions are known, so the posterior factorizes over contexts, and by (K3) it is unchanged by further conditioning on the editing randomness $Q$. The distinct-context mask makes retained positions query distinct vectors. For a single use, write $T_v=-\log U_v/\pi_{s,v}$, independent $\mathrm{Exp}(\pi_{s,v})$, with $T_v=+\infty$ if $\pi_{s,v}=0$. The minimum $E$ is $\mathrm{Exp}(1)$ and independent of the winner, and $U_W=e^{-\pi_{s,W}E}$ gives the Beta law. For $h\ne W$ with $\pi_{s,h}>0$, memorylessness gives $-\log U_h=pE+R_0$ with $E,R_0$ independent $\mathrm{Exp}(1)$, whose tail at $\log(1/r)$ is the stated expression. If $\pi_{s,h}=0$, $U_h$ is unconstrained.
\end{proof}

\begin{remark}[Reused contexts]\label{rem:reuse}
A historical match need not give favourable evidence. Let a two-token context be used twice, with NTP distributions $(2/3,1/3)$ and $(1/2,1/2)$ and winners $A$ then $B$; this can happen in two documents generated with the same key. With $X=-\log U_A$ and $Y=-\log U_B$ the history is $\{X/2<Y<X\}$, of probability $1/6$, and $X$ has density $6(e^{-3x/2}-e^{-2x})$. Querying $A$, a history match, gives $b_A(r)=4r^{3/2}-3r^2$, which equals $5/16>1/4$ at $r=1/4$. Here $b_A(r)>r$ exactly for $1/9<r<1$, so whether a calibrator is affected depends on its whole tail. This does not affect validity, which concerns null positions only.
\end{remark}

Lemma~\ref{lem:condlaw} does not imply Assumption~\ref{ass:power}. That assumption conditions on labels and NTP sequences and takes null pivots uniform and signal pivots with density $f_{\pi_t}$; given $\mathcal W$ the winners are fixed, historical mismatches are super-uniform rather than uniform, and reused contexts need not give a single Beta law. Assumption~\ref{ass:power} remains a separate idealized model. Results that use only independence, known marginals and boundedness can be instantiated under Lemma~\ref{lem:condlaw}; results that use uniform nulls, $f_\pi$ or identical distributions cannot.

\begin{theorem}[Simultaneous Wasserstein band]\label{thm:wband}
Given $\mathcal W$, let $\mathcal S$ be a fixed finite nonempty family of $M$ intervals with thresholds $c_S>0$, the evidence in each $S\in\mathcal S$ be independent with values in $[a_0,b_0]$, $R=b_0-a_0>0$, and $\bar F_S=|S|^{-1}\sum_{t\in S}\PP(\ell_t\le\cdot\mid\mathcal W)$. For $0<\beta<1$ and $L=|S|$ let
\[
r_L=\frac{R}{2\sqrt L}+R\sqrt{\frac{\log(M/\beta)}{2L}} .
\]
(a) With conditional probability at least $1-\beta$, $W_1(\widehat F_S,\bar F_S)\le r_{|S|}$ for all $S\in\mathcal S$.
(b) Let $G_S(j)=L\int_0^{1-j/L}\bar F_S^{-1}(u)\,du$ and, for integers $0\le j\le L$,
\[
D_{\rm lo}(S)=\min\big(\{j:G_S(j)<c_S+Lr_L\}\cup\{L\}\big),\qquad
D_{\rm hi}(S)=\min\big(\{j:G_S(j)<c_S-Lr_L\}\cup\{L\}\big).
\]
On the event in (a), $D_{\rm lo}(S)\le d_{c_S}(S)\le D_{\rm hi}(S)$ for all $S\in\mathcal S$.
\end{theorem}
\begin{proof}
(a) Each $\widehat F_S(x)$ is an average of independent indicators with mean $\bar F_S(x)$ and variance at most $1/(4L)$, so $\EE W_1=\int\EE|\widehat F_S-\bar F_S|\le R/(2\sqrt L)$. Replacing one observation moves $W_1$ by at most $R/L$, so McDiarmid's inequality gives $\PP\{W_1>R/(2\sqrt L)+t\}\le e^{-2Lt^2/R^2}$; take tail $\beta/M$ for each interval and a union bound. The intervals need not be independent of each other. (b) $A_S(j)=L\int_0^{1-j/L}\widehat F_S^{-1}$ and $|A_S(j)-G_S(j)|\le LW_1$. If $j<D_{\rm lo}$ then $A_S(j)\ge c_S$, so $d_{c_S}>j$. If the set defining $D_{\rm hi}$ is nonempty then $A_S(D_{\rm hi})<c_S$, and otherwise $d_{c_S}\le L$.
\end{proof}
The band allows atoms and negative shifted thresholds. If $\bar F_S$ is replaced by an estimate, its $W_1$ error must be added to $r_L$ with its own failure probability.

\begin{corollary}[Explicit rate]\label{cor:condrate}
In Theorem~\ref{thm:wband}, let $h_S=h_{\bar F_S}$ with $h_S(0)>0$, $p_{+,S}=p_+(\bar F_S)$ and $q_S^0=\sup\{q\in[0,p_{+,S}]:h_S(q)>0\}$. Suppose there are $\xi_0>0$ and $0<\eta_0<1-q_S^0$ with $\bar F_S^{-1}(1-q_S^0-\eta_0)\ge\xi_0$ for all $S$. If $r_L\le\xi_0\eta_0$, then on the event in Theorem~\ref{thm:wband}(a), for all $S\in\mathcal S$ with $c=c_S$,
\[
q_S^0-\frac{c/L+r_L}{\xi_0}\ \le\ \frac{d_c(S)}{L}\ \le\ q_S^0+\frac{(r_L-c/L)_+}{\xi_0}+\frac1L .
\]
\end{corollary}
\begin{proof}
Since $\bar F_S^{-1}(u)>0$ only for $u>\bar F_S(0)=1-p_{+,S}$, the condition forces $q_S^0+\eta_0<p_{+,S}$. As $h_S$ is continuous, positive at $0$ and nonincreasing on $[0,p_{+,S}]$, $h_S(q_S^0)=0$. Let $\widehat h(q)=\int_0^{1-q}\widehat F_S^{-1}$, the linear interpolation of $A_S(j)/L$, so $\sup_q|\widehat h-h_S|\le r_L$. Monotonicity of $\bar F_S^{-1}$ gives $h_S(q)\ge\xi_0(q_S^0-q)$ for $q<q_S^0$, and the condition gives $h_S(q_S^0+\eta)\le-\xi_0\eta$ for $0<\eta\le\eta_0$. If $\widehat h(q)<c/L$, one neighbouring grid point has $A_S(j)<c$, so $d_c/L\le q+1/L$. For $t=(r_L-c/L)_+/\xi_0<\eta\le\eta_0$, $\widehat h(q_S^0+\eta)<c/L$; let $\eta\downarrow t$. For the lower bound let $v=q_S^0-(c/L+r_L)/\xi_0$. If $v\ge0$, every $j\le Lv$ has $A_S(j)\ge c$, so $d_c>Lv$; otherwise $d_c\ge0$ suffices.
\end{proof}
If the minimum interval length $L_{\min,n}\to\infty$, $\max_Sc_{n,S}/|S|\to0$, $\log(M_n/\beta_n)/L_{\min,n}\to0$ and $\beta_n\to0$, with uniform range and identification constants, then $\max_S|d(S)/|S|-q_S^0|\to0$ in conditional probability. The convergence is almost sure if, in addition, the events for different $n$ are defined on one probability space and $\sum_n\beta_n<\infty$. Families containing intervals of length one do not have uniform power convergence; simultaneous validity is unaffected. The identification condition excludes evidence that vanishes just below the critical quantile, which is where the pointwise slope condition alone fails.

\section{Proof of Proposition~\ref{prop:gamma-power}: Detection under Single-Winner Histories}
\label{hp:section}
This appendix proves Proposition~\ref{prop:gamma-power}, the detection guarantee that Section~\ref{subsec:single-winner} contrasts with counting.

The distribution of a watermark coordinate depends on how its context was
used before the text was released.  We use this dependence to obtain a
finite-sample power guarantee for a simple certification rule.  The rule
requires the released text and the key, but no request log.  Its validity
holds for the full history class; restrictions on repeated requests enter
only the power analysis.

Consider one reported region with a fixed number $n$ of retained positions.
Write $(\chi_i,w_i)$ for its context--token pairs, whose contexts are
distinct, and use $E_i=-\log(1-P_i)$, with $P_i$ as in Equation~\eqref{eq:pivot}. Let
\begin{equation}
  E_i=-\log U_{\chi_i,w_i},\qquad
  T=\sum_{i=1}^n E_i.
  \label{hp:score}
\end{equation}
Let $H$ be the number of these pairs that occurred as a watermark winner
at least once in the complete history under the same key.  For the $i$th
context, let $J_i$ denote the number of \emph{different} watermark winners
in that history.  Thus $J_i=0$ for an unqueried context, whereas repeated
requests with the same winner still give $J_i=1$.  These are key-wide
historical quantities, rather than counts restricted to the document
currently being examined.  The rule in this section is a Gamma-calibrated binary certificate for a fixed-size region chosen without inspecting key scores.

Logarithmic scores arise naturally in exponential-race watermark detection
\cite{cai2025collision}.  To calibrate \eqref{hp:score} for historical
matching, we first describe its distribution after conditioning on the
generation record.  This uses the same exponential and winner-conditioning
structure that underlies shared-Gumbel sampling and conditional simulation
of max-linear variables \cite{oberst2019counterfactual,wangstoev2011}.

\subsection{Conditioning on the complete history}
\label{hp:conditioning}

Let $\Gamma$ consist of the full output record and key-free randomness in Assumption~\ref{ass:Astar}. It fixes the released text, the score-free region, $H$ and the $J_i$. The rowwise likelihood factorization in Appendix~\ref{app:propK} applies to almost every such record.

\begin{lemma}[Conditional independence and null residuals]
\label{hp:null}
Given $\Gamma$, the key vectors of distinct retained contexts are
independent.  If $(\chi_i,w_i)$ is not a historical winner, its conditional
distribution admits the representation
\begin{equation}
  E_i=B_i+X_i,\qquad B_i\ge0,\qquad X_i\sim\Exp(1).
  \label{hp:nullrepresentation}
\end{equation}
The residuals for distinct null positions can be taken mutually independent
and independent of the corresponding shifts, conditional on $\Gamma$.
In particular, on every branch with $H=0$,
\begin{equation}
  \Gam(n,1)\st T\mid\Gamma.
  \label{hp:nullorder}
\end{equation}
Here $\Gam(a,1)$ has shape $a$ and unit rate, and $\Gam(0,1)$ denotes a
point mass at zero.
\end{lemma}

\begin{proof}
Fix a complete record branch.  Replaying the outputs and the exogenous
randomness fixes each probability vector and each requested context.
For a request at context $\chi$ with recorded winner $a$ and probability
vector $\pi$, the key-dependent restriction is
\[
  E_{\chi,v}\ge
  \frac{\pi_v}{\pi_a}E_{\chi,a}
  \quad\text{for every }v\text{ with }\pi_v>0.
\]
The probability of a recorded winner with $\pi_a=0$ is zero, so such
branches can be discarded.  A zero-probability competing token imposes no
restriction.  All constraints on the fixed branch therefore separate by
physical context row.  Multiplying the independent prior row densities by
these rowwise indicators, and then normalizing, preserves independence
between rows.  Although the retained rows were selected from the released
text, they are fixed and distinct on this branch.

Now fix the other coordinates in a retained row whose queried token has
never won.  Every restriction involving its exponential coordinate is a
lower bound; their maximum is a nonnegative quantity $B_i$.  Its conditional
density is
\[
  e^{-(x-B_i)}\ind\{x\ge B_i\}.
\]
Equivalently, subtracting $B_i$ leaves a unit exponential independent of
the other coordinates under the conditional law.  This follows either
from memorylessness or by factoring the joint density after the change of
variable $X_i=E_i-B_i$.  The rowwise factorization gives independent
residuals across retained contexts.  On a branch with $H=0$, their sum has
distribution $\Gam(n,1)$, and adding the nonnegative shifts proves
\eqref{hp:nullorder}.
\end{proof}

This argument conditions on enough information to fix the adaptive
history.  Independence is not asserted conditional on the released text
alone.  Averaging over the unobserved record will nevertheless preserve
the stochastic bounds used below.

\subsection{The single-winner conditional distribution}
\label{hp:singlewinner}

The number of different winners controls a particularly simple part of
the alternative distribution.  When a retained row has only one winner,
all its past requests can be summarized, for the following calculation,
by the largest probability ratio for each competing token.

\begin{lemma}[A row with one different winner]
\label{hp:kernel}
Condition on $\Gamma$, and suppose a retained context has been queried and
every request at that context returned the same token $a$.  Define
\begin{equation}
  b_v=\max_{j:\,\text{request at this context}}
          \frac{\pi^{(j)}_v}{\pi^{(j)}_a}
       \quad(v\ne a),
  \qquad c=1+\sum_{v\ne a}b_v.
  \label{hp:ratios}
\end{equation}
There are independent unit exponentials $Z$ and $(X_v)_{v\ne a}$ giving
the conditional representation
\begin{equation}
  E_a=\frac{Z}{c},\qquad
  E_v=X_v+\frac{b_v}{c}Z\quad(v\ne a).
  \label{hp:singlekernel}
\end{equation}
If every token probability in every request at this context is at most
$s$, then
\begin{equation}
  \frac1c\le s,\qquad \frac{b_v}{c}\le s\quad(v\ne a).
  \label{hp:coefficients}
\end{equation}
An unqueried row retains its independent unit-exponential coordinates.
\end{lemma}

\begin{proof}
After fixing the record, the row constraints are exactly
$E_v\ge b_vE_a$ for $v\ne a$.  With $x_v=E_v-b_vE_a$, the unnormalized
joint density becomes
\[
  \exp(-cE_a)\prod_{v\ne a}\exp(-x_v),
  \qquad E_a\ge0,\quad x_v\ge0.
\]
Its integral is $1/c$.  Normalization shows that $cE_a$ and the $x_v$
are independent unit exponentials, proving \eqref{hp:singlekernel}.

For any request $j$ at this context,
\[
  c\ge1+\sum_{v\ne a}\frac{\pi^{(j)}_v}{\pi^{(j)}_a}
    =\frac1{\pi^{(j)}_a},
\]
and therefore $1/c\le\pi^{(j)}_a\le s$.  If $b_v>0$, let $j_v$ be a
request attaining the maximum in \eqref{hp:ratios}; it exists because the
history is finite.  The same inequality yields
\[
  \frac{b_v}{c}
  \le\frac{\pi^{(j_v)}_v}{\pi^{(j_v)}_a}\pi^{(j_v)}_a
  =\pi^{(j_v)}_v\le s.
\]
The case $b_v=0$ is immediate.
\end{proof}

The lemma permits repeated requests and changing probability vectors.
It requires one \emph{different winner}, rather than one request.  Its
consequence for the score is direct: a matched position contributes a
scaled exponential with scale at most $s$, while an unmatched position
contributes an additional independent unit exponential.

\subsection{A calibrated certificate and its power}
\label{hp:rule}

For a fixed threshold $t$, consider the verifier
\begin{equation}
  C_t=\ind\{T\le t\}.
  \label{hp:decision}
\end{equation}
Thus the procedure certifies at least one historical match when the sum
of the queried exponential coordinates is sufficiently small.  The
threshold can be selected from a lower quantile of $\Gam(n,1)$, without
knowing the history or the number of matches.

\begin{theorem}[Validity and history-dependent power]
\label{hp:power}
For the full class of permitted finite histories with $n$ retained
distinct contexts,
\begin{equation}
  \sup_P\PP_P(C_t>H)\le F_{\Gam(n,1)}(t).
  \label{hp:size}
\end{equation}
This validity statement does not require a request budget, a bound on the
number of different winners, or a bound on the token probabilities.

Fix $1\le h\le n$ and $0<s<1$.  Consider the power subclass in which,
almost surely, $H\ge h$, every retained row has $J_i\le1$, and all token
probabilities in every historical watermark request at a retained row
are at most $s$.  Uniformly over this subclass,
\begin{equation}
  \begin{aligned}
  T&\st\Gam(n-h,1)+s\Gam(n,1),\\
  \PP_P(C_t\ge1)&\ge F_{\Gam(n-h,1)+s\Gam(n,1)}(t).
  \end{aligned}
  \label{hp:powergeneral}
\end{equation}
where the two Gamma variables are independent.
\end{theorem}

\begin{proof}
If $H\ge1$, the output in \eqref{hp:decision} cannot exceed $H$.  On
each complete branch with $H=0$, Lemma~\ref{hp:null} bounds its rejection
probability by $F_{\Gam(n,1)}(t)$.  Integrating this branchwise bound
proves \eqref{hp:size}, including histories with any number of different
winners.

For power, fix a branch satisfying the subclass conditions and write
$k=H\ge h$.  Combining the independent row representations gives
\begin{equation}
  T\overset{d}{=}
  \sum_{i:\,\text{unmatched}}X_i
  +\sum_{i:\,J_i=1}a_iZ_i,
  \qquad 0\le a_i\le s,
  \label{hp:decomposition}
\end{equation}
conditional on the branch.  All displayed unit exponentials can be taken
independent.  There are $n-k\le n-h$ residuals $X_i$ and
$\sum_iJ_i\le n$ latent variables $Z_i$.  Increase each coefficient
$a_i$ to $s$, then add independent nonnegative terms to make the two
counts equal to $n-h$ and $n$, respectively.  This constructs the
stochastic upper bound in \eqref{hp:powergeneral}.  In particular,
$\sum_iJ_i$ is bounded by $n$, not assumed equal to $n$; unqueried rows
are accounted for by the added terms.

The bounding distribution is the same on every branch.  Averaging over
$\Gamma$ therefore preserves the stochastic order.  Since certification
is a lower-tail event, this order gives the stated lower bound on its
probability.
\end{proof}

For $n=300$, use the threshold $t=272$.  The verifier computes the 300
key coordinates in \eqref{hp:score} and reports one exactly when their
negative logarithms sum to at most 272.  Its worst-case overstatement
probability satisfies
\begin{equation}
  \sup_P\PP_P(C_{272}>H)
    \le F_{\Gam(300,1)}(272)
    \le\HistSizeUpper<0.05.
  \label{hp:numericalsize}
\end{equation}
If at least 180 of the positions are historical matches, each retained
row has at most one different winner, and the request probabilities are
bounded by $1/2$, then
\begin{equation}
  \PP_P(C_{272}\ge1)
   \ge F_{\Gam(120,1)+(1/2)\Gam(300,1)}(272)
   \ge\HistPowerLower.
  \label{hp:numericalpower}
\end{equation}
A protocol may additionally impose a total budget of 420 watermark
requests.  The same guarantee applies under that budget, but the proof
does not use it: the relevant bound is on distinct winners at the
retained rows, together with the match and probability conditions above.
In particular, the validity of the report does not rely on checking these
power conditions from an unavailable log.

The numerical inequalities can be certified by rational enclosures.
For the size calculation,
\begin{equation}
  F_{\Gam(300,1)}(272)
    =1-e^{-272}\sum_{j=0}^{299}\frac{272^j}{j!}.
  \label{hp:sizeformula}
\end{equation}
For power, let $N\sim\operatorname{Poisson}(544)$.  Uniformization at
rate 2 represents the sum as 300 rate-2 stages followed by 120 rate-1
stages.  After the first 300 arrivals, independently retaining arrivals
with probability $1/2$ produces the latter stages.  Consequently,
\begin{equation}
  \begin{aligned}
  &F_{\Gam(120,1)+(1/2)\Gam(300,1)}(272)\\
   &\quad=\sum_{j=420}^{\infty}\PP(N=j)
      \PP\!\left\{\operatorname{Bin}(j-300,1/2)\ge120\right\}\\
   &\quad\simeq\HistPowerDisplay.
  \end{aligned}
  \label{hp:poisson}
\end{equation}
The binomial probabilities are rational.  Truncating at $M=1000$ leaves
a remainder bounded above by
\[
  e^{-544}\frac{544^{M+1}}{(M+1)!}
    \left(1-\frac{544}{M+2}\right)^{-1}.
\]
Alternating Taylor polynomials of degrees 41 and 40 bracket $e^{-1}$;
taking integer powers brackets the exponential terms in
\eqref{hp:sizeformula} and \eqref{hp:poisson}.  These enclosures establish
the directions of the inequalities in \eqref{hp:numericalsize} and
\eqref{hp:numericalpower}, rather than relying on floating-point rounding.

The size guarantee holds for any number of distinct winners. The power bound uses a key-wide restriction, so first-use tracking within each document alone does not impose it.

\subsection{The scan count under single-winner histories}
\label{hp:scan}
The row representation of Lemma~\ref{hp:kernel} also bounds the CountScan count. Fix the C1 calibrator with $\lambda=1/2$ and write its evidence as a function of the exponential coordinate, $\ell_t=\varphi(E_t)$ with $P_t=1-e^{-E_t}$. Since $g_{\rm C1}$ is nonincreasing, $\varphi$ is nonincreasing, and its values lie in $[\log(1/2),\ell_{\max}]$ with $\ell_{\max}=\log\{1+\tfrac12(g_{\rm C1}(0^+)-1)\}$.

For $0<s<1$ and $0\le\kappa\le1$ let $F_{\kappa,s}$ be the law of $\varphi(E)$ when $E$ is distributed as $sZ$ with probability $\kappa$ and as $X+sZ$ otherwise, where $X,Z$ are independent unit exponentials. Let $G_{\kappa,s}(j)=L\int_0^{1-j/L}F_{\kappa,s}^{-1}(u)\,du$ and
\[
q^0_{\kappa,s}=\sup\{q\in[0,1]:\ \textstyle\int_0^{1-q}F_{\kappa,s}^{-1}(u)\,du>0\},
\]
with $q^0_{\kappa,s}=0$ when the mean of $F_{\kappa,s}$ is nonpositive.

\begin{corollary}[Scan count under single-winner histories]\label{hp:scancor}
Let $S\subseteq T$ be a region of $L$ retained positions determined by $\Gamma$, and let $c>0$. Suppose that, almost surely, at least $h$ positions of $S$ are historical matches, every row of $S$ has $J_i\le1$, and every token probability in every historical request at these rows is at most $s$. Put $\kappa=h/L$, $R=\ell_{\max}-\log(1/2)$ and
\[
r_L=\frac{R}{2\sqrt L}+R\sqrt{\frac{\log(1/\beta)}{2L}},\qquad
D_{\rm lo}=\min\big(\{j:\ G_{\kappa,s}(j)<c+Lr_L\}\cup\{L\}\big).
\]
Then $\PP\{d_c(S)\ge D_{\rm lo}\}\ge1-\beta$ for every $0<\beta<1$. If in addition $F^{-1}_{\kappa,s}(u)\ge\xi_0>0$ for all $u>1-q^0_{\kappa,s}$, then with probability at least $1-\beta$,
\[
\frac{d_c(S)}{L}\ \ge\ q^0_{\kappa,s}-\frac{c/L+r_L}{\xi_0}.
\]
\end{corollary}

\begin{proof}
Fix a branch of $\Gamma$ in the subclass. By Lemmas~\ref{hp:null} and~\ref{hp:kernel}, the coordinates $E_t$, $t\in S$, are conditionally independent. A matched position has $E_t=Z_t/c_t\le sZ_t$. An unmatched position in a queried row has $E_t=X_t+(b_t/c_t)Z_t\le X_t+sZ_t$, and an unqueried row has $E_t=X_t$. Hence on one probability space we can construct independent $E^{\rm ref}_t\ge E_t$, distributed as $sZ_t$ at matched positions and as $X_t+sZ_t$ elsewhere. Then $\ell_t\ge\ell^{\rm ref}_t=\varphi(E^{\rm ref}_t)$ for every $t\in S$.

The trimmed count is coordinatewise nondecreasing in the evidence. Raising one value by $\Delta\ge0$ raises the total by $\Delta$ and the sum of the $j$ largest positive parts by at most $\Delta$, so every $R_S(j)$ in Definition~\ref{def:trim} is nondecreasing, and so is $d_c(S)=\min\{j:R_S(j)<c\}$. Thus $d_c(S)\ge d^{\rm ref}_c(S)$ on the coupling.

The reference evidence is independent with values in $[\log(1/2),\ell_{\max}]$. Its averaged distribution is $F_{\kappa',s}$, where $\kappa'\ge\kappa$ is the realized match fraction of $S$ on this branch. Theorem~\ref{thm:wband}(b) with $M=1$ gives $d^{\rm ref}_c(S)\ge\min(\{j:G_{\kappa',s}(j)<c+Lr_L\}\cup\{L\})$ with probability at least $1-\beta$. Since $sZ\le X+sZ$, increasing $\kappa'$ moves mass of $F_{\kappa',s}$ upward, so $G_{\kappa',s}\ge G_{\kappa,s}$ and the right side is at least $D_{\rm lo}$. For the second display, if $q^0_{\kappa,s}=0$ the bound is at most zero and follows from $d_c(S)\ge0$. Otherwise $\int_0^{1-q}F^{-1}_{\kappa,s}>0$ for $q<q^0_{\kappa,s}$, so continuity gives $\int_0^{1-q^0_{\kappa,s}}F^{-1}_{\kappa,s}\ge0$, so $G_{\kappa,s}(j)\ge L\xi_0(q^0_{\kappa,s}-j/L)$ for $j/L<q^0_{\kappa,s}$, as in the proof of Corollary~\ref{cor:condrate}. Every integer $j\le Lq^0_{\kappa,s}-(c+Lr_L)/\xi_0$ therefore has $G_{\kappa,s}(j)\ge c+Lr_L$, and $D_{\rm lo}$ exceeds it. The bounds do not depend on the branch, so averaging over $\Gamma$ completes the proof.
\end{proof}

With $c=\log(|T|/\alpha)$ the bound reads $d(S)\gtrsim q^0_{\kappa,s}L-\{\log(|T|/\alpha)+O(\sqrt L)\}/\xi_0$, so the threshold cost of Section~\ref{subsec:cost} appears in the same form. For a fixed family of $M$ regions, replace $\log(1/\beta)$ by $\log(M/\beta)$.

The reference law is conservative. An unmatched position in a queried row carries the shift $sZ$, and its evidence is worse than under a uniform null. For C1 with $s=1/2$, the root $q^0_{\kappa,s}$ is $\BridgeHalfAllQ$ at $\kappa=1$, $\BridgeHalfEightyQ$ at $\kappa=0.8$ and $\BridgeHalfSixtyQ$ at $\kappa=0.6$; with $s=0.3$ the values are $\BridgeThreeAllQ$, $\BridgeThreeEightyQ$ and $\BridgeThreeSixtyQ$ (deterministic quadrature, code available from the author on request). At $\kappa=0.6$ and $s=1/2$, the parameters of Theorem~\ref{thm:full-key}, the corollary is therefore uninformative, and the Gamma certificate of Theorem~\ref{hp:power} gives the parameter-matched comparison.

\section{Proof of Theorem~\ref{thm:sw-sharp}: Counts with Fixed Menus}\label{app:swprice}
This appendix proves Theorem~\ref{thm:sw-sharp}: the case $r=1$ first, then $r$ menus (Appendix~\ref{app:rmenu}), and finally the finite-length behaviour of CountScan used in Section~\ref{subsec:unified}.
\paragraph{Numerical illustration.} At $n_S=\ProtocolN$, $h=\ProtocolH$ and $s=1/2$, the Gamma certificate detects a match with probability at least \ProtocolPowerLower, while under the policy of Theorem~\ref{thm:sw-sharp}(ii) with $j=\SwConvM$ (the $m$ of Theorem~\ref{sw:converse} below) every valid count certifies at most \SwConvM{} matches with probability at least \SwConvLower, so on that event it misses at least \SwConvShortfall{} of the at least \ProtocolH{} matches. Asymptotically $\rho=\SwRho$ against a match fraction of $0.6$. For each prespecified $q<\rho$, CountScan with two-level evidence designed for that $q$ attains it at its simultaneous start thresholds when $\log|T|=o(n_S)$; this appendix also gives finite-length power bounds.

Appendix~\ref{hp:section} shows that a single-winner history admits a binary certificate. Here we ask how many matches a valid count can certify under such histories. For a region of $n$ positions with match quota $h_n$, $h_n/n\to\kappa$, and token probabilities at most $s$, the supremum of the proportions that a valid count certifies with probability tending to one, uniformly over these histories, is
\[
\rho=\frac{(\kappa-s)_+}{1-s}.
\]
Proposition~\ref{sw:tail} attains every proportion below $\rho$ (Corollary~\ref{sw:sharp}), and Theorem~\ref{sw:converse} excludes every proportion above it; neither statement concerns the critical point itself. At finite $n$ both sides give explicit probabilities.

\paragraph{Setting.} We keep the notation of Appendix~\ref{hp:section}: a region of $n$ retained positions with distinct contexts; $E_i=-\log U_i$ with $U_i=U_{\chi_i,w_i}$; $H$ the number of historical matches among the $n$ positions; $J_i$ the number of different historical winners at the $i$th context. The region is fixed, or determined by the released text and public randomness; a region read off the hidden record could leak information to the verifier and is not covered by the converse. The \emph{single-winner subclass} $\mathcal P_{\rm sw}(h,s)$ consists of histories with, almost surely, $H\ge h$, $J_i\le1$ for every $i$, and every token probability of every historical request at these contexts at most $s$. A count $D\in\{0,\dots,n\}$ is \emph{valid} over a class of histories if $\PP(D>H)\le\alpha$ for every member. Validity is required over the whole class, including members with few matches; the condition $H\ge h$ only selects where power is evaluated and is not available to the count.

\subsection{Two counts that are valid for every history}
For $1\le k\le n$, write $N_k=n-k+1$.

\begin{proposition}[Gamma count]\label{sw:gamma}
Let $X=\sum_{i=1}^nE_i$, $p_k=F_{\Gam(N_k,1)}(X)$ and $d_\Gamma=\max(\{0\}\cup\{k:\ p_k\le\alpha\})$. Then $\PP(d_\Gamma>H)\le\alpha$ for every history permitted by Assumption~\ref{ass:Astar}. On $\mathcal P_{\rm sw}(h,s)$ with $1\le h\le n$,
\[
\PP(d_\Gamma\ge k)\ge F_{\Gam(n-h,1)+s\Gam(n,1)}\big(F^{-1}_{\Gam(N_k,1)}(\alpha)\big),\qquad 1\le k\le n .
\]
\end{proposition}
\begin{proof}
For fixed $x$, $F_{\Gam(N,1)}(x)$ decreases in $N$, so $p_k$ is nondecreasing in $k$ and $\{k:p_k\le\alpha\}$ is an initial segment. Hence $d_\Gamma>H$ holds exactly when $H<n$ and $p_{H+1}\le\alpha$. Fix a record branch with $H<n$. By Lemma~\ref{hp:null}, the $n-H$ unmatched coordinates are $E_i=B_i+X_i$ with independent unit exponentials $X_i$ and $B_i\ge0$, and matched coordinates are nonnegative. Thus $X\ge Y=\sum_{\text{unmatched}}X_i\sim\Gam(n-H,1)$ on one probability space, and since $F_{\Gam(n-H,1)}$ is nondecreasing and continuous,
\[
\PP(p_{H+1}\le\alpha\mid\Gamma)\le\PP\{F_{\Gam(n-H,1)}(Y)\le\alpha\}=\alpha .
\]
Averaging over $\Gamma$ gives validity. For power, $d_\Gamma\ge k$ if and only if $X\le F^{-1}_{\Gam(N_k,1)}(\alpha)$, and Theorem~\ref{hp:power} gives $X\st\Gam(n-h,1)+s\Gam(n,1)$ on $\mathcal P_{\rm sw}(h,s)$.
\end{proof}
At $n=\ProtocolN$, $h=\ProtocolH$ and $s=1/2$, the certified threshold $272$ of \eqref{hp:numericalsize} satisfies $F_{\Gam(300,1)}(272)<\alpha$, so $X\le272$ implies $d_\Gamma\ge1$, and \eqref{hp:numericalpower} gives the certified bound $\PP(d_\Gamma\ge1)\ge\ProtocolPowerLower$. Numerical evaluation of the bound for larger $k$, without interval certification, gives $\PP(d_\Gamma\ge10)\approx\SwGammaHalfTen$ and a reference count with mean about $\SwGammaHalfMean$ and median $\SwGammaHalfMedian$ (median $\SwGammaThreeMedian$ at $s=0.3$). Replacing $X$ by the sum of the $N_k$ largest $E_i$ in $p_k$ and inverting the running maximum of these $p$-values gives a valid count at least as large; we do not use it.

\begin{proposition}[Tail count]\label{sw:tail}
Fix $r\in(0,1)$, let $B_r=\#\{i:\ U_i<r\}$, $q_k=F_{\mathrm{Bin}(N_k,r)}(B_r)$ and $d_r=\max(\{0\}\cup\{k:\ q_k\le\alpha\})$. Then $\PP(d_r>H)\le\alpha$ for every history permitted by Assumption~\ref{ass:Astar}. Let $b_k$ be the largest integer $b\ge0$ with $F_{\mathrm{Bin}(N_k,r)}(b)\le\alpha$, and $b_k=-1$ if there is none. Put
\[
\tau_r=\frac{r-sr^{1/s}}{1-s},\qquad B^{\rm ref}_r\sim\mathrm{Bin}(h,r^{1/s})+\mathrm{Bin}(n-h,\tau_r)\ \text{(independent)} .
\]
On $\mathcal P_{\rm sw}(h,s)$, $\PP(d_r\ge k)\ge\PP(B^{\rm ref}_r\le b_k)$. In particular, with $\mu=\EE B^{\rm ref}_r$ and $g_k=N_kr-\sqrt{N_k\log(1/\alpha)/2}-1-\mu$, $\PP(d_r\ge k)\ge1-\exp(-2g_k^2/n)$ whenever $g_k>0$.
\end{proposition}
\begin{proof}
\emph{Validity.} $F_{\mathrm{Bin}(N,r)}(b)$ decreases in $N$, so $q_k$ is nondecreasing in $k$, and $d_r>H$ holds exactly when $H<n$ and $q_{H+1}\le\alpha$. On a record branch with $H<n$, an unmatched coordinate satisfies $U_i=e^{-B_i-X_i}\le V_i=e^{-X_i}$ with independent uniform $V_i$, so $B_r\ge Y=\#\{i\ \text{unmatched}:V_i<r\}\sim\mathrm{Bin}(n-H,r)$. For any random variable $Y$ with distribution function $F$, $\PP\{F(Y)\le\alpha\}\le\alpha$; monotonicity of $F_{\mathrm{Bin}(n-H,r)}$ then gives $\PP(q_{H+1}\le\alpha\mid\Gamma)\le\alpha$.

\emph{Reference law.} Fix a branch in $\mathcal P_{\rm sw}(h,s)$. The proof of Corollary~\ref{hp:scancor} constructs independent $E^{\rm ref}_i\ge E_i$ distributed as $sZ_i$ at matched rows and as $X_i+sZ_i$ at the others. Choose $h$ of the matched rows by a rule fixed on the branch, and at every other matched row add an independent unit exponential to the envelope; the envelopes remain above the $E_i$. Then $B_r\le\#\{i:e^{-E^{\rm ref}_i}<r\}$, and with $\ell=\log(1/r)$,
\[
\PP(sZ>\ell)=r^{1/s},\qquad
\PP(X+sZ>\ell)=e^{-\ell/s}+\int_0^{\ell/s}e^{-(\ell-sz)}e^{-z}\,dz=\tau_r ,
\]
so the right side has the law of $B^{\rm ref}_r$ on every branch.

\emph{Power.} $d_r\ge k$ if and only if $B_r\le b_k$, which gives the first bound. Hoeffding's inequality gives $F_{\mathrm{Bin}(N_k,r)}(b)\le\alpha$ for every integer $0\le b\le N_kr-\sqrt{N_k\log(1/\alpha)/2}$, so $b_k\ge\mu+g_k$ when $g_k>0$, and Hoeffding's inequality for the $n$ independent indicators in $B^{\rm ref}_r$ gives $\PP(B^{\rm ref}_r>\mu+g_k)\le\exp(-2g_k^2/n)$.
\end{proof}
The Hoeffding form is vacuous at $n=\ProtocolN$, $h=\ProtocolH$, $s=1/2$, since $g_1<0$ for every $r$, but the exact form is not. With $r=1/2$ the null cutoff is $b_1=\SwTailCut$ and the reference law is $\mathrm{Bin}(180,1/4)+\mathrm{Bin}(120,3/4)$; an exact rational computation gives $\PP(d_{1/2}\ge1)\ge\SwTailFirst$.

\subsection{No valid count guarantees more}
\paragraph{A public model for each $n$.} Contexts have four tokens. Let $K=\lceil(1-s)/s\rceil$. The vocabulary consists of the released prefix $h_{-3},\dots,h_0$ and, for each position $i=1,\dots,n$, a block $\mathcal V_i=\{h_i,a_{i,1},\dots,a_{i,K}\}$; all $n(K+1)+4$ tokens are distinct. Let $p_i$ put probability $s$ on $h_i$ and $(1-s)/K\le s$ on each $a_{i,k}$. The model returns $p_1$ on a prompt ending in $h_0$, $p_{j+1}$ on a prompt ending in a token of $\mathcal V_j$ for $1\le j<n$, and $p_1$ on every other prompt. The draft is $x=(h_{-3},\dots,h_0,h_1,\dots,h_n)$, and the audited region consists of positions $1,\dots,n$.

\paragraph{The class $\mathcal G_{\rm sw}$.} A procedure releases the prefix $h_{-3},\dots,h_0$, starts from the draft fixed before the key is drawn, submits prompts to this model, chooses later prompts and released tokens from the draft and the returned winner labels only, may release draft tokens that no request returned, requests each context at most once, and releases at position $i$ a token of $\mathcal V_i$. A count observes the released text, the entire key, the public model and the draft, and may use independent randomization; it does not observe the number of requests, the prompts, the returned labels or the identity of the procedure. The target $H$ counts the context--token pairs of positions $1,\dots,n$ returned by some request of the procedure, as in \eqref{hn:target}. Every released token at position $i$ lies in $\mathcal V_i$, so the $n$ released contexts are distinct and all positions are retained; the region is the same for every procedure.

\begin{theorem}[Converse]\label{sw:converse}
Fix $n$, $s$ and $0\le m\le h\le n$. There is a policy $G_m\in\mathcal G_{\rm sw}\cap\mathcal P_{\rm sw}(h,s)$ with $H\ge h$ on every realization such that every count $D$ valid over $\mathcal G_{\rm sw}$ satisfies
\[
\PP_{G_m}(D>m)\le\alpha+\PP\{\mathrm{Bin}(n-m,s)<h-m\}.
\]
\end{theorem}
\begin{proof}
\emph{Two policies.} The low policy releases $h_1,\dots,h_{n-m}$ without requests; at each later position it requests at its current context and releases the winner. The high policy $G_m$ keeps a miss counter, initially zero. At each position $i\le n-m$ it requests at its current context and observes the winner $W_i$. Before fallback, it releases $h_i$ if $W_i=h_i$; otherwise it releases $h_i$ and adds one miss if the counter stays at most $n-h$, and releases $W_i$ and enters fallback if not. In fallback, and at every position $i>n-m$, it releases the winner of its request. In both policies the prompt at position $i$ ends in the token released at $i-1$, which lies in $\mathcal V_{i-1}$ (or is $h_0$), so the model returns $p_i$. Both policies read only returned labels, request each context once and release a token of $\mathcal V_i$ at position $i$, so both lie in $\mathcal G_{\rm sw}$, and every retained row has $J_i\le1$ with request probabilities at most $s$. Neither depends on $D$.

\emph{Match counts.} The only requests of the low policy are at the contexts of positions $n-m+1,\dots,n$; the contexts of the first $n-m$ positions are never requested, so $H_0=m$ (and $H_0<h$ when $m<h$). Under $G_m$ every position releases either the winner of the only request at its context, which is a match, or a kept draft token after the request at that context returned a different token, which is not. The counter bounds the latter, so $H_1\ge n-(n-h)=h$ on every realization.

\emph{Coupling.} Run both policies on the same key, and give $D$ the same randomization. Let $Y_i=\ind\{W(c_i(x),p_i)=h_i\}$ for $i\le n-m$, where $c_i(x)$ are the draft contexts. These rows are distinct, so the $Y_i$ are independent Bernoulli$(s)$ by the race law \eqref{hn:race}. Let $A=\{\sum_{i\le n-m}Y_i\ge h-m\}$. Before fallback $G_m$ releases the draft, so its context at $i$ is $c_i(x)$, its winner is $W(c_i(x),p_i)$ and its miss counter equals $\#\{j\le i:Y_j=0\}$. On $A$ this count never exceeds $(n-m)-(h-m)=n-h$, so no fallback occurs and both policies release $h_1,\dots,h_{n-m}$. At each later position both policies then have the same context, hence the same key row and the same model distribution, so they return and release the same winner; by induction the two released texts coincide on $A$, and so does the audited region. Therefore
\[
\PP_{G_m}(D>m)\le\PP(A^c)+\PP_{\rm low}(D>m)=\PP\{\mathrm{Bin}(n-m,s)<h-m\}+\PP_{\rm low}(D>H_0),
\]
and validity over $\mathcal G_{\rm sw}$ bounds the last term by $\alpha$.
\end{proof}
At $n=\ProtocolN$, $h=\ProtocolH$, $s=1/2$ and $m=\SwConvM$, the binomial term is at most $\SwConvEps$ (exact rational sum). Under the constructed policy $G_{\SwConvM}$, every count valid over $\mathcal G_{\rm sw}$ certifies at most $\SwConvM$ matches with probability at least $\SwConvLower$, and on that event it misses at least $\SwConvShortfall$ of the at least $\ProtocolH$ matches, although every row has a single winner. The value $m=\SwConvM$ is one point of a family; the bound holds for every $m$.

\begin{remark}[The same comparison in the model of Theorem~\ref{thm:full-key}]\label{sw:samemodel}
At $n=\ProtocolN$ and $s=1/2$ the argument runs unchanged in the public model and draft of Appendix~\ref{hn:model}, using route $R_1$ only. There the target draft word wins each first request with probability $1/2$; the low policy releases the first $n-m$ draft words and then releases $R_1$ winners, and the high policy requests $R_1$ once per position and, on reaching the miss limit, releases winners along the backup chain, whose contexts are distinct as in Lemma~\ref{hn:feasible}. Both use at most $\ProtocolN$ requests and one request per context, so they lie in the class of Appendix~\ref{hn:interface}, and the bound above holds for every count valid over its single-request subclass, in particular for every count valid over the whole class. Within one public model, one request per context thus already forces $\PP(D\le\SwConvM)\ge\SwConvLower$ for a policy with $H\ge\ProtocolH$, while allowing a retry gives the binary bound $\PP(C\ge1)\le\ProtocolPowerUpper$ of Theorem~\ref{thm:full-key} for the policy constructed there.
\end{remark}

\begin{corollary}[The certifiable proportion]\label{sw:sharp}
Let $n\to\infty$ with $h_n/n\to\kappa\in(0,1)$ and fixed $s\in(0,1)$, use the public model above for each $n$, and put $\rho=(\kappa-s)_+/(1-s)$.
(i) If $\rho>0$, for every $q<\rho$ there is $r\in(0,1)$ such that $\inf_{\mathcal P_{\rm sw}(h_n,s)}\PP(d_r\ge qn)\to1$.
(ii) For every $q'>\rho$ there are policies in $\mathcal G_{\rm sw}\cap\mathcal P_{\rm sw}(h_n,s)$ such that every sequence of counts valid over $\mathcal G_{\rm sw}$ has $\PP(D\le q'n)\ge1-\alpha-o(1)$, where the $o(1)$ term does not depend on the counts.
\end{corollary}
\begin{proof}
(i) Since $\mu/n\to\kappa r^{1/s}+(1-\kappa)\tau_r=(1-\rho)r+\rho r^{1/s}$, for $k=\lceil qn\rceil$,
\[
g_k/n\to(1-q)r-(1-\rho)r-\rho r^{1/s}=(\rho-q)r-\rho r^{1/s},
\]
which is positive once $r^{(1-s)/s}<(\rho-q)/\rho$. Proposition~\ref{sw:tail} then gives $\PP(d_r\ge k)\ge1-e^{-cn}$ for some $c>0$, uniformly over the subclass.
(ii) Since $\rho<\kappa$ and $\PP(D\le q'n)$ increases in $q'$, it suffices to treat $\rho<q'<\kappa$. Take $m=\lfloor q'n\rfloor$, so $m\le h_n$ for large $n$ and $\{D\le m\}\subseteq\{D\le q'n\}$. Then $(n-m)s-(h_n-m)=n\{(1-s)q'-(\kappa-s)\}+o(n)$, which is a positive multiple of $n$ in both cases $\kappa>s$ and $\kappa\le s$. Hoeffding's inequality gives $\PP\{\mathrm{Bin}(n-m,s)<h_n-m\}\to0$, and Theorem~\ref{sw:converse} with the policy $G_m$ completes the proof.
\end{proof}
Thus $\rho$ is the asymptotic supremum of the proportion that a valid count certifies with probability tending to one, uniformly over single-winner histories with token probabilities at most $s$. Relative to the match quota $h_n$, the gap is $\kappa-\rho=s(1-\kappa)/(1-s)$ when $\kappa>s$; when $\kappa\le s$ no positive proportion can be guaranteed. This is a gap relative to the quota, not the worst case of the realized shortfall $H-D$ over the class, which the corollary does not compute. At $\kappa=0.6$ and $s=1/2$ the gap is $\SwPriceFrac$, that is, $\SwPriceCount$ of the quota of $\ProtocolH$ at $n=\ProtocolN$. Theorem~\ref{thm:full-key} shows that with retries and draft retention even the binary certificate has power at most $\ProtocolPowerUpper$ for the policy constructed there.

\subsection{Several fixed menus per context}\label{app:rmenu}
This subsection proves Theorem~\ref{thm:sw-sharp} for $r\ge1$; with $r=1$ it reduces to Proposition~\ref{sw:tail}, Theorem~\ref{sw:converse} and Corollary~\ref{sw:sharp}.

\paragraph{The model and the class.} Extend the public model above as follows. Each block is $\mathcal V_i=\{h_i\}\cup\{a_{i,j,k}:1\le j\le r,\ 1\le k\le K\}$, all tokens distinct, $K=\lceil(1-s)/s\rceil$. Menu $p^{(j)}_i$ puts probability $s$ on $h_i$ and $(1-s)/K\le s$ on each $a_{i,j,k}$. A request carries a public route label $j\in\{1,\dots,r\}$ outside the context window, as the routes $R_1,R_2$ of Appendix~\ref{hn:model}, and the model returns $p^{(j)}_i$ on a route-$j$ prompt ending in $h_0$ (for $i=1$) or in a token of $\mathcal V_{i-1}$. The class $\mathcal G_r$ consists of procedures that start from the draft, read only returned labels, request each context and route at most once, and release at position $i$ either $h_i$ or a token returned at that context. Released tokens lie in $\mathcal V_i$, so the $n$ contexts are distinct and all positions are retained. A count observes the released text, the entire key, the public model and the draft, and may use independent randomization.

\begin{lemma}[Hit probability and races]\label{rm:race}
At a draft context, let $E_h=-\log U_{h_i}$ and $C_j=\min_kE_{a_{i,j,k}}/\{(1-s)/K\}$. Then $E_h,C_1,\dots,C_r$ are independent, $E_h\sim\Exp(1)$, each $C_j$ is exponential with rate $1-s$, and the index of the minimizing competitor of menu $j$ is uniform on $\{1,\dots,K\}$ and independent of $C_j$. The draft token wins menu $j$ if and only if $E_h/s<C_j$, so the probability that it wins at least one of the $r$ menus is
\[
q_r=1-\int_0^\infty e^{-x}\big(1-e^{-x(1-s)/s}\big)^r\,dx ,\qquad q_1=s,\qquad q_r=\frac r{r+1}\ \text{at }s=1/2 .
\]
\end{lemma}
\begin{proof}
The competitor sets of different menus are disjoint, so the $C_j$ are functions of disjoint sets of independent coordinates. The minimum of $K$ independent exponentials with rates $(1-s)/K$ is exponential with rate $1-s$ and is independent of its index, which is uniform \citep[Theorem~II.1]{moulton2018}. Given $E_h=x$, the draft token loses all $r$ menus with probability $(1-e^{-x(1-s)/s})^r$; integrate. For $r=1$ the integral is $1-1/(1+(1-s)/s)=1-s$, and at $s=1/2$ it is $\int_0^\infty e^{-x}(1-e^{-x})^rdx=1/(r+1)$.
\end{proof}

\begin{lemma}[Row tails]\label{rm:tails}
Fix a record branch of a procedure in $\mathcal G_r$, and let $c_\ast>0$ be the smallest prior probability of the finitely many row events ``the requested routes $J$ returned the winners $(w_j)_{j\in J}$''. Let $u\in(0,1)$ and $\beta=\min\{2,1/s\}>1$. (a) If position $i$ releases $h_i$ and $h_i$ lost the $k\le r$ menus requested there, then $\PP(U_i<u\mid\Gamma)\le u/(1-q_k)\le u/(1-q_r)$. (b) If position $i$ releases a match, then $\PP(U_i<u\mid\Gamma)\le u^\beta/c_\ast$.
\end{lemma}
\begin{proof}
Given $\Gamma$, the row of position $i$ has the prior law restricted to its row event (Lemma~\ref{hp:null} and the factorization of Appendix~\ref{app:propK}). Write $\ell=\log(1/u)$ and $c=(1-s)/s$. (a) By Lemma~\ref{rm:race}, the row event has probability $\int_0^\infty e^{-x}\{(1-e^{-cx})/K\}^kdx=(1-q_k)/K^k$, and its intersection with $\{E_h>\ell\}$ has probability $\int_\ell^\infty e^{-x}\{(1-e^{-cx})/K\}^kdx\le e^{-\ell}/K^k$; divide, and use that $q_k$ increases in $k$. (b) The row event has probability at least $c_\ast$, and every row event has positive probability because the $C_j$ are independent and every index has positive probability. If the released token is the winner $a=a_{i,j,k}$ of menu $j$, then $E_a/\pi_a<E_h/s$ with $\pi_a=(1-s)/K\le s$, so $\{E_a>\ell\}\cap\{a\text{ wins}\}\subseteq\{E_a>\ell,\ E_h>s\ell/\pi_a\}$, which has probability at most $e^{-\ell}e^{-s\ell/\pi_a}\le u^2$. If the released token is $h_i$ and it won menu $j$, then $\PP(E_h>\ell,\ E_h/s<C_j)=\int_\ell^\infty e^{-x}e^{-cx}dx=s\,u^{1/s}$. Divide by $c_\ast$.
\end{proof}

\begin{proof}[Proof of Theorem~\ref{thm:sw-sharp}(i)]
Use the tail count $d_u$ of Proposition~\ref{sw:tail}, which is valid for every history. Fix a branch with $H\ge h$. Given $\Gamma$ the rows are independent, at least $h$ of them are matches, and if $u^\beta/c_\ast\le u/(1-q_r)$, Lemma~\ref{rm:tails} gives
\[
B_u\ \st\ \mathrm{Bin}\big(h,u^\beta/c_\ast\big)+\mathrm{Bin}\big(n-h,u/(1-q_r)\big),
\]
with independent terms and a law that does not depend on $\Gamma$. Take $k=\lceil qn\rceil$ with $q<\rho_r$. As in Corollary~\ref{sw:sharp}(i), it suffices that $(1-q)u>\kappa u^\beta/c_\ast+(1-\kappa)u/(1-q_r)$, that is, $q<\rho_r-\kappa u^{\beta-1}/c_\ast$, which holds for small $u$ because $\beta>1$ and $\rho_r=1-(1-\kappa)/(1-q_r)$ when $\kappa>q_r$. Hoeffding's inequality for the null cutoff and for the reference law gives $\PP(d_u\ge k)\ge1-e^{-cn}$ for some $c>0$, uniformly over the branches.
\end{proof}

\begin{proof}[Proof of Theorem~\ref{thm:sw-sharp}(ii)]
The low policy releases $h_1,\dots,h_{n-j}$ without requests and, at each later position, requests route $1$ and releases the winner; its $H$ equals $j$. The high policy keeps a miss counter. At each position $i\le n-j$ before fallback it requests routes $1,\dots,r$ in turn until $h_i$ wins; it releases $h_i$ if $h_i$ won; if $h_i$ lost all $r$ menus it releases $h_i$ and adds one miss if the counter stays at most $n-h$, and otherwise releases the route-$1$ winner and enters fallback. In fallback and at every position $i>n-j$ it requests route $1$ and releases the winner. Both policies lie in $\mathcal G_r$ and do not depend on $D$. Under the high policy the only nonmatches are kept draft tokens that lost all menus, at most $n-h$ of them, so $H\ge h$ on every realization.
Run both policies on the same key, and let $Y_i$ indicate that $h_i$ wins at least one of the $r$ menus at the draft context $c_i(x)$, $i\le n-j$. The draft contexts are distinct, so the $Y_i$ are independent Bernoulli$(q_r)$ by Lemma~\ref{rm:race}, and before fallback the high policy hits at $i$ exactly when $Y_i=1$. On $A=\{\sum_{i\le n-j}Y_i\ge h-j\}$ the counter never exceeds $(n-j)-(h-j)=n-h$, so both policies release $h_1,\dots,h_{n-j}$; afterwards they have the same contexts and both use route $1$, so they release the same winners. Hence, as in the proof of Theorem~\ref{sw:converse},
\[
\PP_{\rm high}(D>j)\le\PP(A^c)+\PP_{\rm low}(D>H)\le\PP\{\mathrm{Bin}(n-j,q_r)<h-j\}+\alpha .
\]
For $q'>\rho_r$ take $j=\lfloor q'n\rfloor$: $(n-j)q_r-(h-j)=n\{(1-q_r)q'-(\kappa-q_r)\}+o(n)$ is a positive multiple of $n$, so the binomial term tends to zero. If $\kappa<q_r$, take $j=0$; then $\{D>0\}=\{D\ge1\}$.
\end{proof}

At $n=\ProtocolN$, $h=\ProtocolH$, $s=1/2$ and $\alpha=0.05$, $q_2=\RmQTwo$ and $q_3=\RmQThree$, and with $j=0$ the bound gives $\PP(D\ge1)\le\RmDetUpperTwo$ for $r=2$ and $\le\RmDetUpperThree$ for $r=3$ (exact rational binomial tails, code available from the author on request). The high policy then makes up to $2\ProtocolN$ or $3\ProtocolN$ requests, so these values are not comparable with the budget of \ProtocolBudget{} requests in Theorem~\ref{thm:full-key}. The result concerns fixed menus: if a later menu may depend on the returned token, the hit rate can exceed $q_r$; with menus $(1/2,1/4,1/4)$ on the draft token and two competitors, followed by $(1/2,1/2)$ on the draft token and the competitor that did not win, it is $0.70>q_2=2/3$.

\subsection{Finite-length guarantees for CountScan at its start thresholds}\label{sw:envelope-sec}
The counts above use a fixed-region error level. CountScan instead calibrates each start $a$ at $c_a=\log\{1/(\alpha\pi_a)\}$, which is $\log(|T|/\alpha)$ with equal start weights, and covers all intervals at once. Fix an interval $S$ of $n_S$ retained positions with start $a$, satisfying the conditions of $\mathcal P_{\rm sw}(h,s)$, and write $B^{\rm ref}_r$ as in Proposition~\ref{sw:tail} with $n=n_S$. For $r\in(0,1)$ and $0\le\theta<r$, the two-level factor
\[
e_{r,\theta}(P)=\frac\theta r\,\ind\{U<r\}+\frac{1-\theta}{1-r}\,\ind\{U\ge r\},\qquad U=1-P,
\]
is nonincreasing in the pivot and integrates to one under a uniform null, so CountScan with $e_{r,\theta}$ fixed before the scores is valid by Theorem~\ref{thm:main}; $\theta=0$ uses $\log0=-\infty$.

\begin{proposition}[Finite-length envelope]\label{sw:envelope}
Let $1\le k\le n_S$, $m=n_S-k+1$, and let $B=\#\{t\in S:U_t<r\}$. With $e_{r,\theta}$, $d(S)\ge k$ if and only if $B<m$ and $(m-B)\log\frac{1-\theta}{1-r}+B\log\frac\theta r\ge c_a$. Consequently:
(a) if $b<mr$ is an integer with $m\,\mathrm{kl}(b/m\,\|\,r)\ge c_a$, then $\theta=b/m$ gives $\PP\{d(S)\ge k\}\ge\PP(B^{\rm ref}_r\le b)$;
(b) if $\kappa=h/n_S>s$, $\rho=(\kappa-s)/(1-s)$, $\gamma=1/s-1$, $L=\log(1/\beta)$, $q=(k-1)/n_S$ and
\[
q\le\rho(1-r^\gamma)-\sqrt{\frac{2c_a}{n_Sr}}-\sqrt{\frac{2L}{n_Sr}}-\frac{(2/3)L+1}{n_Sr},
\]
then $\theta=\lceil\mu+\sqrt{2n_SrL}+(2/3)L\rceil/m$, with $\mu=\EE B^{\rm ref}_r$, gives $\PP\{d(S)\ge k\}\ge1-\beta$.
\end{proposition}
\begin{proof}
Every position has evidence $\log\frac{1-\theta}{1-r}>0$ or $\log\frac\theta r<0$, and $B$ positions have the latter. Since $d(S)\ge k$ if and only if $R_S(k-1)\ge c_a$, and trimming removes $k-1$ of the $n_S-B$ positive values when they exist, $R_S(k-1)=(m-B)\log\frac{1-\theta}{1-r}+B\log\frac\theta r$ if $B<m$, while $R_S(k-1)\le0<c_a$ if $B\ge m$. This is the displayed equivalence.

(a) With $\theta=b/m$ and $x=b/m$, the left side at $B=b$ equals $m\{(1-x)\log\frac{1-x}{1-r}+x\log\frac xr\}=m\,\mathrm{kl}(x\,\|\,r)\ge c_a$, and it decreases in $B$, so $d(S)\ge k$ whenever $B\le b$. The coupling in the proof of Proposition~\ref{sw:tail} gives $B\le B^{\rm ref}_r$ on every branch.

(b) The mean of $B^{\rm ref}_r$ is $\mu=n_Sr\{1-\rho(1-r^\gamma)\}$, since $r^{1/s}=r\cdot r^\gamma$, and its variance is at most $\mu\le n_Sr$. Bernstein's inequality gives $\PP(B^{\rm ref}_r\le b)\ge1-\beta$ for every integer $b\ge\mu+\sqrt{2n_SrL}+(2/3)L$, so some such $b\le b^\star=\mu+\sqrt{2n_SrL}+(2/3)L+1$. For $0\le x\le r$, $\mathrm{kl}(x\,\|\,r)\ge(r-x)^2/(2r)$: both sides and their first derivatives vanish at $x=r$, and $\partial_x^2\mathrm{kl}(x\,\|\,r)=1/\{x(1-x)\}\ge1/r$ on $[x,r]$, so Taylor's formula with integral remainder at $r$ gives the inequality. Hence (a) applies once $b^\star\le mr-\sqrt{2rc_am}$. Dividing by $n_Sr$, and using $m/n_S=1-q$ and $\sqrt{2c_a(1-q)/(n_Sr)}\le\sqrt{2c_a/(n_Sr)}$, the displayed condition implies this.
\end{proof}
Put $A=c_a+L+1$. Taking $r\asymp(A/n_S)^{1/(2\gamma+1)}$ in (b) shows that CountScan with a prespecified two-level factor certifies the proportion $\rho-O\{(A/n_S)^{(1-s)/(2-s)}\}$ with probability at least $1-\beta$. Let $h/n_S\to\kappa>s$ and take $\beta_n\to0$ with $\log(1/\beta_n)=o(n_S)$. Since $c_a=\log(|T|/\alpha)=o(n_S)$ when $\log|T|=o(n_S)$, CountScan at its simultaneous thresholds then certifies every proportion below $\rho$ with probability tending to one, so the price of simultaneity does not change the first-order certifiable proportion. Each target proportion uses its own design, fixed before the scores at the start of $S$; one start cannot be tuned to several targets or lengths without splitting its budget. Part (b) is an upper bound on the gap for this construction; Theorem~\ref{sw:converse} matches only the first-order proportion.

Table~\ref{tab:swscan} evaluates part (a) at the actual thresholds. For each target $k$ the design $(r,\theta)$ maximizes the reference probability over the integer cutoff $b$, using only $(n_S,h,s,c_a,k)$, so it is fixed before the scores; reporting several targets with different designs requires splitting the budget, for instance by replacing $c_a$ with $c_a+\log(1/\omega_g)$ for weights $\omega_g$. At $n_S=\ProtocolN$, $h=\ProtocolH$ and $s=1/2$, the best reference probability of certifying one match is only \SwScanHalfFirst, against \ProtocolPowerLower{} for the fixed-region Gamma certificate. The start threshold $\log(|T|/\alpha)$ is the main cause: at $\log(1/\alpha)$ the same optimized two-level design gives \SwScanHalfFixedFirst, and the rest of the gap comes from the statistic, two-level factors against the Gamma sum. For the report rule of Section~\ref{subsec:real} at $n_S=256$, $h=154$ and $|T|\le1000$, the reference probability of the eligibility target $d\ge65$ is \SwScanEligHalf{} at $s=1/2$ and \SwScanEligThree{} at $s=0.3$. These are deterministic floating-point evaluations of reference lower bounds, each selected point checked by an independent convolution (code available from the author on request).

\begin{table}[H]
\centering
\caption{Two-level CountScan at its start threshold $\log(|T|/\alpha)$, $\alpha=0.05$, under the single-winner reference law with match fraction $0.6$. $\PP(d\ge1)$ is the best reference lower bound for certifying one match; the last two columns give the largest target $k$ whose reference lower bound on $\PP(d\ge k)$ is at least $50\%$ or $90\%$, each target with its own prespecified design.}\label{tab:swscan}
\begin{tabular}{rrrrrr}
\toprule
$n_S$ & $|T|$ & $s$ & $\PP(d\ge1)$ & count at $50\%$ & count at $90\%$\\
\midrule
300 & 300 & 0.5 & 0.0030 & 0 & 0\\
300 & 300 & 0.3 & 0.9960 & 37 & 19\\
300 & 300 & 0.1 & $\approx1$ & 123 & 113\\
300 & 1000 & 0.5 & 0.0010 & 0 & 0\\
300 & 1000 & 0.3 & 0.9896 & 32 & 14\\
300 & 1000 & 0.1 & $\approx1$ & 121 & 110\\
256 & 1000 & 0.5 & $7.3\times10^{-4}$ & 0 & 0\\
256 & 1000 & 0.3 & 0.9612 & 23 & 6\\
256 & 1000 & 0.1 & $\approx1$ & 100 & 91\\
\bottomrule
\end{tabular}

\end{table}

\section{Proofs of Theorems~\ref{thm:budget} and~\ref{thm:unified}: Two Menus with a Request Budget}\label{app:budget}
This appendix proves Theorems~\ref{thm:budget} and~\ref{thm:unified}: the protocol (\ref{app:budget-protocol}), the scores and potentials (\ref{app:budget-scores}), the lower and upper limits (\ref{app:budget-lower}, \ref{app:budget-upper}), finite-length bounds (\ref{app:budget-finite}) and the simulations (\ref{app:budget-sim}). Every local inequality, vertex strategy and finite-length certificate is checked in exact rational arithmetic in the author's reproducibility records, available on request.

\subsection{Protocol and information}\label{app:budget-protocol}
For each $n$ there is a public finite pool $\mathcal I_n$ of $M_n\ge n$ addresses, possibly growing with $n$. Each address carries an independent ideal key row and public labels $h,a_1,a_2,o$. Menu $j\in\{1,2\}$ is the equal-weight Gumbel-max race between $h$ and $a_j$; both menus use the same coordinate of $h$. The counterfactual winner pair $\omega=(W_1,W_2)$ has prior
\[
\PP\{\omega=(h,h)\}=\tfrac13,\quad \PP\{\omega=(h,a_2)\}=\PP\{\omega=(a_1,h)\}=\tfrac16,\quad \PP\{\omega=(a_1,a_2)\}=\tfrac13 .
\]

\paragraph{Writer.} The writer reads the key only through requests, which return the winner of the requested menu at the requested address, and uses randomness independent of the key. Addresses, menus, stopping and the final release may depend on all past returns. The writer may interleave addresses, return to an old address, discard anything, and finally keep exactly $n$ distinct addresses in any record-measurable order. Under permission A the release at a kept address is any of its four labels; under R it is $h$ or a label returned at that address. All requests count against $C_{\rm all}\le B_n=\lfloor bn\rfloor$, discarded ones and repeated reads included; revisiting an address does not reset it.

\paragraph{Target and auditor.} At kept position $i$, $H_i=1$ if the released token was returned at that address at any time, and $H(S)=\sum_{i\in S}H_i$. A policy is high if $H([n])\ge h_n=\lceil\kappa n\rceil$ on every realization; this defines power and is not an input of any count. The auditor sees the ordered kept addresses and released tokens, the full key and the menus, and may randomize; it does not see the requests, $C_{\rm all}$, the discarded material or $H$. The complete record $\Gamma$, which contains all returns, decisions and the writer's randomness, is used only in proofs.

\begin{lemma}[Record decomposition]\label{bud:record}
Given $\Gamma$, the key rows of distinct kept addresses are conditionally independent, and the conditional law of each row depends only on the menu--label pairs recorded at that address. The same holds before any request, given the labelled history so far: the address, menu and returned label of every earlier request, and the writer's randomness.
\end{lemma}
\begin{proof}
Fix the writer's randomness. The record lists, for each request, its address, its menu and the returned label, and each decision is a function of earlier entries, so a positive-probability record is the event that, at each address, the requested menus returned the recorded labels. This event is a product over addresses of events about one row each, and the rows are independent a priori. Regular conditional probabilities extend this to continuous randomness for almost every value.
\end{proof}

\paragraph{Text embedding.} The addresses can be realized as contexts of a legal text: give position $i$ a disjoint public block of four tokens $\{h_i,a_{i,1},a_{i,2},o_i\}$ after a fixed public prefix, so that the four-token window before each position is distinct, and route the menu index outside the window. Every edited final text stays in the pool. The policy pairs of Appendix~\ref{app:budget-upper} act only at the window fixed by the released prefix; before the hard-budget truncation, given the full key $K$, both policies of a pair have the same release kernel $q(w\mid\omega)$, so $\Pr(w_{1:n}\mid K)=\prod_iq\{w_i\mid\omega_K(c_i(w_{<i}))\}$ is the same for both, with $c_i(w_{<i})$ the window fixed by the prefix. The full key and the whole text therefore have the same joint law; after truncation the two laws are compared through total variation. This realizes the protocol in a block-text model; it is not a statement about a fixed vocabulary or a deployed model.

\subsection{Record-safe scores and potentials}\label{app:budget-scores}
Let $Z_i=(\omega_i,w_i)$. The scores, listed for the release $w=(h,a_1,a_2,o)$, are
\begin{center}\footnotesize
\begin{tabular}{@{}lcccc@{}}
\toprule
$\omega$ & $(h,h)$ & $(h,a_2)$ & $(a_1,h)$ & $(a_1,a_2)$\\
\midrule
$f_{A1}$ & $(1,-\tfrac12,-\tfrac12,0)$ & $(1,-2,1,0)$ & $(1,1,-2,0)$ & $(-2,1,1,0)$\\
$f_{A2}$ & $(1,-\tfrac12,-\tfrac12,-\tfrac34)$ & $(-\tfrac12,-2,1,0)$ & $(-\tfrac12,1,-2,0)$ & $(-\tfrac12,1,1,0)$\\
$f_R$ & $(1,0,0,0)$ & $(1,0,1,0)$ & $(1,1,0,0)$ & $(-2,1,1,0)$\\
\bottomrule
\end{tabular}
\end{center}
All values lie in $[-2,1]$; $f_{A1}$ and $f_R$ equal $1$ at every match, and $f_{A2}\le1$.
\begin{lemma}[Record safety]\label{bud:safe}
Under the permission to which a score belongs, at every local return state and every legal release with $H_i=0$, $\EE[f(Z_i)\mid\Gamma]\le0$.
\end{lemma}
\begin{proof}
By Lemma~\ref{bud:record} the conditional law of $\omega_i$ given $\Gamma$ is the prior restricted to the patterns consistent with the returns at that address. There are nine local states (each menu unrequested or returning one of its two labels); for each, and each release that was not returned, the conditional mean is a finite rational sum, checked exactly in all cases. Since R allows a subset of the releases of A, the A scores are safe under R as well. The converse fails: with no request, releasing $a_1$ gives $\EE f_R(\omega,a_1)=1/2>0$, so each permission uses its own family.
\end{proof}

Let $s$ denote the local state, $\bar f(s,w)=\EE[f(\omega,w)\mid s]$ and $H(s,w)$ the indicator that $w$ was returned. For constants $a,t,d\ge0$ let $g(s,w)=\bar f(s,w)-aH(s,w)+d$. A query certificate is a function $V$ with
\begin{equation}\label{eq:bud-bellman}
V(\emptyset)=0,\qquad V(s)\le0,\qquad V(s)\le g(s,w),\qquad V(s)\le t+\EE[V(s')\mid s,j],
\end{equation}
for every legal release $w$ and every unrequested menu $j$, where $s'$ is the state after requesting $j$. The second inequality is the discard action, and a repeated read leaves $s$ unchanged and is covered by $t\ge0$. The largest such $V$ follows by backward recursion over the nine states, and its root value must be checked to be zero. The four budget planes are $(a,t,d)=(2,1,0)$ and $(\tfrac32,\tfrac12,\tfrac14)$ for $f_{A1}$ and $f_{A2}$, and $(2,1,0)$ and $(2,\tfrac23,\tfrac13)$ for $f_R$; in addition $f_{A1},f_R\ge3H-2$ pointwise, the plane $(3,0,2)$ with $V\equiv0$. The root values are zero, and, given the history and the requested menu, the possible next values of $V$ lie in an interval of length at most $L=3/2$.

\begin{lemma}[Interleaved queries]\label{bud:query}
For a score with certificate \eqref{eq:bud-bellman} and any policy of the protocol, with probability at least $1-\delta$,
\[
\sum_{i=1}^n\bar f(s_i,w_i)\ \ge\ aH([n])-tB_n-dn-L\sqrt{\tfrac{B_n}{2}\log\tfrac1\delta}.
\]
\end{lemma}
\begin{proof}
Keep the full local state $s_x(q)$ of every address after $q$ requests, even after its draft is discarded, and let $\Psi_q=\sum_{x\in\mathcal I_n}V\{s_x(q)\}$; untouched addresses contribute $V(\emptyset)=0$. The address and menu of request $q$ are measurable with respect to earlier returns, so by Lemma~\ref{bud:record} and \eqref{eq:bud-bellman}, $X_q=t+\Psi_q-\Psi_{q-1}$ has nonnegative conditional mean and conditional range at most $L$. Padding with zeros after the last request up to $B_n$ preserves both, so Azuma--Hoeffding gives $tC_{\rm all}+\Psi_{C_{\rm all}}\ge-L\{(B_n/2)\log(1/\delta)\}^{1/2}$. At the end each kept address has terminal value $g\ge V$ and each other address $0\ge V$, so $\sum_i g(s_i,w_i)\ge\Psi_{C_{\rm all}}$ whatever rows are kept, even if they are chosen after all returns. Rearranging with $C_{\rm all}\le B_n$ gives the claim.
\end{proof}

\subsection{Proof of the lower limit and of Theorem~\ref{thm:unified}}\label{app:budget-lower}
\paragraph{Validity.} Given $\Gamma$, the values $H_i$ are fixed and the kept rows are conditionally independent (Lemma~\ref{bud:record}). For a fixed score, bet $\lambda\in[0,1/2]$ and start $a$, the process $\prod_{i=a}^{b}\{1+\lambda f_j(Z_i)\}/(1+\lambda)^{H_i}$ is a nonnegative supermartingale in $b$: at a nonmatch the factor has conditional mean at most one by Lemma~\ref{bud:safe}, and at a match it is at most one because $f_j\le1$. Integration over $\lambda$ and fixed convex weights preserve this. Since $\mathcal E(S,k)$ decreases in $k$, $D_U(S)>H(S)$ implies $\mathcal E\{S,H(S)\}\ge n/\alpha$. Ville's inequality for each of the $n$ starts and a union bound give $\PP\{\exists S:D_U(S)>H(S)\mid\Gamma\}\le\alpha$, and averaging over $\Gamma$ proves Theorem~\ref{thm:unified}(i). No count threshold receives a separate error budget.

\paragraph{Score sums.} Given $\Gamma$, the scores of the kept rows are independent with range at most $3$, so with probability at least $1-\delta$, $\sum_if(Z_i)\ge\sum_i\bar f(s_i,w_i)-3\{(n/2)\log(1/\delta)\}^{1/2}$. With Lemma~\ref{bud:query} and the planes above, on an event of probability at least $1-\beta$ under every high policy,
\begin{equation}\label{eq:bud-sum}
\max\Big\{0,\ \max_j\frac1n\sum_{i=1}^nf_j(Z_i)\Big\}\ \ge\ \rho_P(\kappa,b)-\epsilon_n,\qquad \epsilon_n=\Big(3+\tfrac32\sqrt b\Big)\sqrt{\frac{\log(6/\beta)}{2n}},
\end{equation}
because on each affine piece $aH-tB_n-dn\ge n\{a\kappa-tb-d\}$ and $a\kappa-tb-d$ is the corresponding term of $\rho_P$; the zero term is covered by the $0$ on the left, since $D_U$ has no zero score. The constant does not depend on the pool size $M_n$.

\paragraph{Power of $D_U$.} Fix $\varepsilon>0$. If $\rho_P\le\varepsilon$, $D_U\ge0$ suffices. Otherwise, for $n$ large, \eqref{eq:bud-sum} holds with a score $j$ whose sum $S_j$ is at least $n(\rho_P-\epsilon_n)$. Let $\lambda_n=(\log n/n)^{1/2}$, take $n$ so large that $\lambda_n+n^{-2}\le1/4$, and keep only $\lambda\in[\lambda_n,\lambda_n+n^{-2}]$ in the integral; on this interval the log-integrand changes by $O(n^{-1})$ for every integer $k\le n$. Using $\log(1+\lambda x)\ge\lambda x-4\lambda^2$ for $x\in[-2,1]$ and $\lambda\le1/4$, and $\log(1+\lambda)\le\lambda$, the log of $\mathcal E([n],k)$ is at least
\[
\lambda_n(S_j-k)-4n\lambda_n^2-2\log n+\log(2w_j)-O(n^{-1}).
\]
Take the integer $k=\lceil n(\rho_P-\varepsilon)\rceil\ge1$ and apply this bound to $\mathcal E([n],k-1)$: since $k-1<n(\rho_P-\varepsilon)$ and $S_j\ge n(\rho_P-\varepsilon/2)$ for large $n$, the leading term is at least $(\varepsilon/2)(n\log n)^{1/2}$, which exceeds $\log(n/\alpha)$ and the $O(\log n)$ terms, the weight $w_j>0$ being fixed. Hence $\mathcal E([n],k-1)\ge n/\alpha$ and $D_U([n])\ge k\ge n(\rho_P-\varepsilon)$ with probability at least $1-\beta_n$, with $\beta_n=n^{-2}$ in \eqref{eq:bud-sum}, uniformly over the high policies. This proves Theorem~\ref{thm:unified}(ii) and the lower half of Theorem~\ref{thm:budget}. For a final interval $S$ with $m=|S|$, Lemma~\ref{bud:query} still applies, since rows outside $S$ have terminal value $0\ge V$, but the concentration of the remaining key must be taken over all intervals. With $K=3$ planes per permission and $K_n=n(n+1)/2$ intervals, on an event of probability at least $1-\beta$, simultaneously for all intervals,
\[
\max\Big\{0,\max_j\frac{F_j(S)}m\Big\}\ge\rho_P\Big(\frac{H(S)}m,\frac{B_n}m\Big)-\frac3{2m}\sqrt{\frac{B_n}2\log\frac{2K}\beta}-3\sqrt{\frac1{2m}\log\frac{2KK_n}\beta},
\]
where $F_j(S)=\sum_{i\in S}f_j(Z_i)$. The budget enters as $B_n/m$, not $b$; turning this into a bound on $D_U(S)$ requires the integral threshold as above.

\paragraph{Relation to existing counts.} Let $Y=\ind\{w\in\{W_1,W_2\}\}$. Under R, $f_R=3Y-2$ on every legal release: a release of $h$ scores $1$ exactly when $h$ wins some menu, and a release other than $h$ must have been returned, so it is a winner and scores $1$. Hence, with $M=\sum_{i\in S}Y_i$ and $m=|S|$, the R integrand is $(1+\lambda)^{M-k}(1-2\lambda)^{m-M}$, and the substitution $q=2(1+\lambda)/3$ turns $\mathcal E$ into the truncated uniform beta-binomial mixture for the counts $M$ \citep{howard2021confidence}. Under A, $f_{A1}\ge3Y-2$ pointwise, so $D_U$ at level $\alpha$ is at least the single-score mixture at level $\alpha/2$. It is not at least that mixture at level $\alpha$: when every row has $\omega=(h,a_2)$ and $w=h$, $f_{A1}=1$ and $f_{A2}=-1/2$, and at length \HBCexN{} the single-score mixture certifies \HBCexBase{} while $D_U$ certifies \HBCexMix.

\subsection{Proof of the upper limit}\label{app:budget-upper}
\paragraph{Equal-law pairs.} Restrict to a simpler legal subclass: the first $n$ addresses in order, an independent local request tree at each, release of the tree's token, no discarding. For single-row mixtures $\mu_0,\mu_1$ of trees, let $\ell_\mu$ be the law of $(\omega,w)$ and $c(\mu)$, $m(\mu)$ the mean cost and match. We need $\ell_{\mu_0}=\ell_{\mu_1}$, $c(\mu_0),c(\mu_1)\le b$, $m(\mu_1)\ge\kappa$ and $m(\mu_0)=\rho_P$. On the polygon $\{0\le\kappa\le1,\ \kappa\le b\le2\}$, split by which term of $\rho_P$ is largest into closed convex cells, whose vertices are exact intersections of rational lines. Writing $Q_j[u,v]$ for ``request menu $j$; if it returns $h$ do $u$, otherwise $v$'', the strategies at the vertices with $\rho_P\ne0$ or with distinct strategies are
\begin{center}\footnotesize
\begin{tabular}{@{}llll@{}}
\toprule
& $(\kappa,b)\mapsto\rho$ & low & high\\
\midrule
A, R & $(\tfrac23,\tfrac32),(\tfrac23,2)\mapsto0$ & $h$ & $Q_1[h,Q_2[h,h]]$\\
A, R & $(1,1),(1,2)\mapsto1$ & $Q_1[h,a_1]$ & $Q_1[h,a_1]$\\
A & $(\tfrac12,1)\mapsto0$ & $Q_2[a_1,a_1]$ & $Q_1[a_1,a_1]$\\
A & $(\tfrac34,\tfrac54)\mapsto\tfrac14$ & $\tfrac12h+\tfrac12Q_2[Q_1[h,a_1],a_1]$ & $\tfrac12Q_1[h,a_1]+\tfrac12Q_2[h,Q_1[a_1,h]]$\\
R & $(\tfrac12,1)\mapsto0$ & $h$ & $Q_1[h,h]$\\
\bottomrule
\end{tabular}
\end{center}
together with the release of $h$ at $(0,0)$ and $(0,2)$. Each pair is checked exactly on the sixteen cells of $(\omega,w)$. Inside a cell the same convex combination of the vertex pairs keeps equal laws and the linear constraints, and $\rho_P$ is affine there, so the low match rate equals $\rho_P$. For $b>2$ the strategies for $b=2$ apply.

\paragraph{Full key and text.} Every $\omega$ has positive prior mass, so $\ell_{\mu_0}=\ell_{\mu_1}$ gives the same conditional release kernel given $\omega$ under both policies; given $\omega$, the rest of the row does not depend on the tree, since the writer reads only winners. Independent rows then give the same joint law of the full key and the released text for the untruncated pair, also in the text embedding of Appendix~\ref{app:budget-protocol}; the hard-budget versions below are compared through total variation.

\paragraph{Hard budget.} Fix $\varepsilon\in(0,1)$, $t_0=\{\kappa+\min(b,1)\}/2$, $\eta=\varepsilon/4$ and $\delta=\eta\min\{b-t_0,t_0-\kappa\}$. Mix each policy of the pair with weight $\eta$ with the common policy that requests once with probability $t_0$ and releases the returned token, and otherwise releases $h$; the laws stay equal, $c\le b-\delta$ on both sides, $m(\tilde\mu_1)\ge\kappa+\delta$ and $m(\tilde\mu_0)+\delta\le\rho_P+\varepsilon/2$. The high policy tracks cost $C_u$, matches $H_u$ and deficit $d_u=(h_n-H_u)_+$ after $u$ positions and keeps $C_u+d_u\le B_n$ and $d_u\le n-u$. (i) If the remaining positions equal the deficit, or $d_u>0$ and $B_n-C_u<2+d_u$, it switches to one request per position, releasing the returned token, until the quota is met, and then releases $h$ without requests. (ii) If $d_u=0$ and $B_n-C_u<2$, it releases $h$ without requests from then on. (iii) Otherwise $B_n-C_u\ge2+d_u$ and it runs a fresh tree, which costs at most two. Each step preserves both invariants, so $C_{\rm all}\le B_n$ and $H\ge h_n$ on every realization. The low policy releases $h$ without requests once fewer than two requests remain.

Let $P_1$ and $P_0$ be the laws of the observation (full key, kept addresses and text) under the hard high and low policies. Up to its first switch each agrees with its pure tree sequence, whose observation laws are equal; a switch of the high policy requires one of $T_1=\{\sum M_i\le h_n\}$, $T_2=\{\sum C_i>B_n-2\}$, $T_3=\{\sum(C_i-M_i)>B_n-h_n-2\}$ for its pure sequence, using $C_i-M_i\ge0$, and a switch of the low policy requires $T_4=\{\sum C_i>B_n-2\}$ for its own. Hence $\mathrm{TV}(P_1,P_0)\le\sum_{r=1}^4\PP(T_r)$. The mixture gives $\EE M\ge\kappa+\delta$ and $\EE(C-M)\le b-\kappa-2\delta$ on the high side and $\EE C\le b-\delta$ on both, so for $n\delta\ge8$ each $T_r$ lies at least $n\delta/2$ from its mean, and with per-row ranges at most $2$ Hoeffding's inequality gives $\PP(T_r)\le e^{-n\delta^2/8}$. Let $j=\lfloor n(\rho_P+\varepsilon/2)\rfloor$ and let $H_0$ be the matches of the hard low policy. Truncation only removes requests, so $H_0$ is stochastically below the pure low matches, whose mean is at most $n(\rho_P+\varepsilon/2-\delta)$; in the text embedding the rows after a switch are fresh, so this comparison holds in distribution. Thus $P_0(H_0>j)\le e^{-2n\delta^2}$. For every count $D$ valid over the protocol, $\{D>j\}\subseteq\{D>H_0\}\cup\{H_0>j\}$ under $P_0$, so
\[
P_1(D>j)\le P_0(D>j)+\mathrm{TV}(P_1,P_0)\le\alpha+P_0(H_0>j)+\sum_{r=1}^4\PP(T_r),
\]
which gives
\begin{equation}\label{eq:bud-converse}
\PP\big[D([n])>\lfloor n(\rho_P+\varepsilon/2)\rfloor\big]\le\alpha+5e^{-n\delta^2/8},\qquad n\ge8/\delta .
\end{equation}
The two sides use the same protocol and information, and the hard policy belongs to the class over which the count is valid. Together with the lower half this proves Theorem~\ref{thm:budget}. The boundary cases are direct: for $\kappa=0$ the zero-request policy is high; for $\kappa=1\le b$ the score $f_{A1}$ or $f_R$ equals one at every match; for $b=\kappa$ the high class is nonempty only along $n$ with $\lfloor bn\rfloor\ge\lceil\kappa n\rceil$, and there every request must match, so the limit is $\kappa$.

\subsection{Finite-length bounds}\label{app:budget-finite}
The two concentration steps above can be merged. For a score $f$, $\theta>0$, $a,t\ge0$ and $v>0$, let $U$ be positive on local states with
\[
U(\emptyset)=1,\quad U(s)\ge1,\quad U(s)\ge v^{-1}\EE\big[e^{\theta\{aH(s,w)-f(Z)\}}\mid s\big],\quad U(s)\ge e^{-\theta t}\EE[U(s')\mid s,j].
\]
The product $e^{-\theta tq}\prod_xU\{s_x(q)\}$ is then a nonnegative supermartingale along all requests, and the terminal factors of the kept rows, conditionally independent given $\Gamma$, are controlled by the third inequality, those of the other rows by $U\ge1$. Hence, for every policy,
\[
\EE\big[v^{-n}\exp\{-\theta(F_n-aH+tC_{\rm all})\}\big]\le1,\qquad F_n=\sum_{i=1}^nf(Z_i),
\]
and on a high policy within budget $\PP(F_n<L)\le v^ne^{\theta(L-ah_n+tB_n)}$. We take $\theta=12\log r$ for a rational $r>1$ and rational $a,t,v$ and $U$ chosen so that $12\{aH(s,w)-f(Z)\}$ and $12t$ are integers at every local state; every exponent is then an integer power of $r$, and the inequalities are checked in exact arithmetic. By concavity, $\log(1+\lambda f)\ge\frac{f+2}3\log(1+\lambda)+\frac{1-f}3\log(1-2\lambda)$ on $[-2,1]$. Hence, for score $j$ with weight $w_j$ and on the event $\{F_j\ge L\}$ for its sum $F_j$, and for every integer $k\in\{1,\dots,n\}$,
\[
\mathcal E([n],k-1)\ge w_jI(n,L,k-1),\qquad I(n,L,h)=2\int_0^{1/2}(1+\lambda)^{(L+2n)/3-h}(1-2\lambda)^{(n-L)/3}\,d\lambda,
\]
so $w_jI(n,L,k-1)\ge n/\alpha$ implies $D_U([n])\ge k$; here $w_{A1}=w_{A2}=1/2$ and $w_R=1$. Each such inequality is verified by outward-rounded lower sums. On the converse side, the actual tail probabilities of the five events above (three switching tails of the high policy, one of the low policy, and the low match tail) are bounded by rational Chernoff certificates instead of the uniform $\delta$.

\begin{table}[h]
\centering\footnotesize
\caption{Finite-length bands at $\kappa=\HBKappa$, $b=\HBBudget$, $\alpha=\HBAlpha$. Lower: share that $D_U$ certifies with probability at least $\HBLowProb$ under every high policy. Upper: for every valid count some high policy keeps it at or below this share with probability at least $0.85$. Shares are certified integer counts divided by $n$, lower bounds rounded down and upper bounds rounded up. Deterministic certificates, no simulation.}\label{tab:bud-finite}
\begin{tabular}{@{}lcccccc@{}}
\toprule
& & $n=300$ & $10^3$ & $3\cdot10^3$ & $10^4$ & $10^5$\\
\midrule
A ($\rho_A=\HBRhoA$) & lower & \HBLowAThree & \HBLowAThousand & \HBLowAThreeK & \HBLowATenK & \HBLowAHundredK\\
 & upper & \HBUpAThree & \HBUpAThousand & \HBUpAThreeK & \HBUpATenK & \HBUpAHundredK\\
R ($\rho_R=\HBRhoR$) & lower & \HBLowRThree & \HBLowRThousand & \HBLowRThreeK & \HBLowRTenK & \HBLowRHundredK\\
 & upper & \HBUpRThree & \HBUpRThousand & \HBUpRThreeK & \HBUpRTenK & \HBUpRHundredK\\
\bottomrule
\end{tabular}
\end{table}
The quantifiers differ: the lower bound holds for every high policy, the upper bound for one hard policy per count, and for $D_U$ on that policy both hold jointly with probability at least $0.75$, and at least $0.70$ together with validity over all intervals, which alone holds with probability at least $1-\alpha$. The bands narrow as $n$ grows, and the lower edge is positive from $n=10^4$ on.

\subsection{Simulations in the protocol}\label{app:budget-sim}
\paragraph{Registered study.} The design, seeds, cells and criteria were frozen before any simulation. Validity (E1): \HBEoneCells{} cells of $2000$ documents each, under low-match policies that release the draft, release unqueried competitors (A only), discard, reread and select, or interleave and select, for the fixed-bet count: CountScan with the scores of Appendix~\ref{app:budget-scores}, one bet $\lambda_n$ fixed in advance and a union over the scores. Power (E2, E3): \HBEtwoCells{} cells of $500$ documents, $n\in\{300,10^3,3\cdot10^3,10^4\}$, $\kappa\in\{0.6,0.7,0.8\}$, a grid of $b$, A and R, and three high histories with the hard-budget fallback: (a) the vertex strategies of Appendix~\ref{app:budget-upper}, (b) retries, (c) interleaving with discarding and selection. Each method reads the same histories, so the comparisons are paired and not independent evidence. (E4) pairs A and R with common random numbers. Selection by visible labels was found after the run to be covered more weakly than specified; a stress check for it was frozen separately and is reported apart from the registered results. The single-score mixture is the beta-binomial mixture applied to $3Y-2$ under both permissions; it is valid under A as well, since $3Y-2\le f_{A1}$.

\paragraph{Results.} The fixed-bet count had \HBEoneEvents{} overcounts (largest Wilson upper endpoint \HBEoneWilson). It is not uniformly better than existing counts: the single-score mixture had a higher mean in \HBMixBetterCells{} of \HBEtwoCells{} full-document cells. At $\kappa=\HBKappa$, $b=\HBBudget$, $n=10^4$ and history (a), the fixed bet certified \HBRFixed{} under R, and the R mixture, which equals $D_U$, \HBRUnified. The R$-$A difference of the fixed bet was \HBRminusA{} points \HBRminusACI{} against an asymptotic \HBRminusATheory; outside $1<b<1.3$, where the asymptotic difference is zero, five finite-length intervals excluded zero.

\paragraph{Follow-up for $D_U$ under A.} After the registered study, a follow-up with fresh seeds was frozen before running, with the equal weights fixed in advance. Its validity part used \HBAEoneCells{} cells at lengths $128$ and $512$; the frequency of an upper envelope of the overcount event, which bounds the actual overcounts, had largest Wilson upper endpoint \HBAEoneWilson. On the primary slice $D_U$ certified \HBAMix{} against \HBAFixed{} for the fixed bet on the same histories (paired difference \HBAGain{} points, Bonferroni-corrected $t$ interval \HBAGainCI), \HBACone{} for C1 and \HBABinMix{} for the single-score mixture. The registered short-passage criterion failed: on fixed passages of $128$ and $512$ positions every method certified \HBShortMax{} on average. Over the \HBAMixCells{} full-document cells $D_U$ had a lower mean than the single-score mixture in \HBAMixBelowBin, than C1 in \HBAMixBelowCone{} and than the fixed bet in \HBAMixBelowFixed. The gain belongs to the whole frozen implementation (mixture over bets, score family and inversion); no ablation separates them.

\subsection{When the certificate becomes useful}\label{app:budget-usable}
Table~\ref{tab:bud-usable} reads the per-document counts already stored for the registered study (R) and the follow-up (A); no new documents were simulated. It reports, at $\kappa=\HBKappa$ and $b=\HBBudget$ with $500$ documents per cell, the probability that $D_U$ certifies at least one match and at least $2.5\%$ of the positions. The thresholds $1\%$, $2.5\%$ and $5\%$ and the target probability $0.9$ were chosen on 2026-11-14, after the results of both studies were known, and serve as a sensitivity grid rather than a registered criterion; the full grid, including $b=1.2$, is in the author's records, available on request. All counts use the all-interval threshold $n/\alpha$; a whole-document audit could use the single-start threshold $1/\alpha$, which was not run here. The three histories have different match fractions above $\kappa$, so the rows describe the histories, not a ranking of permissions.
\begin{table}[h]
\centering\footnotesize
\caption{Usability of $D_U$ by length at $\kappa=\HBKappa$, $b=\HBBudget$ ($500$ documents per cell; thresholds set after the runs).}\label{tab:bud-usable}
\begin{tabular}{@{}llrrrrrrrr@{}}
\toprule
& & \multicolumn{4}{c}{$\PP(D_U\ge1)$} & \multicolumn{4}{c}{$\PP(D_U/n\ge2.5\%)$}\\
\cmidrule(lr){3-6}\cmidrule(lr){7-10}
Perm. & History & $300$ & $10^3$ & $3\cdot10^3$ & $10^4$ & $300$ & $10^3$ & $3\cdot10^3$ & $10^4$\\
\midrule
A & vertex & 0\% & 10\% & 83\% & 100\% & 0\% & 4\% & 32\% & 100\%\\
A & retry & 14\% & 96\% & 100\% & 100\% & 6\% & 78\% & 100\% & 100\%\\
A & interleave & 100\% & 100\% & 100\% & 100\% & 100\% & 100\% & 100\% & 100\%\\
R & vertex & 1\% & 10\% & 89\% & 100\% & 0\% & 4\% & 51\% & 100\%\\
R & retry & 0\% & 27\% & 100\% & 100\% & 0\% & 5\% & 100\% & 100\%\\
R & interleave & 100\% & 100\% & 100\% & 100\% & 100\% & 100\% & 100\% & 100\%\\
\bottomrule
\end{tabular}

\end{table}

\section{Proofs for Section~\ref{subsec:limits-end}: Histories that Hide Every Match}
\label{hn:section}
This appendix states and proves Proposition~\ref{prop:draft-only} and Theorem~\ref{thm:full-key}; the construction for the theorem follows the statements.

\subsection{Statements from Section~\ref{subsec:limits-end}}\label{hn:statements}
\begin{proposition}[Draft retention]\label{prop:draft-only}\label{rem:draft-only}
Let the draft $x$ be a function of key-free randomness, and let a writer request a suggestion at every audited position and release $x$. This writer satisfies Assumption~\ref{ass:Astar}, and every certificate $C\in\{0,1\}$ from the final text and the entire key with $\PP(C>H)\le\alpha$ under Assumption~\ref{ass:Astar} has $\PP(C\ge1)\le\alpha$ for this writer. Yet, given the key-free randomness $R$ that fixes $x$, $\EE[H\mid R]=\sum_{t\in S\cap T}\pi_t(x_t)$, where $\pi_t$ is the next-token distribution at $t$, itself a function of $R$.
\end{proposition}
\begin{proof}
The release is $x$ whatever the requests return, so the audit input $(x,\mathrm{Key})$ has the same law as for the allowed writer who makes no requests and releases $x$, for whom $H=0$; hence $\PP(C\ge1)\le\alpha$. Retained contexts are distinct, so each requested row is fresh and returns $x_t$ with probability $\pi_t(x_t)$ (Lemma~\ref{lem:coincidence}).
\end{proof}
Such coincidental matches concentrate on likely tokens, whose pivots are close to null (Lemma~\ref{lem:coincidence}). Even under a finite request budget, retries defeat detection at finite length:
\begin{theorem}[A full-key power limit]\label{thm:full-key}
There is a fixed public model with two menus per context, every menu probability at most $1/2$, and a public draft of \ProtocolN{} audited positions with distinct contexts. Allow at most \ProtocolBudget{} requests, at most two per context, and release of draft tokens even when no request returned them. Any certificate based on the final text and the entire key with $\PP(C>H)\le0.05$ uniformly over the allowed policies has $\PP(C\ge1)\le\ProtocolPowerUpper$ for some policy with $H\ge\ProtocolH$ on every realization.
\end{theorem}
\begin{proof}[Proof idea]
A policy that makes no requests and releases the draft $x$ has $H=0$. A second policy requests each draft position, retries up to \ProtocolRetry{} times after a near miss, keeps the draft word whether or not it was returned, and falls back to returned winners only when keeping the draft would leave fewer than \ProtocolH{} matches; it leaves the draft with probability about $\ProtocolDraftFailureApprox$ \citep[Theorem~II.1]{moulton2018}, and Lemma~\ref{lem:indist} gives $\PP_1(C\ge1)\le0.05+\ProtocolDraftFailureApprox$ (Appendix~\ref{hn:probability}).
\end{proof}
\begin{corollary}[Count form]\label{cor:full-key-count}
Any count $d$ from the final text and the entire key with $\PP(d>H)\le0.05$ uniformly over the allowed policies, CountScan included, has $\PP(H-d\ge\ProtocolH)\ge\PP(d=0)\ge\ProtocolMissLower$ under the policy of Theorem~\ref{thm:full-key}.
\end{corollary}

The information available to a certification rule and the actions available to a text-producing service must be specified together. A service may return several suggestions for the same context while an editor keeps an existing draft. The complete history can then contain many matches to the released text even though the release is unchanged. We construct such a service under explicit local and total query budgets. The construction provides a finite benchmark for checking whether a proposed certification guarantee remains attainable under this interface. For comparison, the single-winner power guarantee remains valid when restricted to the same text size, token-probability cap and query limits used below. Within that common background, it applies uniformly to the single-winner subclass, while the present construction supplies one difficult history mechanism for every uniformly valid rule. These are complementary statements about history classes, rather than performances of two algorithms on one distribution.

The distinction between scalar-pivot and full-vector observation is consequential for watermark proportion estimation \cite{chai2026sample}. Our observation includes the entire ideal key. The difficulty studied here instead arises from unobserved requests sharing the same context key, together with the permission to retain an external draft. The target remains the number of historical context--token matches.

\subsection{Interface and certification requirement}
\label{hn:interface}

Let the key contain mutually independent variables
\[
 U_{c,v}\sim\operatorname{Unif}(0,1),\qquad
 E_{c,v}=-\log U_{c,v}\sim\Exp(1),
 \qquad (c,v)\in\mathcal V^4\times\mathcal V,
\]
where $\mathcal V$ is finite. A request with context $c$ and next-token distribution $p$ returns
\begin{equation}
 W(c,p)=\arg\min_{v:p_v>0}\frac{E_{c,v}}{p_v}.
 \label{hn:race}
\end{equation}
Repeated requests at the same context use the same key row. The shared exponential-race law \citep[Theorem~II.1]{moulton2018} gives the joint return probabilities for the two requests.

Consider the following class of text-producing procedures. A procedure may start with a draft fixed before the key is drawn, issue requests to a publicly fixed model, and choose subsequent prompts and released tokens from the draft and the observed winner labels. It does not read key values or functions of the key other than these returned labels. Before releasing the token at a position, it may request a second suggestion for the same four-token context and retain the first suggestion. It may also retain a draft token that neither request returned. Every procedure in the class uses at most $420$ requests and at most two requests at any one context; every requested next-token distribution has maximum probability at most $1/2$. There is no positive lower bound on the number of requests.

Write a released text as $Z=(z_{-3},\ldots,z_0,z_1,\ldots,z_{300})$ and let $c_i(Z)=(z_{i-4},\ldots,z_{i-1})$. We require the $300$ contexts $c_i(Z)$ to be distinct. If $\mathcal H$ is the set of all context--winner pairs returned during generation, the target is
\begin{equation}
 H(Z,\mathcal H)
 =\sum_{i=1}^{300}\ind\{(c_i(Z),z_i)\in\mathcal H\}.
 \label{hn:target}
\end{equation}
This count concerns the complete request history, including discarded suggestions.

A certification rule $C(Z,U;\xi)\in\{0,\ldots,300\}$ observes the released text, the entire key, and the public model, prompt menu, draft, and budget limits. Its auxiliary randomization $\xi$ is independent of generation and the key. It does not observe the actual number of requests, their prompts, their returned labels, or the identity of the producing procedure. For the allowed class $\mathcal G$, the coverage requirement is
\begin{equation}
 \sup_{G\in\mathcal G}\PP_G\{C>H\}\le\alpha.
 \label{hn:coverage}
\end{equation}
In particular, $\mathcal G$ includes the procedure that makes no requests and releases the external draft. The condition $H\ge180$ will specify the class on which power is evaluated; it is not additional information supplied to $C$.

\subsection{A fixed model on a finite vocabulary}
\label{hn:model}

Fix a draft
\[
 x=(h_{-3},\ldots,h_0,h_1,\ldots,h_{300})
\]
whose $304$ tokens are distinct. Let $c_i=c_i(x)$. Add two competitor tokens $A,B$, two routing tokens $R_1,R_2$, and $300$ backup pairs $(C_j,D_j)$, $j=0,\ldots,299$. All these tokens are mutually distinct and are distinct from the draft tokens. Thus $908$ vocabulary elements suffice.

For each draft context, fix the two distributions
\begin{equation}
 p_i(h_i,A,B)=(1/2,2/5,1/10),\qquad
 q_i(h_i,B)=(1/2,1/2),
 \label{hn:menus}
\end{equation}
with zero probabilities outside the displayed supports. These are two preassigned menus at each context, not two globally identical probability vectors across different target tokens.

A single time-homogeneous model depending only on the last five submitted tokens implements all requests. On input suffix $(R_1,c_i)$ it uses $p_i$, and on $(R_2,c_i)$ it uses $q_i$. On any other suffix ending in $A$ or $B$, it assigns probability $1/2$ to each of $C_0,D_0$. On any other suffix ending in $C_j$ or $D_j$, for $0\le j<299$, it assigns probability $1/2$ to each token in the next backup pair. Every remaining suffix, including one ending in $C_{299}$ or $D_{299}$, receives the default distribution $(1/2,1/2)$ on $(A,B)$. Inputs shorter than five tokens may use the same default. These rules define the model before the key is drawn and do not consult a position counter or a returned winner.

The client submits the five-token prompt $(R_k,c)$ when using route $k$ at its current context $c$. The model sees all five tokens, whereas the key in \eqref{hn:race} is indexed only by the last four. Thus the two routes share a key row without requiring any modification of the fixed model after a request. All specified distributions satisfy the maximum-probability restriction.

\subsection{An immediate retry policy}
\label{hn:policy}

Reserve one first request for each of the $300$ released positions and a separate pool of $120$ second-request coupons. Start with the four-token draft prefix. Before a fallback occurs, the released prefix agrees with the draft. At position $i$, perform the following steps.
\begin{enumerate}
 \item Submit route $R_1$ and retain the identity of its winner $w$. If $w=h_i$, the position is matched.
 \item If $w=A$ and a second-request coupon remains, spend one coupon and submit route $R_2$ at the same context. The position is matched if this request returns $h_i$. If $w=B$, if this retry fails, or if $w=A$ and no coupon remains, treat the position as unmatched.
 \item If the position is matched, release $h_i$. If it is unmatched and this would leave at most $120$ unmatched positions so far, also release the draft token $h_i$ and record one permanent miss.
 \item If releasing $h_i$ would create the $121$st permanent miss, release the first-request winner $w$ instead and enter fallback mode. At every remaining position, make one route-$R_1$ request at the current context and release its actual winner. Make no further second requests.
\end{enumerate}
The first-request budget remains available after the retry coupons are exhausted. This separation of budgets permits later positions to succeed on their first requests. The policy never revisits a released position and never observes a future request outcome. Keeping an unmatched draft token and retaining the first winner after a failed retry are essential permissions of this interface.

\begin{lemma}[Pathwise feasibility]
\label{hn:feasible}
The immediate retry policy releases $304$ tokens with $300$ distinct four-token contexts. On every realization, it makes at most $420$ requests, visits each context at most twice, and satisfies $H\ge180$.
\end{lemma}
\begin{proof}
If no fallback occurs, there are at most $120$ permanent misses, so at least $180$ released context--token pairs have appeared as actual winners.

Suppose fallback occurs at position $i$. Exactly $120$ preceding positions are permanent misses. Of the first $i-1$ positions, the other $i-121$ are historical matches. The first winner retained at position $i$ is a winner at the current, unchanged context. Each of the remaining $300-i$ positions retains an actual winner at its own context. Consequently the number of historical matches is at least
\[
 (i-121)+1+(300-i)=180.
\]

Before fallback, all released tokens are distinct draft tokens. The fallback token is $A$ or $B$, and each subsequent token belongs to the next, previously unused backup pair. Hence all $304$ released tokens are distinct. In particular, the last tokens of their $300$ four-token contexts are distinct, so the contexts are distinct. Every actual request is at the context of its corresponding released position; only an immediate retry can reuse that context. There are exactly $300$ first requests and at most $120$ retries. These assertions hold separately on every realization, including those on which the unchanged draft cannot be retained.
\end{proof}

\subsection{The probability of retaining the draft}
\label{hn:probability}

For the probability calculation, define both request outcomes at all draft contexts using the full key, whether or not the online policy reaches those contexts. This is a counterfactual fixed-draft plan, not additional information available to the policy. Let $X$ be the number of first-request target winners and $J$ the number of first-request $A$ winners. Let $Y$ count successful second requests among the earliest $\min(J,120)$ of these $A$ positions, in document order.

\begin{lemma}[Fixed-draft event]
\label{hn:event}
Under the independent-row key model,
\begin{align}
 (X,J,300-X-J)&\sim\operatorname{Multinomial}(300;1/2,2/5,1/10),
 \label{hn:multinomial}\\
 Y\mid(X,J)&\sim\operatorname{Bin}(\min(J,120),5/14).
 \label{hn:conditional-binomial}
\end{align}
The immediate retry policy releases $x$ if and only if $X+Y\ge180$.
\end{lemma}
\begin{proof}
At one draft context, write the relevant independent exponential coordinates as $E_h,E_A,E_B$. The first-request probabilities are $1/2,2/5,1/10$. For both requests to return $h$, it is necessary and sufficient that
\[
 E_A\ge(4/5)E_h,\qquad E_B\ge E_h.
\]
The probability of this event is therefore
\begin{equation}
 \int_0^\infty e^{-t}e^{-(4/5)t}e^{-t}\,dt=\frac5{14}.
 \label{hn:joint-win}
\end{equation}
The second request returns $h$ with marginal probability $1/2$. If the first request returns $B$, then $E_B<E_h/5<E_h$, so the second request cannot return $h$. It follows that
\begin{equation}
 \PP\{W(c,q_i)=h_i\mid W(c,p_i)=A\}
 =\frac{1/2-5/14}{2/5}=\frac5{14}.
 \label{hn:rescue}
\end{equation}
The draft contexts have independent key rows. Conditional on all first-request labels, the second-request success indicators at the $A$ positions are independent, each with probability $5/14$. Selection of the earliest $\min(J,120)$ such positions uses only these first-request labels. This proves \eqref{hn:multinomial}--\eqref{hn:conditional-binomial}, including after conditioning only on $(X,J)$.

In the counterfactual plan, the retries are spent at exactly these positions. A missed position cannot subsequently be rescued: its retry has failed, its first winner was $B$, or the coupon pool has already been exhausted. Thus the number of permanent misses is nondecreasing and ends at $300-X-Y$. It stays at most $120$ exactly when no prefix reaches $121$. Until that first crossing, the actual policy and the counterfactual plan coincide. A crossing triggers fallback and changes the current token from $h_i$ to $A$ or $B$. Therefore the actual release equals $x$ exactly when $300-X-Y\le120$.
\end{proof}

Write $F_{m,p}(k)=\PP\{\operatorname{Bin}(m,p)\le k\}$, with value zero for $k<0$ and one for $k\ge m$. Lemma~\ref{hn:event} gives the following finite failure probability:
\begin{equation}
\begin{split}
 \varepsilon
 &=\PP\{X+Y<180\}\\
 &=\sum_{x=0}^{179}\sum_{j=0}^{300-x}
 \frac{300!}{x!j!(300-x-j)!}
 \left(\frac12\right)^x
 \left(\frac25\right)^j
 \left(\frac1{10}\right)^{300-x-j}
 F_{\min(j,120),\,5/14}(179-x).
\end{split}
\label{hn:epsilon}
\end{equation}
Every term is rational. In particular, using $X\sim\operatorname{Bin}(300,1/2)$ and $J\mid X=x\sim\operatorname{Bin}(300-x,4/5)$ permits integer accumulation with common denominator $2^{300}5^{300}14^{120}$. The resulting value is
\[
 \varepsilon\simeq\HistFailureDisplay.
\]
The formula is exact for this retry policy and these menus.

\subsection{A limit for certification from the entire key}
\label{hn:limit}

\begin{theorem}[Finite-budget certification bound]
\label{hn:theorem}
For the fixed model and interface above, there exists a procedure $G_1$ with $H\ge180$ on every realization, at most $420$ requests, and at most two requests per context, such that every certification rule satisfying \eqref{hn:coverage} obeys
\begin{equation}
 \PP_{G_1}\{C\ge1\}\le\alpha+\varepsilon,
 \label{hn:power-bound}
\end{equation}
where $\varepsilon$ is given by \eqref{hn:epsilon}. The rule may use the entire key and independent randomization. At $\alpha=0.05$, the right-hand side is at most $\HistPowerUpper$.
\end{theorem}
\begin{proof}
Take $G_1$ to be the immediate retry policy. Its pathwise properties follow from Lemma~\ref{hn:feasible}. Define $G_0$ to make no requests and release the fixed draft $x$. This procedure satisfies all upper-budget restrictions and has $H=0$. It belongs to the same coverage class, with the same publicly specified model, available menus, draft, and budget limits.

Couple the two procedures using the same entire key $U$ and the same independent certification randomization $\xi$. Let $Z_1$ denote the release under $G_1$. On $\{Z_1=x\}$, both procedures supply exactly the same inputs to $C$, including every coordinate of the key. Pointwise under this coupling,
\[
 \ind\{C(Z_1,U;\xi)\ge1\}
 \le
 \ind\{C(x,U;\xi)\ge1\}+\ind\{Z_1\ne x\}.
\]
Taking expectations, applying coverage to $G_0$, and using Lemma~\ref{hn:event} yields
\[
 \PP_{G_1}\{C\ge1\}
 \le\PP_{G_0}\{C\ge1\}+\PP\{Z_1\ne x\}
 \le\alpha+\varepsilon.
\]
The last numerical assertion follows by exact evaluation of \eqref{hn:epsilon} and upward rounding. In particular, the probability bound is unconditional: no conditioning on successful retention of the draft is used to establish \eqref{hn:power-bound}.
\end{proof}

\section{Algorithm and Runtime}\label{app:algosec}
This appendix supports the algorithmic remark at the end of Section~\ref{sec:theorem}: it gives the full procedure, including the reporting rule, its CPU cost, and an online variant.

\subsection{Algorithm}\label{app:algo}
\begin{algorithm}[!ht]
\caption{CountScan with disjoint reports}\label{alg:countscan}
\begin{algorithmic}[1]
\Require text $w_{1:n}$, key, calibrator $g$, $\lambda$, level $\alpha$, $\eta$, lengths $L_{\min}\le L_{\max}$, $\bar n\ge|T|$
\State $T\gets\{t>m:\chi_s\ne\chi_t\text{ for all }s>m,\ s\ne t\}$; re-index $T$ as $1,\dots,|T|$
\State $\ell_t\gets\log(1-\lambda+\lambda g(P_t))$ for $t\in T$; $c\gets\log(\bar n/\alpha)$ \Comment{$\bar n=|T|$ in Theorem~\ref{thm:main}}
\For{$a=1,\dots,|T|$}
  \State $s\gets0$; empty Fenwick tree over the ranks of the $\ell_t$
  \For{$b=a,\dots,|T|$}
    \State $s\gets s+\ell_b$; \textbf{if} $\ell_b>0$ \textbf{then} insert $\ell_b$
    \State $d([a,b])\gets0$ if $s<c$, else $1+\max\{j:\Top_j([a,b])\le s-c\}$ \Comment{equals Definition~\ref{def:trim}}
  \EndFor
\EndFor
\State $f(0)\gets0$
\For{$b=1,\dots,|T|$}
  \State $f(b)\gets f(b-1)$; $\mathrm{choice}(b)\gets\varnothing$
  \For{$L=L_{\min},\dots,\min(L_{\max},b)$} \Comment{shortest first}
    \State $a\gets b-L+1$
    \If{$d([a,b])\ge\lfloor\eta L\rfloor+1$ \textbf{and} $f(a-1)+d([a,b])>f(b)$}
      \State $f(b)\gets f(a-1)+d([a,b])$; $\mathrm{choice}(b)\gets a$
    \EndIf
  \EndFor
\EndFor
\State $b\gets|T|$; $\mathcal R\gets\varnothing$
\While{$b>0$}
  \If{$\mathrm{choice}(b)=\varnothing$} $b\gets b-1$ \Else\ add $[\mathrm{choice}(b),b]$ to $\mathcal R$; $b\gets\mathrm{choice}(b)-1$ \EndIf
\EndWhile
\State \Return all $d([a,b])$ and the reports $\mathcal R$, each with its bound
\end{algorithmic}
\end{algorithm}

Algorithm~\ref{alg:countscan} computes every interval bound and the reports used in \S\ref{sec:experiments}. The reports maximize $\sum_id(S_i)$ over disjoint qualifying intervals; ties keep the earlier solution, and among intervals ending at the same position the shortest strictly better one is chosen, so the output is unique. An independent implementation of this pseudocode reproduces the reports of the reference implementation on $40$ holdout documents and $40$ synthetic sequences\ifanon{} (anonymized supplementary code)\else{} (test code available from the author on request)\fi.

In the real-data experiments $\bar n$ is the final document length, which is at least $|T|$ and so gives a more conservative threshold.

\paragraph{CPU runtime and memory.}\label{app:runtime}
We froze the resource-measurement plan before running either warmups or timed trials. Synthetic sequences have between \RuntimeMinN{} and \RuntimeMaxN{} retained positions, with every planned length shown in Table~\ref{tab:task26runtime}. The null workload has independent uniform pivots. The mixed workload replaces the central half by Gumbel-max pivots with selected-token probability $\pi_W=\RuntimeSignalPi$. We use C1 with $\lambda=\RuntimeLambda$, truncation level $\RuntimeTau$ and $\alpha=\RuntimeAlpha$. Reporting uses the existing A1 rule with $\eta=\RuntimeEta$ and lengths from $\RuntimeReportMin$ to $\RuntimeReportMax$, capped by the sequence length. Each workload and length has \RuntimeRepeats{} measured repetitions in a fixed randomized order, following an excluded warmup per cell.

Measurements used an OS-scheduled, single-thread process on Apple M2 (macOS-14.8.7-arm64-arm-64bit), with Python 3.12.4, NumPy 2.5.4 and SciPy 1.18.1.

The unchanged \texttt{scan\_check} interface evaluates every interval bound through a Fenwick sweep and retains scalar diagnostics. It streams the computation without storing a matrix of all interval counts. The \texttt{report\_intervals} interface separately ranks and rescans its capped candidate family before scheduling disjoint reports. The sum of these two operation times includes both scans and their Python wrappers. Input generation and calibration are recorded separately. A subprocess timer additionally includes startup, imports and output. The benchmark excludes the additional superadditive and interpolated bounds computed by \texttt{all\_bounds}.

\begin{table}[H]
\centering\footnotesize
\caption{CPU resource measurements over \RuntimeRepeats{} repetitions per cell. Times and baseline HWM are medians [interquartile ranges]; peak RSS is median/maximum. Baseline HWM is the lifetime process high-water mark immediately before scanning. Peak RSS includes the interpreter, imported libraries and inputs as well as both operations, and is not an exclusive allocation measure.}\label{tab:task26runtime}
\resizebox{\linewidth}{!}{\begin{tabular}{llrrrrr}
\toprule
Workload & $n$ & Scan (s) & Report (s) & Sum (s) & Baseline HWM (MiB) & Peak RSS (MiB) \\
\midrule
Mixed & 256 & 0.0010 [0.0010, 0.0012] & 0.0012 [0.0012, 0.0013] & 0.0022 [0.0022, 0.0025] & 55.3 [55.2, 55.6] & 55.4/56.4 \\
Mixed & 512 & 0.0045 [0.0044, 0.0050] & 0.0053 [0.0052, 0.0060] & 0.0098 [0.0096, 0.0110] & 55.3 [55.1, 55.4] & 55.7/56.3 \\
Mixed & 1024 & 0.0201 [0.0196, 0.0233] & 0.0164 [0.0159, 0.0186] & 0.0365 [0.0356, 0.0419] & 55.4 [55.2, 55.8] & 56.5/57.4 \\
Mixed & 2048 & 0.0930 [0.0915, 0.1017] & 0.0351 [0.0348, 0.0373] & 0.1283 [0.1265, 0.1396] & 55.4 [55.3, 55.6] & 57.5/57.9 \\
Mixed & 4096 & 0.4359 [0.4302, 0.4728] & 0.0720 [0.0716, 0.0731] & 0.5087 [0.5022, 0.5482] & 55.4 [55.1, 55.5] & 59.4/60.1 \\
Mixed & 8192 & 1.9104 [1.8957, 2.0351] & 0.1527 [0.1514, 0.1604] & 2.0621 [2.0492, 2.1896] & 55.5 [55.2, 55.6] & 59.5/63.4 \\
Null & 256 & 0.0003 [0.0003, 0.0003] & 0.0004 [0.0004, 0.0005] & 0.0007 [0.0006, 0.0008] & 55.1 [55.1, 55.3] & 55.4/56.6 \\
Null & 512 & 0.0007 [0.0007, 0.0008] & 0.0013 [0.0013, 0.0015] & 0.0020 [0.0019, 0.0022] & 55.2 [54.8, 55.5] & 55.6/56.4 \\
Null & 1024 & 0.0027 [0.0024, 0.0029] & 0.0043 [0.0039, 0.0045] & 0.0070 [0.0063, 0.0075] & 55.2 [55.1, 55.6] & 56.3/57.2 \\
Null & 2048 & 0.0101 [0.0099, 0.0106] & 0.0096 [0.0092, 0.0102] & 0.0197 [0.0191, 0.0206] & 55.2 [55.0, 55.4] & 57.3/58.0 \\
Null & 4096 & 0.0447 [0.0438, 0.0491] & 0.0217 [0.0215, 0.0250] & 0.0671 [0.0655, 0.0764] & 55.5 [55.2, 55.6] & 59.4/61.2 \\
Null & 8192 & 0.2557 [0.2472, 0.2752] & 0.0492 [0.0487, 0.0565] & 0.3045 [0.2976, 0.3308] & 55.3 [55.1, 55.6] & 63.2/63.7 \\
\bottomrule
\end{tabular}
}
\end{table}

\begin{figure}[H]
\centering
\includegraphics[width=\linewidth]{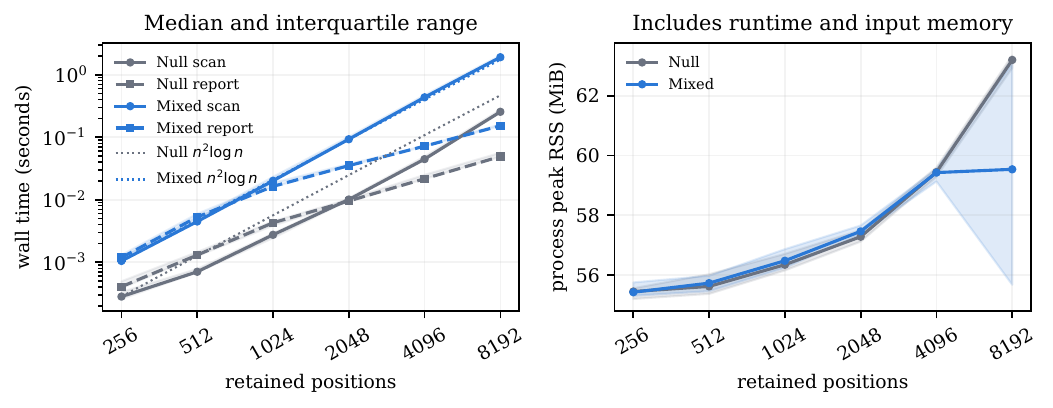}
\caption{CPU wall time and lifetime process peak RSS for the two fixed synthetic workloads. Lines show medians and shading shows interquartile ranges. Dotted time references are proportional to $n^2\log n$ and separately anchored at the smallest length. No full interval-count matrix is stored.}\label{fig:task26runtime}
\end{figure}
Log--log fits to the six median scan times give slopes $\RuntimeNullScanSlope$ (null) and $\RuntimeMixedScanSlope$ (mixed). The corresponding scan-plus-report slopes are $\RuntimeNullPipelineSlope$ and $\RuntimeMixedPipelineSlope$. These fits describe this machine and these workloads. The reference curve in Figure~\ref{fig:task26runtime} is proportional to $n^2\log n$; each workload reference is anchored to its scan median at $n=\RuntimeMinN$.

\section{Scope and Comparisons with Existing Methods}\label{app:related}
This appendix covers the scope statements after Corollary~\ref{cor:select}, and the comparisons with FissionGuard and SeqE-Guard.

\subsection{Scope of the Simultaneous Guarantee}\label{app:scopeguarantee}
Under Assumption~\ref{ass:A}, the null pivots are independent and uniform given $\G$. Under Assumption~\ref{ass:Astar}, the retained null pivots are independent and super-uniform given $\mathcal H$ (Proposition~\ref{lem:superunif}); nonincreasing calibrators then turn super-uniformity into $e$-calibration (Corollary~\ref{cor:mono}). The guarantee holds per document over all intervals; a joint guarantee over several documents audited with one key needs an additional error allocation. The mask keeps contexts that occur once in the full text.

\begin{proposition}[Arbitrary subsets]\label{prop:subset}
For $A\subseteq T$ let $d_{\rm ext}(A)=\max\sum_{i=1}^rd(I_i)$ over $r\ge0$ and pairwise disjoint intervals $I_1,\dots,I_r\subseteq A$, with the empty sum equal to zero. On the event of Theorem~\ref{thm:main}, $d_{\rm ext}(A)\le\delta(A)$ for all $A\subseteq T$ simultaneously, including data-selected $A$.
\end{proposition}
\begin{proof}
On the event, $d(I_i)\le\delta(I_i)$ for each $i$, and $\sum_i\delta(I_i)\le\delta(A)$ because the $I_i$ are disjoint subsets of $A$ and $\delta$ is nonnegative and additive.
\end{proof}
The extension uses the original simultaneous event.

\subsection{Comparison with FissionGuard}\label{app:fgcomparison}

\subsubsection{Synthetic Comparison}

Table~\ref{tab:e9} (Appendix~\ref{app:fgcomparison}) uses documents of $2048$ tokens with one edited $384$-token generated passage, $\alpha=0.1$ and $\eta=0.25$ (\PEnineReps{} documents per cell). Our reporting rule (maximize $\sum_id(S_i)$ over disjoint intervals) was adopted after one earlier run, and our maximum interval length was $512$, against $256$ for FissionGuard. Coverage without edits is \PEnineMidCovA{} versus \PEnineMidCovFG{} at mid entropy, and both methods lose most power once $20\%$ of tokens are edited or the entropy is low.

\begin{table}[H]
\centering
\caption{Synthetic comparison (E9). Cover and purity of reported intervals; ``certified'' is our simultaneous count as a fraction of all watermarked tokens. Largest frequency of any false report: \PEnineAnyFalseA{} (CountScan), \PEnineAnyFalseFG{} (FissionGuard).}\label{tab:e9}\scriptsize
\setlength{\tabcolsep}{3pt}
\begin{tabular}{llcccccc}
\toprule
 & & \multicolumn{2}{c}{FissionGuard} & \multicolumn{3}{c}{Interval scan ($-\log p$, $\lambda=0.5$)} \\
\cmidrule(lr){3-4}\cmidrule(lr){5-7}
entropy & edit rate & cover & purity & cover & purity & certified \\
\midrule
mid & 0.0 & 0.794 & 0.976 & 0.996 & 0.979 & 0.492 \\
mid & 0.1 & 0.100 & 0.788 & 0.657 & 0.657 & 0.271 \\
mid & 0.2 & 0.006 & 0.628 & 0.021 & 0.619 & 0.009 \\
skewed & 0.0 & 0.567 & 0.984 & 0.995 & 0.982 & 0.413 \\
skewed & 0.1 & 0.044 & 0.809 & 0.389 & 0.692 & 0.146 \\
skewed & 0.2 & 0.002 & 0.847 & 0.009 & 0.781 & 0.003 \\
low & 0.0 & 0.009 & 1.000 & 0.148 & 0.994 & 0.038 \\
low & 0.1 & 0.001 & 0.906 & 0.002 & 0.821 & 0.001 \\
low & 0.2 & 0.000 & 0.000 & 0.000 & 0.000 & 0.000 \\
\bottomrule
\end{tabular}

\end{table}

\subsubsection{Holdout Comparison and Registered Statements}

Two registered statements concern this comparison. The first, registered with the original holdout, compared CountScan reports of up to $512$ tokens with FissionGuard on cover. The second is a paired reanalysis of the same documents, frozen after the first results had been seen, so it is not independent evidence: with reports of up to $256$ tokens ($\tau=1.0$, $\alpha=0.05$, $\eta=0.25$, C1, denominator $\sum KH$), CountScan (A2) has larger cover$^\ast$ than FissionGuard (A6), with paired confidence interval above $0$ in each cell. Both cells pass: OPT, \TfOptSOne{} \TfOptSOneCI{}; Qwen, \TfQwenSOne{} \TfQwenSOneCI{} (Table~\ref{tab:main}).

\begin{figure}[H]
\centering
\includegraphics[width=\linewidth]{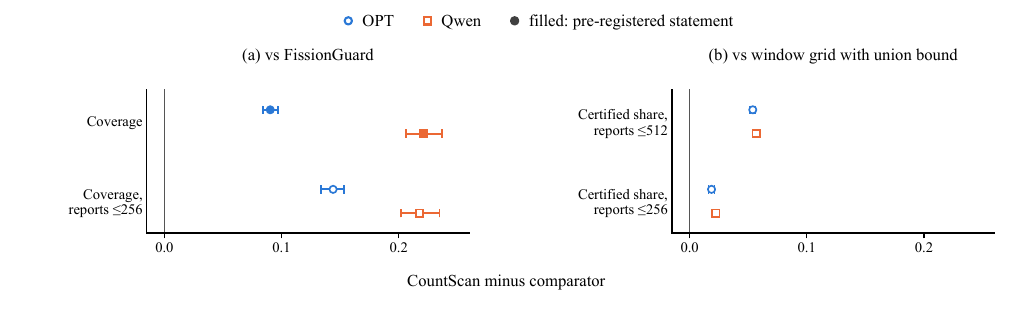}
\caption{Paired differences, CountScan minus comparator, with $95\%$ bootstrap intervals on $500$ unedited mixed documents per model ($\tau=1.0$, C1, $\alpha=0.05$, $\eta=0.25$). Statement~1: cover against FissionGuard (pre-registered). cover$^\ast$, reports $\le256$: CountScan with reports capped at $256$ tokens against FissionGuard. Certified share: CountScan against a window grid with a union bound, with reports of up to $512$ and of up to $256$ tokens (Appendix~\ref{app:further}). Panel (a) gives the FissionGuard comparisons and (b) the window-grid comparisons; intervals narrower than the markers retain their stored widths.}\label{fig:paired}
\end{figure}

\paragraph{Matched paragraph reporting.}
Table~\ref{tab:main} compares CountScan with FissionGuard under a common maximum report length of $256$ tokens. Both methods run on the masked sequence. CountScan reports disjoint intervals with $d(S)\ge\lfloor\eta|S|\rfloor+1$, so each report rejects the partial-conjunction null tested by FissionGuard. We use our re-implementation of FissionGuard from its Appendix~C, checked against its Table~1 (Appendix~\ref{app:real}). The coverage target is cover$^\ast$, the share of $\sum KH$ inside reports. The paired advantage is \TfOptSOne{} for OPT and \TfQwenSOne{} for Qwen; both bootstrap intervals lie above zero (Figure~\ref{fig:paired}). FissionGuard returns purer and more numerous reports, while CountScan covers more of the target tokens.

The first registered statement (reports of up to $512$ tokens, cover, whose denominator also includes history matches removed by masking) holds in both cells (Figure~\ref{fig:paired}, Table~\ref{tab:labels-cover}). Statement~2, that at most $5\%$ of documents contain a false report, also holds: neither method made one, with Wilson upper bound \FgSTwoWilson. The matched-length comparison above retains a positive advantage under the theorem's coverage target. Appendix~\ref{app:labels} explains the effect of masked history matches on cover, and Appendix~\ref{app:further} gives the matched baselines. Figure~\ref{fig:paired} also compares certified shares with a fixed window grid and union-bound threshold.

\begin{table}[H]
\centering
\caption{Comparison on the real holdout ($\tau=1.0$, C1, $\alpha=0.05$, $\eta=0.25$, $500$ documents per cell). $d/1000$ and $q^*$: generated-only documents. Other columns: unedited mixed documents, CountScan with reports of at most $256$ tokens (CS) against FissionGuard (FG), with the paired difference in cover$^\ast$ and its $95\%$ bootstrap interval. The low-temperature settings in the text rows beneath the model comparison fail gate Q0. The original denominator comparison is in Table~\ref{tab:labels-cover}.}\label{tab:main}
\begin{minipage}{.60\linewidth}
\begingroup\scriptsize
\setlength{\tabcolsep}{2.2pt}
\renewcommand{\arraystretch}{1.08}
\begin{tabular}{@{}lcccccc@{}}
\toprule
Model & \shortstack{$d/1000$\\$[95\%\ \mathrm{CI}]$} & $q^*$ & \shortstack{cover$^*$\\CS / FG} & \shortstack{paired $\Delta$\\$[95\%\ \mathrm{CI}]$} & \shortstack{purity\\CS / FG} & \shortstack{false docs\\CS / FG} \\
\midrule
OPT & \shortstack{0.586\\{}[0.568,0.603]} & 0.702 & \shortstack{0.934\\0.790} & \shortstack{0.144\\{}[0.134,0.154]} & \shortstack{0.955\\0.984} & \shortstack{0/500\\0/500} \\
Qwen & \shortstack{0.617\\{}[0.604,0.630]} & 0.691 & \shortstack{0.910\\0.692} & \shortstack{0.218\\{}[0.202,0.235]} & \shortstack{0.957\\0.987} & \shortstack{0/500\\0/500} \\
\midrule
\multicolumn{7}{@{}p{\columnwidth}@{}}{OPT, $\tau=0.7$ (fails Q0): $d/1000=0.038$ [0.032,0.044], $q^*=0.252$; cover$^*$ CS/FG 0.487/0.051; purity 0.980/0.989; false 0/500, 0/500.} \\
\multicolumn{7}{@{}p{\columnwidth}@{}}{Qwen, $\tau=0.7$ (fails Q0): $d/1000=0.082$ [0.073,0.090], $q^*=0.228$; cover$^*$ CS/FG 0.470/0.039; purity 0.982/0.990; false 0/500, 0/500.} \\
\bottomrule
\end{tabular}
\par\vspace{2pt}CS: CountScan with reports of at most $256$ tokens; FG: FissionGuard. Columns 2--3 use the $1000$-token generated-only documents; the others use the unedited mixed documents. Paired CIs are not stored for the low-temperature cells.
\endgroup

\end{minipage}
\end{table}

Table~\ref{tab:labels-cover} records the effect of the coverage denominator. Appendix~\ref{app:further} gives the matched-baseline definitions and all configurations. The certified-share rows of Figure~\ref{fig:paired} compare CountScan with a fixed window grid and a union-bound threshold; they are internal design comparisons, not comparisons with FissionGuard.

\subsection{Same-input comparison with restarted SeqE-Guard}\label{app:backend-review}
At each start, Algorithm~1 of \citet{fischer2025admissible} includes every retained position of that prefix. At a crossing it deletes the largest retained factor. The deleted count is the minimum number of deletions needed to keep every prefix below the threshold, by Theorem~3.1 and the prefix characterization in their equation~(6). Removing the actual historical matches is feasible on $\Omega_\alpha$. The terminal trimming rule minimizes deletions with only the endpoint constraint, so its count is no larger. This applies separately to each start with its allocated budget.

The paired B1 and E3 archive check holds fixed final text, whole-text unique-context mask, historical target, C1 factors and order, threshold $\log(n_{\rm final}/0.05)$ and the same unrestricted disjoint aggregation. This aggregation maximizes the sum over disjoint intervals. It differs from the length-capped paragraph reporting used in the holdout comparison above. B1 has six temperature-one OPT/Qwen groups with \CellDocs{} parents per group. On four edited versions per parent, all \EditedComparisons{} final disjoint counts agree between the backends. Across the six unedited cells, mean gains range from \UneditedGainRange{} matches. Agreement on the final aggregate does not imply equality on all intervals.

\begin{table}[!ht]
\centering
\caption{Matched B1 archive comparison under four token edits. Each column reports the common mean disjoint count for trimming and restarting. The equality count pools four paired edit arms per group. The final column lists restarted document-level overcount events in unedited, random, rare, deletion and insertion arms.}\label{tab:backend-b1}
\small\begin{tabular}{lrrrrll}
\toprule Group & Random & Rare & Delete & Insert & Same/800 & Errors/200 \\ \midrule
OPT 1 & 23.010 & 15.200 & 29.810 & 34.215 & 800/800 & 0/0/0/0/0 \\
OPT 2 & 23.410 & 15.265 & 29.515 & 34.355 & 800/800 & 0/0/0/0/0 \\
OPT 3 & 25.940 & 16.640 & 31.345 & 36.055 & 800/800 & 0/0/0/0/0 \\
Qwen 1 & 20.145 & 13.090 & 28.445 & 31.680 & 800/800 & 0/0/0/0/0 \\
Qwen 2 & 22.040 & 12.380 & 28.695 & 32.140 & 800/800 & 1/1/1/1/1 \\
Qwen 3 & 22.510 & 13.385 & 29.265 & 33.635 & 800/800 & 0/0/0/0/0 \\
\bottomrule
\end{tabular}

\end{table}
\begin{table}[!ht]
\centering
\caption{Matched E3 archive comparison using one B1 Qwen group of \EThreeDocs{} parents. Trim and Restart are mean disjoint counts. Same counts equal document pairs. The gain interval is the archived paired document-bootstrap percentile interval.}\label{tab:backend-e3}
\small\begin{tabular}{lrrrrl}
\toprule Condition & \shortstack{Used nonmatches/\\retained} & Trim & Restart & Same & Gain interval \\ \midrule
Unedited & 3719/186054 & 156.135 & 156.185 & 190/200 & [0.020, 0.080] \\
10\% rewrite & 4491/183930 & 123.375 & 123.380 & 199/200 & [0.000, 0.015] \\
30\% rewrite & 5072/182601 & 96.160 & 96.160 & 200/200 & [0.000, 0.000] \\
\bottomrule
\end{tabular}

\end{table}

The E3 archive reuses one B1 group. It adds two paraphrasing conditions, with the unedited row repeated for reference. Table~\ref{tab:backend-e3} reports the paired difference. The distributed aggregates rebuild the tables and verify the stored equality totals. Independent re-estimation of the bootstrap intervals or document-level errors requires the controlled paired records.

\subsubsection{Same paragraph reporter on the original holdout}\label{app:matched-reporter}
We rerun both backends on the original mixed-document holdout used in Table~\ref{tab:main}. Both rules use the same retained sequence, C1 factors and order, start threshold $\log(n_{\rm final}/0.05)$, and candidate lengths of $32$ to $256$ retained tokens. Both apply the same count requirement $\lfloor |S|/4\rfloor+1$ and maximize the sum of counts over disjoint reports. On ties the dynamic program keeps its previous solution, and among candidates with the same end it considers shorter lengths first. The endpoint backend reproduces the archived counts and report intervals for every document--condition pair. This comparison reuses the holdout and makes no new model or key draws.

In the two unedited temperature-one cells, restarting increases mean certified counts by \MatchedCountRange{} per document and changes pooled cover$^*$ by \MatchedCoverRange{}. The paired document-bootstrap intervals for both coverage differences include zero. It changes the interval list in \MatchedPrimaryReportChanges{} of \MatchedPrimaryDocs{} documents and the reported union in \MatchedPrimaryUnionChanges{}. Across the edited temperature-one variants, the report list changes in \MatchedEditedReportChanges{} of \MatchedEditedDocs{} pairs. Table~\ref{tab:matched-reporter} also gives the unedited low-temperature results. The pointwise count dominance thus translates to small changes in aggregate coverage on this holdout, while some report locations change.

\begin{table}[H]
\centering
\small
\setlength{\tabcolsep}{3.5pt}
\caption{Endpoint trimming (T) and restarted SeqE-Guard (R) with the same reporter. Each row contains $500$ paired documents. Count is the mean sum of certified counts in reported intervals. Cover$^*$ and purity are pooled ratios. The last three columns give document counts for any change in report intervals, any change in their union, and any selected interval with $d>H$. Edited variants reuse the same parent documents within each cell.}
\label{tab:matched-reporter}
\begin{tabular}{@{}llrrrrrr@{}}
\toprule
Cell & Edit & Count T/R & Cover$^*$ T/R & Purity T/R & Reports changed & Union changed & $d>H$ T/R \\
\midrule
OPT 1.0 & Unedited & 148.808/148.828 & 0.934/0.934 & 0.955/0.955 & 24 & 23 & 0/0 \\
Qwen 1.0 & Unedited & 146.784/146.800 & 0.910/0.910 & 0.957/0.958 & 26 & 26 & 0/0 \\
OPT 0.7 & Unedited & 17.888/17.888 & 0.487/0.487 & 0.980/0.980 & 2 & 2 & 0/0 \\
Qwen 0.7 & Unedited & 33.216/33.228 & 0.470/0.470 & 0.982/0.982 & 7 & 7 & 0/0 \\
OPT 1.0 & Random replace & 3.496/3.514 & 0.073/0.073 & 0.533/0.534 & 1 & 1 & 0/0 \\
OPT 1.0 & Rare replace & 1.946/1.946 & 0.045/0.045 & 0.564/0.564 & 0 & 0 & 0/0 \\
OPT 1.0 & Deletion & 9.416/9.416 & 0.186/0.186 & 0.498/0.498 & 0 & 0 & 0/0 \\
OPT 1.0 & Insertion & 5.388/5.388 & 0.086/0.086 & 0.499/0.499 & 1 & 1 & 0/0 \\
Qwen 1.0 & Random replace & 1.806/1.806 & 0.033/0.033 & 0.522/0.522 & 0 & 0 & 0/0 \\
Qwen 1.0 & Rare replace & 0.914/0.914 & 0.019/0.019 & 0.608/0.608 & 0 & 0 & 0/0 \\
Qwen 1.0 & Deletion & 6.558/6.558 & 0.117/0.117 & 0.501/0.501 & 0 & 0 & 0/0 \\
Qwen 1.0 & Insertion & 4.232/4.232 & 0.063/0.063 & 0.512/0.512 & 0 & 0 & 0/0 \\
\bottomrule
\end{tabular}

\end{table}
The candidate-band check, before selecting the reports, finds one overstatement document under each backend, both in the OPT insertion condition. Neither backend overstates on any selected report in this comparison.

\section{Synthetic Experiments}\label{app:synthetic}
This appendix supports Section~\ref{subsec:synthetic} and the synthetic statements in Sections~\ref{sec:theorem}--\ref{sec:price}.

\paragraph{Validity.} Over \PEoneConfigs{} configurations with 2000
\unskip{} documents each (lengths, watermark layouts, $\lambda$, proxy calibrators), the largest frequency of \emph{any} interval exceeding its true count is 0.70
\unskip\% against a nominal $5\%$.

\paragraph{Simulator.} Next-token distributions are Zipf laws over $V=2000$ tokens, $\pi_v\propto v^{-s}$, with exponent $s$ drawn per token from a grid on $[0.9,3.2]$. Watermarked pivots are drawn exactly from the Gumbel-max law given $\pi$, and human pivots are $\Unif(0,1)$. The entropy profiles are ``mid'' ($s\sim\Unif[1.0,2.6]$), ``low'' ($s\sim\Unif[1.8,3.2]$) and ``skewed'' (an assumed mixture: $40\%$ near-deterministic tokens with $s\in[2.6,3.2]$ and $60\%$ with $s\in[1.0,2.0]$). Proxy entropies add Gaussian noise to $s$. Contexts are independent at every position, so masking is not exercised. These results describe the simulator, not real text.

\paragraph{E1: validity.} Frequencies (in percent) of documents in which \emph{some} interval bound exceeds its true count, over 2000
\unskip{} documents per cell; nominal $\alpha=5\%$ (Table~\ref{tab:e1}).

\begin{table}[H]
\centering
\caption{E1: Frequency (\%) of Any Violated Interval Bound. Pure-human documents of length $n$ and mixed documents with watermarked segments, for several $e$-values.}\label{tab:e1}
\scriptsize
\resizebox{\linewidth}{!}{\begin{tabular}{lrrrrrrrr}
\toprule
 & \multicolumn{4}{c}{mid} & \multicolumn{4}{c}{low} \\
\cmidrule(lr){2-5}\cmidrule(lr){6-9}
$e$-value & human $n=300$ & human $n=1000$ & human $n=4000$ & mixed $n=1000$ & human $n=300$ & human $n=1000$ & human $n=4000$ & mixed $n=1000$ \\
\midrule
$\log$, $\lambda=0.05$ & 0.00 & 0.00 & 0.00 & 0.00 & 0.00 & 0.00 & 0.00 & 0.00 \\
$\log$, $\lambda=0.1$ & 0.00 & 0.00 & 0.05 & 0.00 & 0.00 & 0.00 & 0.05 & 0.00 \\
$\log$, $\lambda=0.2$ & 0.00 & 0.05 & 0.10 & 0.00 & 0.00 & 0.05 & 0.05 & 0.00 \\
$\log$, $\lambda=0.5$ & 0.15 & 0.10 & 0.30 & 0.00 & 0.15 & 0.25 & 0.30 & 0.15 \\
proxy-entropy calib., $\lambda=0.5$ & 0.35 & 0.70 & 0.60 & 0.15 & 0.05 & 0.10 & 0.25 & 0.10 \\
true-entropy calib., $\lambda=0.5$ & 0.25 & 0.45 & 0.35 & 0.05 & 0.05 & 0.15 & 0.15 & 0.05 \\
\bottomrule
\end{tabular}
}
\end{table}

\paragraph{E4: convergence to $q^*$.} Table~\ref{tab:e4} reports the mean $d/L$ against the limit $q^*$ of Proposition~\ref{prop:qstar}.

\begin{table}[H]
\centering
\caption{E4: Mean Certified Proportion $d/L$ by Segment Length, against the Theoretical Limit $q^*$.}\label{tab:e4}
\scriptsize
\begin{tabular}{llrrrrrr}
\toprule
entropy & watermark fraction $\kappa$ & $\lambda$ & $L=128$ & $L=512$ & $L=2048$ & $L=8192$ & theory $q^*$ \\
\midrule
mid & 1.0 & 0.05 & 0.079 & 0.329 & 0.475 & 0.541 & 0.569 \\
mid & 1.0 & 0.1 & 0.219 & 0.415 & 0.515 & 0.550 & 0.566 \\
mid & 1.0 & 0.2 & 0.335 & 0.473 & 0.532 & 0.552 & 0.560 \\
mid & 1.0 & 0.5 & 0.413 & 0.498 & 0.523 & 0.529 & 0.534 \\
mid & 1.0 & 0.8 & 0.421 & 0.476 & 0.492 & 0.497 & 0.498 \\
mid & 0.75 & 0.05 & 0.015 & 0.211 & 0.329 & 0.377 & 0.402 \\
mid & 0.75 & 0.1 & 0.117 & 0.281 & 0.357 & 0.385 & 0.397 \\
mid & 0.75 & 0.2 & 0.206 & 0.319 & 0.365 & 0.382 & 0.386 \\
mid & 0.75 & 0.5 & 0.259 & 0.318 & 0.342 & 0.345 & 0.350 \\
mid & 0.75 & 0.8 & 0.229 & 0.277 & 0.289 & 0.295 & 0.296 \\
low & 1.0 & 0.05 & 0.000 & 0.092 & 0.210 & 0.257 & 0.280 \\
low & 1.0 & 0.1 & 0.018 & 0.159 & 0.236 & 0.262 & 0.274 \\
low & 1.0 & 0.2 & 0.087 & 0.199 & 0.243 & 0.256 & 0.263 \\
low & 1.0 & 0.5 & 0.133 & 0.193 & 0.213 & 0.219 & 0.220 \\
low & 1.0 & 0.8 & 0.113 & 0.144 & 0.158 & 0.161 & 0.163 \\
low & 0.75 & 0.05 & 0.000 & 0.040 & 0.136 & 0.178 & 0.196 \\
low & 0.75 & 0.1 & 0.001 & 0.094 & 0.156 & 0.180 & 0.190 \\
low & 0.75 & 0.2 & 0.033 & 0.122 & 0.160 & 0.171 & 0.177 \\
low & 0.75 & 0.5 & 0.065 & 0.106 & 0.125 & 0.128 & 0.130 \\
low & 0.75 & 0.8 & 0.031 & 0.055 & 0.062 & 0.065 & 0.066 \\
\bottomrule
\end{tabular}

\end{table}

\paragraph{E13--E15.} The interpolated bound gave no practical gain (E13, pre-specified, negative). The finite-length target $q_c(L)$ predicts $d/L$ (E14; main text). On synthetic low entropy, no pre-specified design reaches $d/128\ge0.2$ (E15; best \EfifteenLowBest).

\section{Real-Data Experiments}\label{app:realsec}
This appendix gives the designs and the results behind Section~\ref{subsec:real} and Table~\ref{tab:checks}. Configuration-level tables, earlier holdouts, filter settings and registration ledgers are in the reproducibility material.

\subsection{Holdout Design and Results}\label{app:real}
Each cell generates $1000$ new tokens from PG-19 prompts with a new key, watermarks a context at its first occurrence and samples without the key on repeats, which avoids the looping of an earlier holdout. The calibrator C1 uses truncation level $2.5$ and hedge $\lambda=0.5$; since $\int_0^1\min\{-\log p,2.5\}\,dp=1-e^{-2.5}$, $g_{\mathrm{C1}}$ integrates to one and is nonincreasing. Mixed documents place $300$ generated tokens between two blocks of $350$ human tokens and are evaluated before and after random replacement, rare-token replacement, deletion and insertion. All four cells pass the registered validity gates: the largest human-text scan false-positive rate is \TtScanMax{}, and the largest mixed-document FWER is \TtMixFwerMax{}. At $\tau=1.0$ both models pass the generation-quality gate Q0, with whole-document bounds $d/1000$ of \TtOptOneDthousand{} (OPT) and \TtQwenOneDthousand{} (Qwen). At $\tau=0.7$ mean generation-time entropy is only \TtOptSevenH{} and \TtQwenSevenH{} nats, both cells fail Q0, and their results are descriptive, as registered.

\begin{table}[h]
\centering
\caption{Prespecified holdout with C1. Generated-only columns give the mask drop rate, mean entropy $\bar H$, and certified proportions. Mixed-document reports are capped at $256$ kept tokens. Cover$^\ast$ uses retained historical matches as its denominator. Purity is the match fraction within reports, and false counts documents with a false report. Intervals resample documents at fixed key and history. Rows marked $\ddagger$ are descriptive because they miss the prespecified representativeness criterion Q0. See Appendix~\ref{app:real} for the 7B replication and design.}\label{tab:real}
\begingroup\small
\setlength{\tabcolsep}{3.5pt}
\renewcommand{\arraystretch}{1.13}
\begin{tabular}{@{}lcccccccccc@{}}
\toprule
Model & $\tau$ & Q0 & drop & $\bar H$ & $q^*$ & $d/128$ & $d/1000$ [$95\%$ CI] & cover$^\ast$ & purity & false \\
\midrule
OPT & 1.0 & pass & 0.14 & 3.37 & 0.702 & 0.614 & 0.586 [0.568, 0.603] & 0.934 & 0.955 & 0/500 \\
Qwen & 1.0 & pass & 0.08 & 3.51 & 0.691 & 0.622 & 0.617 [0.604, 0.630] & 0.910 & 0.957 & 0/500 \\
OPT & 0.7 & fail$^\ddagger$ & 0.82 & 0.29 & 0.252 & 0.186 & 0.038 [0.032, 0.044] & 0.487 & 0.980 & 0/500 \\
Qwen & 0.7 & fail$^\ddagger$ & 0.60 & 0.57 & 0.228 & 0.259 & 0.082 [0.073, 0.090] & 0.470 & 0.982 & 0/500 \\
Qwen-7B & \BTwoTemperature{} & \BTwoQZero{} & \BTwoDropTwo{} & \BTwoHTwo{} & \BTwoQStar{} & \BTwoDOneTwoEight{} & \BTwoDThousand{} \BTwoDThousandCI{} & \BTwoCSCover{} & \BTwoPurity{} & \BTwoMixedCOneEvents{}/\BTwoMixedDocuments{} \\
\bottomrule
\end{tabular}
\endgroup

\end{table}

\begin{figure}[h]
\centering
\includegraphics[width=\textwidth]{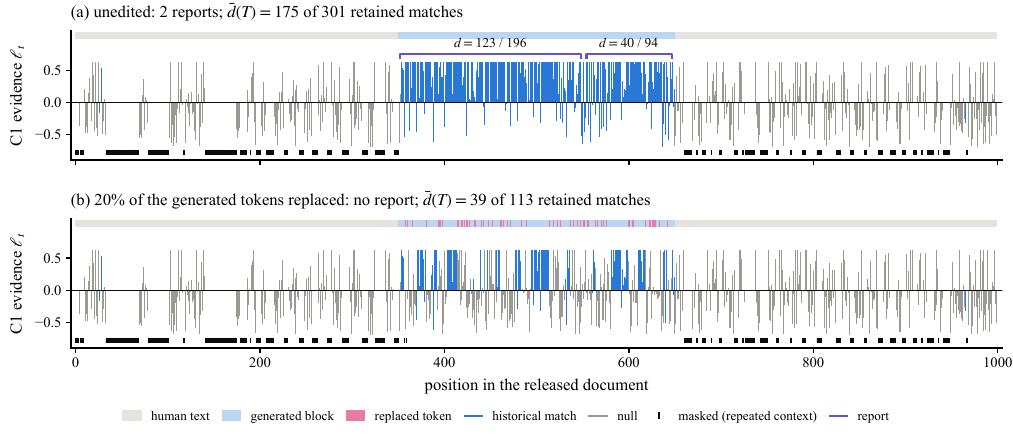}
\caption{The document of Figure~\ref{fig:overview} before (a) and after (b) replacing $20\%$ of the generated tokens, which also breaks pairs of unchanged tokens whose contexts change: retained matches fall from \OvUnMatches{} to \OvEdMatches{} and no interval meets the report rule, although the disjoint-interval bound still certifies $\bar d(T)=\OvEdBard$.}\label{fig:overview-full}
\end{figure}

\paragraph{Labels.}\label{app:labels} Three labels are recorded at each position $t$: $K_t=\ind\{t\in T\}$; the history label $H_t$, whether the context--token pair of $t$ occurs at a watermarked step of the key-wide history; and the origin label $O_t$, whether $t$ is an unedited output of a watermarked step of this document. The guarantees concern $\sum K_tH_t$. Human-only documents contain kept history matches without origin, for example Project Gutenberg boilerplate that the models also generated under the same key; they are correct history matches, not failures of the guarantee. Table~\ref{tab:labels-cover} shows that the sign of the CountScan minus FissionGuard coverage difference does not depend on the label used as denominator.

\begin{table}[H]
\centering
\caption{Coverage in the Statement~1 configuration (unedited mixed documents, $\tau=1.0$, $\alpha=0.05$, $\eta=0.25$, C1) under four denominators, with $95\%$ document-bootstrap intervals. The difference is paired (shared index matrix). The first row of each cell uses the original Statement~1 denominator.}\label{tab:labels-cover}
\scriptsize
\begin{tabular}{llccc}
\toprule
model, temp. & denominator & C1 & FissionGuard & difference \\
\midrule
OPT, 1.0 & $\sum H$ (cover) & 0.540 [0.519, 0.560] & 0.449 [0.431, 0.468] & 0.090 [0.084, 0.097] \\
 & $\sum KH$ (cover$^\ast$) & 0.949 [0.944, 0.954] & 0.790 [0.779, 0.801] & 0.159 [0.149, 0.169] \\
 & $\sum O$ & 0.959 [0.953, 0.964] & 0.795 [0.782, 0.807] & 0.164 [0.154, 0.174] \\
 & $\sum KO$ & 0.981 [0.977, 0.985] & 0.813 [0.802, 0.825] & 0.168 [0.157, 0.178] \\
\midrule
Qwen, 1.0 & $\sum H$ (cover) & 0.859 [0.845, 0.872] & 0.638 [0.617, 0.659] & 0.221 [0.206, 0.237] \\
 & $\sum KH$ (cover$^\ast$) & 0.932 [0.923, 0.939] & 0.692 [0.673, 0.710] & 0.240 [0.224, 0.257] \\
 & $\sum O$ & 0.950 [0.940, 0.958] & 0.702 [0.682, 0.721] & 0.248 [0.231, 0.265] \\
 & $\sum KO$ & 0.973 [0.965, 0.980] & 0.720 [0.700, 0.738] & 0.254 [0.237, 0.272] \\
\bottomrule
\end{tabular}

\end{table}

\subsection{Further Analyses of the Holdout Documents}\label{app:further}
\paragraph{Matched baselines.} All methods run on the masked sequence. A1 is CountScan with reports of $32$ to $512$ tokens and A2 the same bound with reports of at most $256$ tokens, as in FissionGuard. A3 replaces the per-start Ville thresholds by a union bound over all intervals of $32$ to $512$ tokens; A4 uses a grid of half-overlapping windows with a union bound. A6 is FissionGuard, whose count of a report $S$ is $\lfloor\eta|S|\rfloor+1$, implied by its FDR rejection rather than a simultaneous bound. A1 to A3 produced no document with a false report in any configuration, and the fraction of documents in which some interval bound of A1 to A4 exceeded its true count was at most \TfSimKHMax{}.

\paragraph{Reference-family baseline.}\label{app:reffamily} A further baseline, specified and frozen before any of its outputs were computed, asks how much of CountScan's advantage a structured post hoc method with exact interpolation already attains. This paired reanalysis uses the same documents and followed the earlier analyses. On the kept coordinates $0,\dots,n-1$, the regions $R_{L,j}=[jL,(j+1)L-1]$, $L\in\{32,64,128,256,512\}$ up to the cap and $0\le j<\lfloor n/L\rfloor$, form a forest $\mathcal F$: any two are disjoint or nested. Each region gets the single-interval bound $\zeta_R=|R|-d_{c_R}(R)$ on its null positions, with $c_R=\log(|\mathcal F|/\alpha)$ (A8); on the event $E=\{\forall R\in\mathcal F:\ d_{c_R}(R)\le\delta(R)\}$, of probability at least $1-\alpha$ by a union bound, $|R\cap N|\le\zeta_R$ for every region. For any interval $S$ the bounds are interpolated exactly over the forest \citep{durand2018posthocspatial}: bottom-up,
\[
f(R)=\min\Big\{\min\big(\zeta_R,|S\cap R|\big),\ \sum_{C\in\mathrm{ch}(R)}f(C)+\Big|S\cap R\setminus\textstyle\bigcup_{C\in\mathrm{ch}(R)}C\Big|\Big\},
\]
with $V(S)=\sum_{R\ \mathrm{root}}f(R)+|S\setminus\bigcup_{\mathrm{roots}}R|$ and $d_{\mathrm{RF}}(S)=|S|-V(S)$. On $E$, $V(S)\ge|S\cap N|$ for all $S$, hence $d_{\mathrm{RF}}(S)\le\delta(S)$ simultaneously. A8L splits the budget by level, $c_R=\log(k|\mathcal F_L|/\alpha)$ with $\mathcal F_L$ the regions of length $L$ and $k$ the number of nonempty levels within the cap. Reports use the same rule as CountScan. Table~\ref{tab:reffamily-256} lists the methods at cap $256$; the cap-$512$ table is in the reproducibility material.

Both registered statements hold. The paired difference in certified share, CountScan minus A8, is \TrfOptSOne{} \TrfOptSOneCI{} (OPT) and \TrfQwenSOne{} \TrfQwenSOneCI{} (Qwen) with reports of up to $512$ tokens, and \TrfOptSTwo{} \TrfOptSTwoCI{} and \TrfQwenSTwo{} \TrfQwenSTwoCI{} with reports of up to $256$; at that cap the purity of the reports is \TrfOptMainCapShortATwoPurityValue{} and \TrfQwenMainCapShortATwoPurityValue{} for CountScan against \TrfOptMainCapShortAEightPurityValue{} and \TrfQwenMainCapShortAEightPurityValue{} for A8. Interpolation adds little to the grid it refines: A8 minus A4 in certified share is \TrfOptMainCapShortAEightMinusAFourEfficiency{} (OPT) and \TrfQwenMainCapShortAEightMinusAFourEfficiency{} (Qwen) at cap $256$, while cover$^\ast$ rises by \TrfOptMainCapShortAEightMinusAFourCoverage{} and \TrfQwenMainCapShortAEightMinusAFourCoverage{}. The grid alone exceeds FissionGuard in certified share by \TrfOptMainCapShortAFourMinusASixEfficiency{} and \TrfQwenMainCapShortAFourMinusASixEfficiency{}, so a simultaneous count baseline already captures a substantial part of the observed gap to FissionGuard. The ranking changes with the cap: the all-interval Bonferroni bound A3 is below A8 at cap $256$ (\TrfOptMainCapShortAThreeMinusAEightEfficiency{} and \TrfQwenMainCapShortAThreeMinusAEightEfficiency{}) and above it at cap $512$ (\TrfOptMainCapLongAThreeMinusAEightEfficiency{} and \TrfQwenMainCapLongAThreeMinusAEightEfficiency{}), so these paired contrasts do not identify separate causal contributions to the gain. Over the \TrfRfSimConfigs{} configurations of A8 and A8L on unedited mixed, generated-only and human-source documents, the largest fraction of the \TrfRfSimDocs{} documents with some interval bound above its $\sum KH$ was \TrfRfSimMax{}, with Wilson upper limits at most \TrfRfSimUpperMax{}.

\begin{table}[H]
\centering
\caption{Reference-family comparison with reports of at most $256$ tokens: certified share, cover$^\ast$ and purity with $95\%$ document-bootstrap intervals. A3, A4 and A8 use only intervals and regions of $32$ to $256$ tokens.}\label{tab:reffamily-256}
\begin{minipage}{\linewidth}\centering\begingroup\scriptsize
\begin{tabular}{@{}llccc@{}}
\toprule
Model, temperature & Method & Count share$^\dagger$ & cover$^\ast$ & Purity \\
\midrule
OPT, $\tau=1.0$ & A6 FissionGuard, native$^\dagger$ & \TrfOptMainCapShortASixEfficiency & \TrfOptMainCapShortASixCoverage & \TrfOptMainCapShortASixPurity \\
 & A4 window grid & \TrfOptMainCapShortAFourEfficiency & \TrfOptMainCapShortAFourCoverage & \TrfOptMainCapShortAFourPurity \\
 & A8 RF, uniform budget & \TrfOptMainCapShortAEightEfficiency & \TrfOptMainCapShortAEightCoverage & \TrfOptMainCapShortAEightPurity \\
 & A8L RF, level budget (same level) & \TrfOptMainCapShortAEightLayerEfficiency & \TrfOptMainCapShortAEightLayerCoverage & \TrfOptMainCapShortAEightLayerPurity \\
 & A3 all-interval Bonferroni & \TrfOptMainCapShortAThreeEfficiency & \TrfOptMainCapShortAThreeCoverage & \TrfOptMainCapShortAThreePurity \\
 & A2 CountScan & \TrfOptMainCapShortATwoEfficiency & \TrfOptMainCapShortATwoCoverage & \TrfOptMainCapShortATwoPurity \\
\midrule
Qwen, $\tau=1.0$ & A6 FissionGuard, native$^\dagger$ & \TrfQwenMainCapShortASixEfficiency & \TrfQwenMainCapShortASixCoverage & \TrfQwenMainCapShortASixPurity \\
 & A4 window grid & \TrfQwenMainCapShortAFourEfficiency & \TrfQwenMainCapShortAFourCoverage & \TrfQwenMainCapShortAFourPurity \\
 & A8 RF, uniform budget & \TrfQwenMainCapShortAEightEfficiency & \TrfQwenMainCapShortAEightCoverage & \TrfQwenMainCapShortAEightPurity \\
 & A8L RF, level budget (same level) & \TrfQwenMainCapShortAEightLayerEfficiency & \TrfQwenMainCapShortAEightLayerCoverage & \TrfQwenMainCapShortAEightLayerPurity \\
 & A3 all-interval Bonferroni & \TrfQwenMainCapShortAThreeEfficiency & \TrfQwenMainCapShortAThreeCoverage & \TrfQwenMainCapShortAThreePurity \\
 & A2 CountScan & \TrfQwenMainCapShortATwoEfficiency & \TrfQwenMainCapShortATwoCoverage & \TrfQwenMainCapShortATwoPurity \\
\midrule
OPT, $\tau=0.7^{\ddagger}$ & A6 FissionGuard, native$^\dagger$ & \TrfOptSecondaryCapShortASixEfficiency & \TrfOptSecondaryCapShortASixCoverage & \TrfOptSecondaryCapShortASixPurity \\
 & A4 window grid & \TrfOptSecondaryCapShortAFourEfficiency & \TrfOptSecondaryCapShortAFourCoverage & \TrfOptSecondaryCapShortAFourPurity \\
 & A8 RF, uniform budget & \TrfOptSecondaryCapShortAEightEfficiency & \TrfOptSecondaryCapShortAEightCoverage & \TrfOptSecondaryCapShortAEightPurity \\
 & A8L RF, level budget (same level) & \TrfOptSecondaryCapShortAEightLayerEfficiency & \TrfOptSecondaryCapShortAEightLayerCoverage & \TrfOptSecondaryCapShortAEightLayerPurity \\
 & A3 all-interval Bonferroni & \TrfOptSecondaryCapShortAThreeEfficiency & \TrfOptSecondaryCapShortAThreeCoverage & \TrfOptSecondaryCapShortAThreePurity \\
 & A2 CountScan & \TrfOptSecondaryCapShortATwoEfficiency & \TrfOptSecondaryCapShortATwoCoverage & \TrfOptSecondaryCapShortATwoPurity \\
\midrule
Qwen, $\tau=0.7^{\ddagger}$ & A6 FissionGuard, native$^\dagger$ & \TrfQwenSecondaryCapShortASixEfficiency & \TrfQwenSecondaryCapShortASixCoverage & \TrfQwenSecondaryCapShortASixPurity \\
 & A4 window grid & \TrfQwenSecondaryCapShortAFourEfficiency & \TrfQwenSecondaryCapShortAFourCoverage & \TrfQwenSecondaryCapShortAFourPurity \\
 & A8 RF, uniform budget & \TrfQwenSecondaryCapShortAEightEfficiency & \TrfQwenSecondaryCapShortAEightCoverage & \TrfQwenSecondaryCapShortAEightPurity \\
 & A8L RF, level budget (same level) & \TrfQwenSecondaryCapShortAEightLayerEfficiency & \TrfQwenSecondaryCapShortAEightLayerCoverage & \TrfQwenSecondaryCapShortAEightLayerPurity \\
 & A3 all-interval Bonferroni & \TrfQwenSecondaryCapShortAThreeEfficiency & \TrfQwenSecondaryCapShortAThreeCoverage & \TrfQwenSecondaryCapShortAThreePurity \\
 & A2 CountScan & \TrfQwenSecondaryCapShortATwoEfficiency & \TrfQwenSecondaryCapShortATwoCoverage & \TrfQwenSecondaryCapShortATwoPurity \\
\bottomrule
\end{tabular}
\par\smallskip
\begin{tabular}{@{}llccc@{}}
\toprule
Model, temperature & Adjacent step (later minus earlier) & $\Delta$ count share & $\Delta$ cover$^\ast$ & $\Delta$ purity \\
\midrule
OPT, $\tau=1.0$ & A4$-$A6 & \TrfOptMainCapShortAFourMinusASixEfficiency & \TrfOptMainCapShortAFourMinusASixCoverage & \TrfOptMainCapShortAFourMinusASixPurity \\
 & A8$-$A4 & \TrfOptMainCapShortAEightMinusAFourEfficiency & \TrfOptMainCapShortAEightMinusAFourCoverage & \TrfOptMainCapShortAEightMinusAFourPurity \\
 & A3$-$A8 & \TrfOptMainCapShortAThreeMinusAEightEfficiency & \TrfOptMainCapShortAThreeMinusAEightCoverage & \TrfOptMainCapShortAThreeMinusAEightPurity \\
 & A2$-$A3 & \TrfOptMainCapShortATwoMinusAThreeEfficiency & \TrfOptMainCapShortATwoMinusAThreeCoverage & \TrfOptMainCapShortATwoMinusAThreePurity \\
\midrule
Qwen, $\tau=1.0$ & A4$-$A6 & \TrfQwenMainCapShortAFourMinusASixEfficiency & \TrfQwenMainCapShortAFourMinusASixCoverage & \TrfQwenMainCapShortAFourMinusASixPurity \\
 & A8$-$A4 & \TrfQwenMainCapShortAEightMinusAFourEfficiency & \TrfQwenMainCapShortAEightMinusAFourCoverage & \TrfQwenMainCapShortAEightMinusAFourPurity \\
 & A3$-$A8 & \TrfQwenMainCapShortAThreeMinusAEightEfficiency & \TrfQwenMainCapShortAThreeMinusAEightCoverage & \TrfQwenMainCapShortAThreeMinusAEightPurity \\
 & A2$-$A3 & \TrfQwenMainCapShortATwoMinusAThreeEfficiency & \TrfQwenMainCapShortATwoMinusAThreeCoverage & \TrfQwenMainCapShortATwoMinusAThreePurity \\
\midrule
OPT, $\tau=0.7^{\ddagger}$ & A4$-$A6 & \TrfOptSecondaryCapShortAFourMinusASixEfficiency & \TrfOptSecondaryCapShortAFourMinusASixCoverage & \TrfOptSecondaryCapShortAFourMinusASixPurity \\
 & A8$-$A4 & \TrfOptSecondaryCapShortAEightMinusAFourEfficiency & \TrfOptSecondaryCapShortAEightMinusAFourCoverage & \TrfOptSecondaryCapShortAEightMinusAFourPurity \\
 & A3$-$A8 & \TrfOptSecondaryCapShortAThreeMinusAEightEfficiency & \TrfOptSecondaryCapShortAThreeMinusAEightCoverage & \TrfOptSecondaryCapShortAThreeMinusAEightPurity \\
 & A2$-$A3 & \TrfOptSecondaryCapShortATwoMinusAThreeEfficiency & \TrfOptSecondaryCapShortATwoMinusAThreeCoverage & \TrfOptSecondaryCapShortATwoMinusAThreePurity \\
\midrule
Qwen, $\tau=0.7^{\ddagger}$ & A4$-$A6 & \TrfQwenSecondaryCapShortAFourMinusASixEfficiency & \TrfQwenSecondaryCapShortAFourMinusASixCoverage & \TrfQwenSecondaryCapShortAFourMinusASixPurity \\
 & A8$-$A4 & \TrfQwenSecondaryCapShortAEightMinusAFourEfficiency & \TrfQwenSecondaryCapShortAEightMinusAFourCoverage & \TrfQwenSecondaryCapShortAEightMinusAFourPurity \\
 & A3$-$A8 & \TrfQwenSecondaryCapShortAThreeMinusAEightEfficiency & \TrfQwenSecondaryCapShortAThreeMinusAEightCoverage & \TrfQwenSecondaryCapShortAThreeMinusAEightPurity \\
 & A2$-$A3 & \TrfQwenSecondaryCapShortATwoMinusAThreeEfficiency & \TrfQwenSecondaryCapShortATwoMinusAThreeCoverage & \TrfQwenSecondaryCapShortATwoMinusAThreePurity \\
\bottomrule
\end{tabular}
\par\smallskip\noindent
Displayed configuration: D1 unedited, C1, $\alpha=\TrfPrimaryAlpha$, $\eta=\TrfPrimaryEta$. Entries are stored estimates and marginal document-bootstrap intervals; paired differences use the shared document resamples. No joint coverage is claimed for the collection of intervals.
\par\noindent $\dagger$ FG count share uses its native FDR-implied count and the same native FG row in both cap tables. Count methods use simultaneous FWER bounds under the stated assumptions. The quantities do not have the same guarantee.
\par\noindent $\ddagger$ These archived cells fail Q0 and are descriptive. A8L is a same-level budget variant; it is not an extra step in the primary ladder.
\par\noindent -- denotes an undefined aggregate ratio or unavailable interval. Zero-denominator document counts are retained in JSON; nulls are not imputed as zero or one. Values are rounded for display. All registered configurations and negative results remain in the full JSON.
\endgroup
\end{minipage}
\end{table}

\begin{figure*}[ht]
\centering
\includegraphics[width=\textwidth]{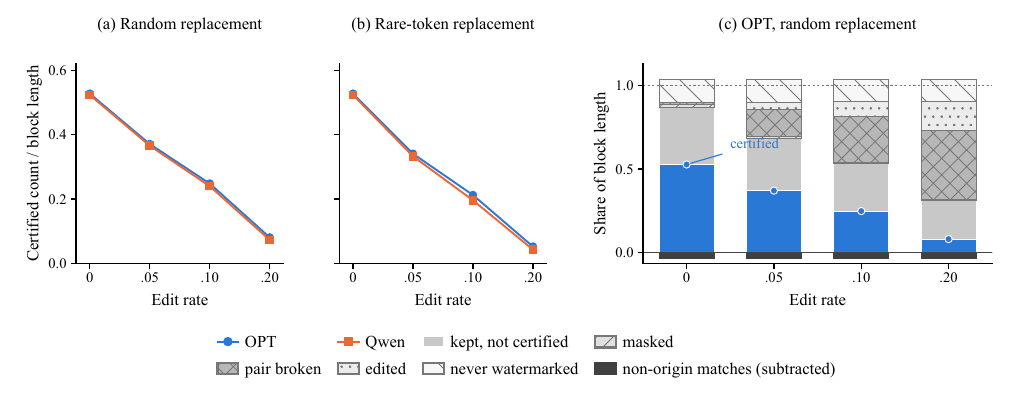}
\caption{CountScan under editing on mixed documents (C1). Panels (a) and (b) show $\bar d(T)$ divided by the original generated-block length of $300$. Panel (c) decomposes the OPT counts from (a). The algebraic decomposition offsets the terms above the dashed line by the non-origin matches below zero. Table~\ref{tab:further-edits} gives coverage and the full breakdown.}\label{fig:edits-main}
\end{figure*}

\paragraph{Edit breakdown.} Figure~\ref{fig:edits-main} and Table~\ref{tab:further-edits} follow the $300$ generated tokens of a mixed document through editing at rates $r\in\{0.05,0.1,0.2\}$ for random and rare-token replacement and at $r=0.2$ for deletion and insertion. Edits at $0.05$ and $0.1$ are new reconstructions with the original editing code and new seeds; the others reuse the holdout edits. $O_0$ counts origin tokens before editing, $O$ those left after editing, and ``$O$ without $H$'' those whose (context, token) pair no longer matches the key-wide history because an edit changed their context. Random replacement leaves \TfPairRatioMin--\TfPairRatioMax{} times as many origin tokens without a history match as it edits (at $\tau=1.0$); insertion edits no origin token but leaves \TfOptInsPairs{} (OPT) and \TfQwenInsPairs{} (Qwen) per document without one. At $r=0.2$ the retained share $OH/O$ is \TfOptRetain{} (OPT) and \TfQwenRetain{} (Qwen) against $(1-r)^4=\TfRetainRef$ for independent edits with context length $m=4$. The loss of power under editing therefore comes mainly from broken pairs, not from the edited tokens themselves.

\begin{table}[!htbp]
\centering
\caption{Edit breakdown, per-document means over $500$ mixed documents ($\alpha=0.05$, $\eta=0.25$). $C$: kept history matches without origin in the whole document. $\bar d(T)$: whole-document bound of A1 with C1. Last column: cover$^\ast$ of A1 (C1) and of FissionGuard. $\ddagger$: fails gate Q0.}\label{tab:further-edits}
\resizebox{\linewidth}{!}{\begin{tabular}{lllrrrrrrrcc}
\toprule
model, temp. & edit & $r$ & $O_0$ & edited & $O$ & $O$ without $H$ & $OH$ & $\sum KH$ & $C$ & $\bar d(T)$ & cover$^\ast$ (A1/A6) \\
\midrule
OPT, 1.0 & unedited & -- & 259.4 & 0.0 & 259.4 & 3.4 & 256.0 & 260.8 & 10.4 & 158.2 & 0.949/0.790 \\
 & E-rand & 0.05 & 259.4 & 13.0 & 246.4 & 48.6 & 197.7 & 204.7 & 11.1 & 111.2 & 0.909/0.569 \\
 & E-rand & 0.10 & 259.4 & 26.0 & 233.4 & 82.6 & 150.8 & 159.3 & 11.4 & 74.4 & 0.796/0.262 \\
 & E-rand & 0.20 & 259.4 & 52.0 & 207.3 & 123.4 & 83.9 & 94.1 & 11.6 & 24.1 & 0.073/0.018 \\
 & E-rare & 0.05 & 259.4 & 14.2 & 245.2 & 51.5 & 193.7 & 199.7 & 10.9 & 102.2 & 0.895/0.504 \\
 & E-rare & 0.10 & 259.4 & 28.4 & 231.0 & 84.3 & 146.7 & 153.6 & 11.3 & 63.7 & 0.736/0.180 \\
 & E-rare & 0.20 & 259.4 & 56.5 & 202.9 & 122.9 & 80.0 & 88.5 & 11.9 & 15.6 & 0.045/0.010 \\
 & E-del & 0.20 & 259.4 & 51.9 & 207.5 & 123.2 & 84.3 & 94.2 & 11.4 & 29.4 & 0.186/0.043 \\
 & E-ins & 0.20 & 259.4 & 0.0 & 259.4 & 154.5 & 104.9 & 114.7 & 11.6 & 34.6 & 0.087/0.028 \\
\midrule
Qwen, 1.0 & unedited & -- & 285.3 & 0.0 & 285.3 & 4.0 & 281.3 & 289.5 & 15.0 & 157.0 & 0.932/0.692 \\
 & E-rand & 0.05 & 285.3 & 14.2 & 271.1 & 53.5 & 217.5 & 228.2 & 15.6 & 109.7 & 0.881/0.425 \\
 & E-rand & 0.10 & 285.3 & 28.5 & 256.8 & 90.8 & 165.9 & 178.4 & 15.9 & 71.9 & 0.691/0.166 \\
 & E-rand & 0.20 & 285.3 & 57.1 & 228.3 & 136.3 & 91.9 & 106.3 & 16.1 & 21.8 & 0.033/0.013 \\
 & E-rare & 0.05 & 285.3 & 14.7 & 270.6 & 54.1 & 216.6 & 226.0 & 15.5 & 99.8 & 0.831/0.331 \\
 & E-rare & 0.10 & 285.3 & 29.4 & 256.0 & 90.4 & 165.5 & 176.1 & 15.9 & 58.8 & 0.530/0.089 \\
 & E-rare & 0.20 & 285.3 & 58.9 & 226.5 & 132.2 & 94.3 & 106.9 & 16.6 & 12.8 & 0.019/0.002 \\
 & E-del & 0.20 & 285.3 & 57.1 & 228.2 & 136.0 & 92.2 & 106.7 & 15.8 & 28.8 & 0.117/0.022 \\
 & E-ins & 0.20 & 285.3 & 0.0 & 285.3 & 169.8 & 115.5 & 129.6 & 15.8 & 32.3 & 0.063/0.013 \\
\midrule
OPT, 0.7$^\ddagger$ & unedited & -- & 115.5 & 0.0 & 115.5 & 3.3 & 112.2 & 111.7 & 15.3 & 22.2 & 0.516/0.051 \\
 & E-rand & 0.05 & 115.5 & 5.8 & 109.7 & 22.8 & 86.9 & 92.4 & 17.2 & 13.7 & 0.172/0.009 \\
 & E-rand & 0.10 & 115.5 & 11.6 & 103.9 & 37.4 & 66.4 & 76.0 & 18.1 & 7.4 & 0.012/0.002 \\
 & E-rand & 0.20 & 115.5 & 23.3 & 92.2 & 55.3 & 36.9 & 51.1 & 18.4 & 1.1 & 0.000/0.000 \\
 & E-rare & 0.05 & 115.5 & 10.1 & 105.4 & 28.7 & 76.7 & 83.2 & 17.1 & 10.1 & 0.099/0.003 \\
 & E-rare & 0.10 & 115.5 & 18.5 & 96.9 & 41.5 & 55.4 & 65.7 & 17.9 & 3.5 & 0.000/0.000 \\
 & E-rare & 0.20 & 115.5 & 33.2 & 82.2 & 52.5 & 29.7 & 44.3 & 18.9 & 0.1 & 0.000/0.000 \\
 & E-del & 0.20 & 115.5 & 23.2 & 92.2 & 54.8 & 37.5 & 51.1 & 17.6 & 2.0 & 0.000/0.000 \\
 & E-ins & 0.20 & 115.5 & 0.0 & 115.5 & 69.5 & 45.9 & 58.7 & 17.8 & 1.9 & 0.000/0.000 \\
\midrule
Qwen, 0.7$^\ddagger$ & unedited & -- & 222.6 & 0.0 & 222.6 & 4.0 & 218.7 & 220.0 & 27.3 & 43.2 & 0.494/0.039 \\
 & E-rand & 0.05 & 222.6 & 11.1 & 211.6 & 41.5 & 170.1 & 179.9 & 29.4 & 26.6 & 0.162/0.008 \\
 & E-rand & 0.10 & 222.6 & 22.2 & 200.5 & 70.2 & 130.3 & 147.0 & 30.7 & 14.0 & 0.017/0.001 \\
 & E-rand & 0.20 & 222.6 & 44.4 & 178.2 & 104.7 & 73.5 & 98.4 & 31.7 & 2.1 & 0.000/0.000 \\
 & E-rare & 0.05 & 222.6 & 13.4 & 209.3 & 48.2 & 161.1 & 171.6 & 30.3 & 19.1 & 0.093/0.002 \\
 & E-rare & 0.10 & 222.6 & 26.6 & 196.1 & 76.0 & 120.0 & 137.5 & 32.5 & 6.5 & 0.004/0.000 \\
 & E-rare & 0.20 & 222.6 & 52.3 & 170.3 & 102.4 & 67.9 & 93.9 & 34.7 & 0.6 & 0.009/0.001 \\
 & E-del & 0.20 & 222.6 & 44.5 & 178.1 & 105.8 & 72.3 & 97.5 & 30.9 & 3.7 & 0.000/0.000 \\
 & E-ins & 0.20 & 222.6 & 0.0 & 222.6 & 133.1 & 89.6 & 113.0 & 30.7 & 4.1 & 0.001/0.000 \\
\bottomrule
\end{tabular}
}
\end{table}

\subsubsection{Selection Demonstration}\label{app:seldemo}
This post hoc demonstration reuses the $\tau=1.0$ holdout documents and adds synthetic sequences; it is not an independent replication. The plan, code and fixed-interval lookups were frozen before the pivots were read. A fixed-interval bound N1 trims with the single-interval threshold $c_0=\log(1/\alpha)$; it is valid for one prespecified interval. Applied to the interval of $32$ to $256$ kept tokens with the largest N1 count, it overclaims, $d>\delta_H$, on \SelDemoSelOpt{} (OPT) and \SelDemoSelQwen{} (Qwen) of human-source documents. These documents still contain retained history matches, \SelDemoHumanMatchOpt{} and \SelDemoHumanMatchQwen{} per document on average, and overclaiming is judged against $\delta_H$. On prespecified random intervals N1 overclaims on at most \SelDemoFixedMax{} of documents, and CountScan's all-interval overclaim rate is at most \SelDemoCSMax{} on these document types. On mixed documents the highest-count interval lies in the watermarked block and rarely overclaims, but some candidate interval overclaims on \SelDemoMixAnyOpt{} (OPT) and \SelDemoMixAnyQwen{} (Qwen) of documents; the risk arises when weak regions are selected. In synthetic null sequences of length $1000$ the selected N1 interval overclaims in \SelDemoSynNullSel{} of runs, while CountScan's all-interval rate stays at most \SelDemoSynCSMax{} across lengths. On the prespecified $300$-token generated block, CountScan certifies on average \SelDemoCostOpt{} (OPT) and \SelDemoCostQwen{} (Qwen) tokens fewer than N1. Full tables are in the \ifanon anonymized supplementary data\else repository folder code available from the author on request\fi.

\subsection{Assumption Audit}\label{app:audit}
The audit (AUD-1, \AudNdocs{} documents, pre-specified) uses independent contexts at every position, $n=1000$ and the offline threshold. Key-blind content-dependent edits (S1, S2) keep validity for the nonincreasing calibrators $-\log p$ and C1: the largest FWER is \AudMonoFwerMax{}, and $\max_p(\hat F(p)-p)\le\AudMonoEcdfMax$ at edited positions. A detector-querying editor (S3) gives FWER \AudOracleFwerMin{}, and a fraction \AudOracleDnMin--\AudOracleDnMax{} of pure human text is certified. The increasing control $g=2\cdot\ind\{p>1/2\}$ has FWER \AudGthreeFwerMax{} with interleaved edits. In the post hoc block-edit check AUD-1b, it reaches \AudBHundredGthreetwo{} ($\lambda=0.2$) and \AudBHundredGthreefive{} ($\lambda=0.5$) with blocks of $100$ tokens, and \AudBFourHundredGthreefive{} with blocks of $400$; $-\log p$ stays at \AudBGoneMax.

\subsection{History-Match Audits and Text-Only Filtering}\label{app:historyaudit}
Within the reports of the two main-temperature cells (unedited A1, C1, $\alpha=0.05$, $\eta=0.25$), non-origin matches account for at most \TaskAuditReportCMax{} of the history matches, and \TaskAuditOriginMax{} a report exceeding its retained origin-match count. These source comparisons are descriptive; The full record and Appendix~\ref{app:historyaudit} give the decomposition and paired filter results.

\subsubsection{History Matches and Provenance}\label{subsec:history-origin}
Write $\delta_H(S)=\delta(S)$ for the original retained history-match target. Let $O_t$ indicate that position $t$ is an unedited output of a watermarked step of the document being audited. The history label $H_t$ instead records whether the final context--token pair occurs at any watermarked step in the key-wide generation history. Thus, without a probabilistic assumption,
\begin{equation}\label{eq:history-origin-decomp}
\begin{split}
\delta_H(S)&=\underbrace{\sum_{t\in S}K_tH_tO_t}_{\text{retained origin matches}}+C(S),\\
C(S)&=\sum_{t\in S}K_tH_t(1-O_t).
\end{split}
\end{equation}
The second term counts retained non-origin matches. They can arise from copying or public templates as well as coincidence. The identity separates historical association from provenance; $d(S)$ bounds the former. It does not by itself establish authorship, plagiarism or use of an AI system.

\begin{remark}[An origin bound requires another simultaneous bound]\label{rem:origin-bound}
Suppose an additional procedure provides $U_C(S)$ with
\[
\PP\bigl(\forall S:\ C(S)\le U_C(S)\bigr)\ge1-\alpha'.
\]
On its intersection with the original simultaneous validity event, which has probability at least $1-\alpha-\alpha'$, every interval satisfies
\[
\sum_{t\in S}O_t\ \ge\ \sum_{t\in S}K_tH_tO_t
\ \ge\ \max\{0,d(S)-U_C(S)\}.
\]
Indeed, \eqref{eq:history-origin-decomp} gives $d(S)\le\sum_{t\in S}K_tH_tO_t+C(S)$ on the original event; subtract the additional upper bound and use nonnegativity. A union bound controls the two failure events, without requiring independence. Such a $U_C$ needs an additional text model, contamination budget or deployment structure, and is not supplied here. An empirical mean of $C$ is not a simultaneous upper bound. Document- or session-specific keys narrow the relevant generation history and are a possible deployment choice; the present results do not quantify that choice.
\end{remark}

\subsection{Replication: New Keys, a Larger Model and News-Like Text}\label{app:newkeylarger}\label{app:etwo}
Three registered replications change one factor at a time. B1 uses \BOneNumKeys{} sampled new keys each for OPT-1.3B and Qwen2.5-1.5B and disjoint books; the paired cover$^\ast$ advantage of CountScan over FissionGuard has a positive lower bootstrap endpoint in all \BOneNumCells{} key/model cells, while the registered equivalence of the whole-document bound with the original key was not confirmed (the difference intervals leave the band $\pm\BOneEquivalenceMargin$). B2 evaluates Qwen2.5-7B base with one new key and new books. E2 changes the domain to C4 \texttt{realnewslike} prompts with one new key per model and $300$ formal documents per cell. All cells pass the quality and validity gates, and no document has a false report under either method. Each replication describes its key, books and accumulated history; comparisons across them do not identify the effect of model size or domain.

\begin{table}[h]
\centering
\caption{Replications with C1, $\alpha=0.05$, reports capped at $256$ retained positions. $d/1000$: mean whole-document bound on generated-only documents; cover$^\ast$ and purity on unedited mixed documents; Diff: CountScan minus FissionGuard (FG) cover$^\ast$ with $95\%$ document-bootstrap interval at a fixed key. A dash marks a quantity not recorded for that study.}\label{tab:etwo}
\small
\begin{tabular}{@{}lccccc@{}}
\toprule
Study & $d/1000$ & CS cover$^\ast$ & FG cover$^\ast$ & Diff & Purity CS/FG\\
\midrule
B2, Qwen2.5-7B, PG-19 & \BTwoDThousand & \BTwoCSCover & \BTwoFGCover & \BTwoDiff{} \BTwoDiffCI & --\\
E2, OPT-1.3B, C4 & \EtwoOptDThousand & \EtwoOptCoverCS & \EtwoOptCoverFG & \EtwoOptDiff{} \EtwoOptDiffCI & \EtwoOptPurityCS/\EtwoOptPurityFG\\
E2, Qwen2.5-1.5B, C4 & \EtwoQwenDThousand & \EtwoQwenCoverCS & \EtwoQwenCoverFG & \EtwoQwenDiff{} \EtwoQwenDiffCI & \EtwoQwenPurityCS/\EtwoQwenPurityFG\\
\bottomrule
\end{tabular}
\end{table}

After random replacement of $20\%$ of the tokens, coverage collapses on C4 as on PG-19: for OPT, cover$^\ast$ is \EtwoOptCoverRandCS{} for CountScan and \EtwoOptCoverRandFG{} for FissionGuard.

\subsection{Paraphrase Editing by a Key-Blind Instruction Model}\label{app:ethree}
\paragraph{Purpose.} The editing results above use synthetic edits (replacement, deletion, insertion), which are key-independent by construction. This registered experiment asks whether the validity conditions hold when a real language model paraphrases watermarked text without access to the key, and how much power remains. Validity is the primary question; power was registered as secondary, with the explicit possibility that it would be close to zero.

\paragraph{Design.} The documents are the \EtDocs{} mixed documents of one sampled new key in the Qwen2.5-1.5B, $\tau=1.0$ cell of Appendix~\ref{app:newkeylarger}, reused rather than newly generated, so the comparisons are paired. In the generated block of each document a frozen rule splits the text into sentences and selects about $30\%$ of them, with the $10\%$ arm a nested subset; the selected sentences cover on average \EtTenTokenFrac{} and \EtThirtyTokenFrac{} of the generated tokens (over the \EtThirtyTokenFracDocs{} documents where the share is resolved). Each selected sentence is rewritten once by Qwen2.5-1.5B-Instruct (fixed revision) with a fixed paraphrasing instruction, greedy decoding, FP32 and at most $256$ new tokens. Every output is used as returned: of the \EtThirtyOutputs{} rewrites, \EtThirtyUnchanged{} are unchanged, \EtThirtyNewline{} contain line breaks and \EtThirtyTruncated{} reach the token limit. Masks, history labels and pivots are recomputed on the complete edited document with the frozen key-wide history, and CountScan and FissionGuard run on the same retained pivots with the configuration of Section~\ref{subsec:real}.

\paragraph{Key blindness and registration.} The editor ran on a separate machine that never held the key, received only the plaintext of the selected sentences and was offline while rewriting. Data, sentence rule, editor, prompt, decoding and all statistics were fixed in a hashed pre-registration before any edit was generated. Null positions are retained positions with $H_t=0$ under the original key-wide history, on the edited document.

\paragraph{Observed calibration under paraphrasing.} In no arm and for none of the three registered calibrators (G1 with $\lambda=0.2$ and $0.5$, and C1) did any of the \EtDocs{} documents contain a retained interval with $d(S)>\sum_{t\in S}H_t$: \EtMaxEvents{} events throughout, with Wilson $95\%$ upper limits of at most \EtWilsonUpperMax. Neither CountScan nor FissionGuard emitted a false report. For the empirical null-pivot diagnostic, the largest excess $\max_p(\widehat F(p)-p)$ on \EtTenNullN{} and \EtThirtyNullN{} deduplicated null pairs is \EtTenExcess{} and \EtThirtyExcess{} at the two rates, against \EtZeroExcess{} without editing. These diagnostics report the observed departures for this editor and these documents.

\paragraph{Coverage under paraphrasing.} Rewriting removes history matches, since rewritten sentences produce new (context, token) pairs; the mean whole-document bound falls from \EtZeroBard{} to \EtTenBard{} and \EtThirtyBard{}. Relative to the history matches that remain, however, CountScan's cover$^\ast$ declines only from \EtZeroCsCover{} to \EtTenCsCover{} and \EtThirtyCsCover{}, a paired drop of \EtTenCsDrop{} \EtTenCsDropCI{} and \EtThirtyCsDrop{} \EtThirtyCsDropCI{}, while FissionGuard's declines from \EtZeroFgCover{} to \EtTenFgCover{} and \EtThirtyFgCover{}, a drop of \EtTenFgDrop{} \EtTenFgDropCI{} and \EtThirtyFgDrop{} \EtThirtyFgDropCI{}. The paired CountScan$-$FissionGuard difference in cover$^\ast$ therefore widens with the editing rate, from \EtZeroDiff{} \EtZeroDiffCI{} to \EtTenDiff{} \EtTenDiffCI{} and \EtThirtyDiff{} \EtThirtyDiffCI{} (Table~\ref{tab:ethree}). The two methods target different quantities, and cover$^\ast$ is not a certified share.

\begin{table}[h]
\centering
\caption{Paraphrase editing by a key-blind instruction model (\EtDocs{} paired documents, C1, $\alpha=0.05$, $\eta=0.25$). Errors: documents with some interval bound above its $\sum KH$, largest over the three registered calibrators. Excess: $\max_p(\widehat F(p)-p)$ on deduplicated null pairs. cover$^\ast$ with $95\%$ paired document-bootstrap intervals ($5000$ resamples); $\bar d(T)$: mean whole-document bound.}\label{tab:ethree}
\small
\begin{tabular}{@{}lccccc@{}}
\toprule
Rewritten & Errors & Excess & CountScan cover$^\ast$ & FissionGuard cover$^\ast$ & $\bar d(T)$\\
\midrule
none & $\EtMaxEvents/\EtDocs$ & \EtZeroExcess & \EtZeroCsCover{} \EtZeroCsCoverCI & \EtZeroFgCover{} \EtZeroFgCoverCI & \EtZeroBard\\
$10\%$ & $\EtMaxEvents/\EtDocs$ & \EtTenExcess & \EtTenCsCover{} \EtTenCsCoverCI & \EtTenFgCover{} \EtTenFgCoverCI & \EtTenBard\\
$30\%$ & $\EtMaxEvents/\EtDocs$ & \EtThirtyExcess & \EtThirtyCsCover{} \EtThirtyCsCoverCI & \EtThirtyFgCover{} \EtThirtyFgCoverCI & \EtThirtyBard\\
\bottomrule
\end{tabular}
\end{table}

\paragraph{Scope.} After paraphrasing, the original tokens cannot always be aligned to the rewritten text, so the registered counts of copied and changed tokens are undetermined in \EtTenFTwoUnknown{} and \EtThirtyFTwoUnknown{} documents; no average over the resolved subset is substituted. One editor, from the same model family as the watermarked model, one key and one cell are used.

\section{Simulated Writing Policies}\label{app:hhexp}

\subsection{Design}
OPT-1.3B and Qwen2.5-1.5B use Gumbel-max watermarking at $\tau=1.0$ with the context width of the main experiments and a new key per model. Each model receives \HXDocs{} passages from PG-19 books not used in earlier experiments: a $100$-token prompt and, as the draft $x_{1:300}$, the next $300$ human tokens of the book. The audited region $S$ consists of the retained positions of the final text that correspond to the $300$ draft positions under the unique-context mask; it is fixed before generation.

At draft position $i$ the current context $\chi_i$ is set by the released prefix, and a request returns the Gumbel-max winner $W_i$. A position is requested with probability $q$. If $W_i=x_i$, the draft token is released and is a match. Otherwise up to $R$ retries are made, each with a fixed system-prefix variant that changes the next-token distribution but lies outside the context window, so it leaves the key row unchanged; if a retry returns $x_i$, the draft token is released as a match. If no request returned $x_i$, the writer releases the first winner $W_i$ with probability $a$ (a match) and the draft token with probability $1-a$ (not a match). Unrequested positions release $x_i$. Request and adoption decisions use key-independent seeds. All requests, retries included, enter a controlled log used only to compute $H=\delta(S)$; every audit statistic reads only the final text and the key. Table~\ref{tab:hhexp} lists the six policies with the main results.

\begin{table}[h]
\centering
\caption{Policies and main results; cells give OPT/Qwen means in percent. $q$: request probability; $a$: adoption probability of a disagreeing suggestion; $R$: retries. $\kappa=H/n_S$; Detect: Gamma test, $\alpha=0.05$; $d_{\mathrm{CS}}$: CountScan on the region; $d_{\mathrm{tail}}$: tail count of Proposition~\ref{sw:tail}. Empty N0 regions count as non-detections and are excluded from ratios.}\label{tab:hhexp}
\small
\begin{tabular}{@{}lccccccc@{}}
\toprule
Policy & $q$ & $a$ & $R$ & $\kappa$ & Detect & $d_{\mathrm{CS}}/H$ & $d_{\mathrm{tail}}/H$\\
\midrule
N0 & $1$ & $1$ & $0$ & 100/100 & 76/93 & 57/54 & 62/59 \\
N1 & $1$ & $\tfrac12$ & $0$ & 58/55 & 98/100 & 48/56 & 54/63 \\
N2 & $1$ & $0$ & $0$ & 40/40 & 6/7 & 0/0 & 1/0 \\
N3 & $1$ & $\tfrac12$ & $3$ & 59/56 & 98/100 & 46/55 & 52/62 \\
N4 & $1$ & $\tfrac14$ & $0$ & 45/43 & 86/92 & 10/14 & 18/25 \\
N5 & $\tfrac12$ & $\tfrac12$ & $0$ & 32/31 & 80/90 & 14/18 & 24/32 \\
\bottomrule
\end{tabular}

\end{table}

Two detection rules and five counts were applied, all at $\alpha=0.05$: D1, the Gamma certificate $X\le F^{-1}_{\Gam(n_S,1)}(\alpha)$ of Proposition~\ref{prop:gamma-power}; D2, the fixed-region $e$-value test $\sum_{t\in S}\ell_t\ge\log(1/\alpha)$ with calibrator C1; C-a, CountScan on the fixed region with threshold $\log(1/\alpha)$; C-b, the whole-document scan at threshold $\log(|T|/\alpha)$ restricted to $S$; C-c, the Gamma count of Proposition~\ref{sw:gamma}; C-d, the tail count of Proposition~\ref{sw:tail} with $r=1/2$; and C-e, the all-interval union bound of Appendix~\ref{app:reffamily}. Uncertainty uses \HXReps{} document-bootstrap resamples at fixed key and history, and overcount frequencies use Wilson $95\%$ upper endpoints.

The design, the books, the seeds and four statements were committed before generation:
(S1) every overcount frequency has Wilson upper endpoint at most $0.10$;
(S2) in N1, N3, N4 and N5, D1 detects at least $80\%$ of documents while the mean $d/H$ of C-a is at most $0.5$;
(S2$'$) in N2, D1 and D2 detect at most $\alpha$ plus three binomial standard errors, \HXStwoPrimeThr;
(S3) the paired mean $d/H$ of C-a is lower under N3 than under N1.

\subsection{Results}
\begin{figure}[h]
\centering
\includegraphics[width=\textwidth]{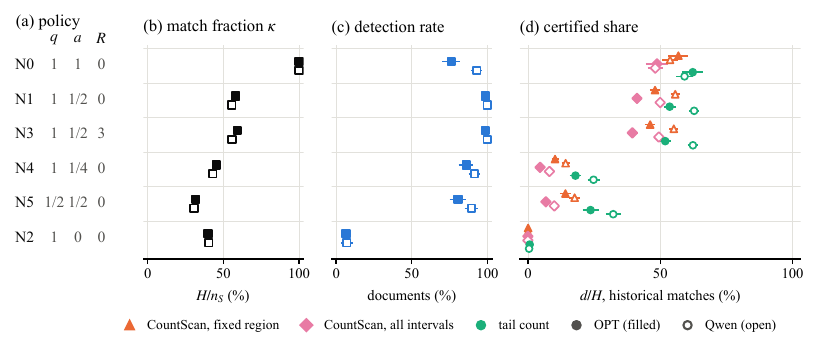}
\caption{Simulated writing policies, \HXDocs{} documents per model (filled: OPT; open: Qwen), ordered by how much of the draft the writer keeps. (a) Request probability $q$, adoption probability $a$ of a suggestion that differs from the draft, and retries $R$. (b) Match fraction $\kappa=H/n_S$, where $H$ is the number of historical matches among the $n_S$ audited positions. (c) Share of documents in which the Gamma test of Proposition~\ref{prop:gamma-power} detects at $\alpha=0.05$. (d) Mean per-document share $d/H$ certified by CountScan on the fixed region (threshold $\log(1/\alpha)$), by the whole-document CountScan scan restricted to the region (threshold $\log(|T|/\alpha)$, simultaneous over all intervals) and by the tail count of Proposition~\ref{sw:tail}. Lines: $95\%$ document-bootstrap intervals. Under N0, \HXOptEmptyMask{} OPT and \HXQwenEmptyMask{} Qwen documents have empty regions; they count as non-detections and are excluded from (b) and (d). }\label{fig:hhexp-full}
\end{figure}

Table~\ref{tab:hhexp-full} gives every cell. S1 holds in all \HXSoneCells{} cells, with largest Wilson upper endpoint \HXSoneWilsonMax. S2 holds in \HXStwoPass{} of \HXStwoTotal{} cells; it fails for Qwen under N1 and N3, where C-a certifies \HXQwenNoneCa{} \HXQwenNoneCaCI{} and \HXQwenNthreeCa{} \HXQwenNthreeCaCI{} of $H$. S2$'$ holds for both rules and both models. For S3 the paired difference N3 minus N1 is \HXOptSthreeDiff{} percentage points for OPT, with interval \HXOptSthreeCI, and \HXQwenSthreeDiff{} for Qwen, with interval \HXQwenSthreeCI{} containing zero. The difference measures the whole retry protocol, not retries in isolation.

Two diagnostics bear on the theory. First, the single-winner results assume token probabilities at most $s$; among N2 matches, \HXOptNtwoCapShare{} (OPT) and \HXQwenNtwoCapShare{} (Qwen) lie in contexts whose largest next-token probability exceeds $1/2$, so the runs do not satisfy the cap with $s=1/2$ and do not test Theorem~\ref{thm:sw-sharp}. Second, retries produce contexts with several distinct winners: under N3, \HXOptNthreeMultiWinner{} retained OPT contexts and \HXQwenNthreeMultiWinner{} Qwen contexts have at least two, the situation of Theorem~\ref{thm:full-key}. Relative to the three levels of Section~\ref{sec:history-protocol}, N2 is the writer of Proposition~\ref{prop:draft-only}, and S2$'$ tests its prediction that detection stays at the level; the identity $\EE[H\mid R]=\sum_t\pi_t(x_t)$ is not checked, because the runs did not record the model probability of the draft token at every position. N1, N4 and N5 request each context at most once and are nearly single-winner histories (at most \HXMultiWinnerNoRetryMax{} of retained contexts have two winners, from other documents under the same key), but the cap fails and $H\ge h$ is not enforced, and the converse policy of Theorem~\ref{thm:sw-sharp}(ii) is not simulated, so these runs do not test the sharpness of $\rho$. Under N3, \HXMultiWinnerRetryMin--\HXMultiWinnerRetryMax{} of retained contexts have two or more winners, but N3 is not the near-miss retry policy of Theorem~\ref{thm:full-key}, which is an existence result over histories and does not predict the average effect of natural retries.

\begin{table}[h]
\centering
\caption{All cells of the simulated writing policies, means in percent. $\kappa=H/n_S$; D1, D2: detection rates; C-a to C-e: mean $d/H$; Wilson: the largest Wilson $95\%$ upper endpoint of the overcount frequency over the five counts; Empty: documents with an empty audited region.}\label{tab:hhexp-full}
\small
\begin{tabular}{@{}llrrrrrrrrrr@{}}
\toprule
Model & Policy & $\kappa$ & D1 & D2 & C-a & C-b & C-c & C-d & C-e & Wilson & Empty\\
\midrule
OPT & N0 & 100.0 & 76.0 & 73.5 & 56.8 & 48.7 & 65.5 & 62.3 & 43.4 & 1.88 & 41 \\
OPT & N1 & 58.2 & 98.5 & 99.5 & 48.0 & 41.2 & 53.1 & 53.6 & 35.4 & 1.88 & 0 \\
OPT & N2 & 39.8 & 6.5 & 1.0 & 0.1 & 0.0 & 0.9 & 0.7 & 0.0 & 1.88 & 0 \\
OPT & N3 & 59.1 & 98.5 & 99.5 & 46.2 & 39.5 & 51.2 & 51.9 & 33.8 & 1.88 & 0 \\
OPT & N4 & 45.2 & 86.0 & 74.5 & 10.2 & 4.6 & 18.1 & 18.0 & 1.8 & 1.88 & 0 \\
OPT & N5 & 31.6 & 80.5 & 67.0 & 14.2 & 6.8 & 24.8 & 23.7 & 3.3 & 1.88 & 0 \\
\midrule
Qwen & N0 & 100.0 & 93.0 & 91.5 & 53.7 & 48.1 & 58.4 & 59.2 & 43.8 & 1.88 & 4 \\
Qwen & N1 & 55.4 & 100.0 & 100.0 & 55.7 & 50.0 & 61.6 & 62.8 & 44.9 & 2.78 & 0 \\
Qwen & N2 & 40.2 & 7.0 & 0.0 & 0.0 & 0.0 & 0.5 & 0.4 & 0.0 & 1.88 & 0 \\
Qwen & N3 & 55.7 & 100.0 & 100.0 & 55.1 & 49.5 & 61.2 & 62.3 & 44.5 & 2.78 & 0 \\
Qwen & N4 & 42.9 & 91.5 & 86.5 & 14.2 & 8.1 & 25.4 & 24.8 & 4.2 & 1.88 & 0 \\
Qwen & N5 & 30.6 & 89.5 & 81.5 & 17.6 & 10.0 & 31.1 & 32.2 & 5.1 & 1.88 & 0 \\
\bottomrule
\end{tabular}

\end{table}

\paragraph{Deviation.} Under N0 some documents fall into repeated contexts, and the unique-context mask then leaves no retained position in $S$: \HXOptEmptyMask{} OPT and \HXQwenEmptyMask{} Qwen documents, none under the other policies. The pilot did not contain such a document. The formal run stopped at the first one, and the rule used here was added after the formal run had started: an empty region receives count zero and no detection, stays in the denominators of detection and overcount frequencies, and is excluded from $\kappa$ and the count ratios, which are then averaged over \HXOptNzeroDefined{} OPT and \HXQwenNzeroDefined{} Qwen documents. The rule cannot create an overcount, and it lowers the N0 detection rates.

\subsection{Why retained drafts carry weak evidence}
\begin{lemma}[Evidence at a coincidental match]\label{lem:coincidence}
Consider a context row not read by any earlier request, with next-token distribution $\pi$ and the coordinates $U_v$ i.i.d.\ $\Unif(0,1)$ as in (K1). Let $W$ be the Gumbel-max winner and $x$ a token with $\pi_x>0$. Then $\PP(W=x)=\pi_x$ and, conditional on $W=x$, the variable $-\log U_x$ has the law of $\pi_x Z$ with $Z\sim\mathrm{Exp}(1)$; in particular $\PP(U_x\le u\mid W=x)=u^{1/\pi_x}$ and $\EE[-\log U_x\mid W=x]=\pi_x$.
\end{lemma}
\begin{proof}
For $\pi_v>0$ put $T_v=-\log U_v/\pi_v$, independent exponential variables with rates $\pi_v$, so that $W$ is the token with the smallest $T_v$; tokens with $\pi_v=0$ do not compete. With $T_*=\min_vT_v$ and $t\ge0$,
\[
\PP(W=x,\,T_*>t)=\int_t^\infty\pi_xe^{-\pi_xz}\prod_{v\ne x}e^{-\pi_vz}\,dz=\pi_xe^{-t},
\]
because $\sum_v\pi_v=1$. Hence $\PP(W=x)=\pi_x$ and $T_*\mid\{W=x\}\sim\mathrm{Exp}(1)$. On $\{W=x\}$ we have $-\log U_x=\pi_xT_x=\pi_xT_*$, which gives the claims.
\end{proof}
In a fresh row read at a token chosen independently of that row, $-\log U$ is $\mathrm{Exp}(1)$; this is the reference law, not the law at every nonmatch position, which after editing is only super-uniform. Against it, a match at a token with probability $\pi_x$ scales the evidence by the factor $\pi_x$. A writer who keeps the draft creates a match exactly when the winner equals the draft token, which by the lemma happens with probability $\pi_x$; such matches concentrate on likely tokens, whose evidence is close to null. The mean model probability of the released token at matches is \HXOptNtwoPiMatched{} (OPT) and \HXQwenNtwoPiMatched{} (Qwen) under N2, against \HXOptNzeroPiMatched{} and \HXQwenNzeroPiMatched{} under N0. Proposition~\ref{prop:draft-only} gives the sharper statement that with draft retention alone no valid test exceeds its level.

\section{A Green-List Diagnostic}\label{app:greenlist}
This diagnostic asks whether the behaviour of CountScan depends on the continuous Gumbel-max pivot; it does not claim a validity theorem for green-list watermarking. A soft green list \citep{kirchenbauer2023watermark} with green fraction $\gamma=0.25$ and logit shift $2.0$ is keyed by the same context width, with a new key per model and new books; sampling uses key-independent randomness. With $g_t=\ind\{w_t\in G(\chi_t)\}$, two evidence rules were fixed in advance: E-GL, $e_t=1-\lambda+\lambda g_t/\gamma$ with $\lambda=1/2$, an $e$-value whenever a null position is green with probability at most $\gamma$; and E-RP, the randomized pivot $P_t=\gamma V_t$ for green and $\gamma+(1-\gamma)V_t$ for red tokens, with $V_t$ uniform and key-independent, followed by C1. Both run CountScan at threshold $\log(|T|/0.05)$ with the report rule of Table~\ref{tab:real}. Condition G1 uses unedited mixed documents of $350$ human, $300$ generated and $350$ human tokens; G2 replaces $20\%$ of the generated block at random; G3 keeps the draft as in N2.

Two of the three registered statements are tests: the null green rate has Wilson upper endpoint at most $0.30$ in each of \GLToneCells{} null categories (T1), and every overcount frequency has Wilson upper endpoint at most $0.10$ in all \GLTtwoCells{} cells (T2); both hold, with largest overcount endpoint \GLWilsonMax. The third (T3) is descriptive: under G1, E-GL certifies \GLOptGoneGL{} (OPT) and \GLQwenGoneGL{} (Qwen) of $H$, below the Gumbel-max values \GLOptGoneHist{} and \GLQwenGoneHist{} of the earlier unedited runs. Those runs used other documents, keys and thresholds, so the comparison is not paired. Random replacement lowers the green-list counts to at most \GLGtwoMax{} of $H$, and draft retention leaves at most \GLGthreeMax, as for Gumbel-max.

\begin{table}[h]
\centering
\caption{Green-list diagnostic, \HXDocs{} documents per model and condition. $\bar d/H$: mean whole-document count over retained historical matches; cover$^\ast$ and purity as in Table~\ref{tab:real}; overcount: documents with any interval count above its truth. A dash marks cells without reports.}\label{tab:greenlist}
\small
\begin{tabular}{@{}lllrrrr@{}}
\toprule
Model & Condition & Evidence & $\bar d/H$ & cover$^\ast$ & purity & overcount\\
\midrule
OPT & G1 & E-GL & 28.6 & 81.9 & 99.4 & 1/200 \\
OPT & G1 & E-RP & 25.2 & 63.5 & 98.5 & 2/200 \\
OPT & G2 & E-GL & 2.7 & 0.3 & 80.0 & 1/200 \\
OPT & G2 & E-RP & 3.4 & 0.0 & -- & 0/200 \\
OPT & G3 & E-GL & 0.0 & 0.0 & -- & 0/200 \\
OPT & G3 & E-RP & 0.0 & 0.0 & -- & 0/200 \\
\midrule
Qwen & G1 & E-GL & 27.6 & 76.3 & 99.3 & 1/200 \\
Qwen & G1 & E-RP & 25.0 & 57.1 & 99.2 & 1/200 \\
Qwen & G2 & E-GL & 2.5 & 0.3 & 91.1 & 2/200 \\
Qwen & G2 & E-RP & 3.6 & 0.0 & -- & 1/200 \\
Qwen & G3 & E-GL & 0.0 & 0.0 & -- & 0/200 \\
Qwen & G3 & E-RP & 0.0 & 0.0 & -- & 0/200 \\
\bottomrule
\end{tabular}

\end{table}

\end{document}